\documentclass[prd,onecolumn,showpacs,floatfix,superscriptaddress,10pt,a4]{revtex4}%
\usepackage{graphicx}
\usepackage{placeins}
\usepackage{subfigure}
\usepackage{epsf}
\usepackage{bm}
\usepackage{amsmath}
\usepackage{amsthm}
\theoremstyle{definition}
\newtheorem{definition}{Definition}
\theoremstyle{plain}
\newtheorem{theorem}{Theorem}
\theoremstyle{remark}

\theoremstyle{plain}
\newtheorem{corollary}{Corollary}
\usepackage{tasks}
\usepackage{amsfonts}
\usepackage{amssymb}
\usepackage{graphicx}
\usepackage{tabularx}
\usepackage{color}
\usepackage{tikz,xcolor,hyperref}%
\providecommand{\U}[1]{\protect\rule{.1in}{.1in}}

\newcommand{\be}{\begin{equation}}
\newcommand{\ee}{\end{equation}}

\newcommand{\mincir}{\raise
-3.truept\hbox{\rlap{\hbox{$\sim$}}\raise4.truept\hbox{$<$}\ }}
\newcommand{\magcir}{\raise
-3.truept\hbox{\rlap{\hbox{$\sim$}}\raise4.truept\hbox{$>$}\ }}

\providecommand{\U}[1]{\protect\rule{.1in}{.1in}}
\definecolor{lime}{HTML}{A6CE39}
\DeclareRobustCommand{\orcidicon}{	\begin{tikzpicture}
	\draw[lime, fill=lime] (0,0)
	circle [radius=0.16]
	node[white] {{\fontfamily{qag}\selectfont \tiny ID}};
	\draw[white, fill=white] (-0.0625,0.095)
	circle [radius=0.007];
	\end{tikzpicture}
	\hspace{-2mm}
}
\foreach \x in {A, ..., Z}{
	\expandafter\xdef\csname orcid\x\endcsname{\noexpand\href{https://orcid.org/\csname orcidauthor\x\endcsname}{\noexpand\orcidicon}}
}

\begin{document}
\title{Dynamics of a higher-dimensional Einstein-Scalar-Gauss-Bonnet cosmology}
\author{Alfredo D. Millano\orcidC{}}
\email{alfredo.millano@ce.ucn.cl}
\affiliation{Departamento de Matem\'{a}ticas, Universidad Cat\'{o}lica del Norte, Avenida.
Angamos 0610, Casilla 1280 Antofagasta, Chile}
\author{Claudio Michea\orcidD{}}
\email{claudio.ramirez@ce.ucn.cl}
\address{Departamento de F\'{i}sica, Universidad Cat\'{o}lica del Norte, Avenida
Angamos 0610, Casilla 1280 Antofagasta, Chile}
\author{Genly Leon\orcidA{}}
\email{genly.leon@ucn.cl}
\affiliation{Departamento de Matem\'{a}ticas, Universidad Cat\'{o}lica del Norte, Avenida.
Angamos 0610, Casilla 1280 Antofagasta, Chile}
\affiliation{Institute of Systems Science, Durban University of Technology, PO Box 1334,
Durban 4000, South Africa}
\author{Andronikos Paliathanasis\orcidB{}}
\email{anpaliat@phys.uoa.gr}
\affiliation{Institute of Systems Science, Durban University of Technology, PO Box 1334,
Durban 4000, South Africa}
\affiliation{School for Data Science and Computational Thinking, Stellenbosch University, 44 Banghoek Rd, Stellenbosch 7600, South Africa}
\affiliation{Departamento de Matem\'{a}ticas, Universidad Cat\'{o}lica del Norte, Avenida.
Angamos 0610, Casilla 1280 Antofagasta, Chile}

\begin{abstract}
We study the dynamics of the field equations in a five-dimensional spatially flat Friedmann-Lemaître-Robertson-Walker metric in the context of a Gauss-Bonnet-Scalar field theory where the quintessence scalar field is coupled to the Gauss-Bonnet scalar. Contrary to the four-dimensional Gauss-Bonnet theory, where the Gauss-Bonnet term does not contribute to the field equations, in this five-dimensional Einstein-Scalar-Gauss-Bonnet model, the Gauss-Bonnet term contributes to the field equations even when the coupling function is a constant. Additionally, we consider a more general coupling described by a power-law function. For the scalar field potential, we consider the exponential function. For each choice of the coupling function, we define a set of dimensionless variables and write the field equations into a system of ordinary differential equations. We perform a detailed analysis of the dynamics for both systems and classify the stability of the equilibrium points. We determine the presence of scaling and super-collapsing solutions using the cosmological deceleration parameter. This means that our models can explain the Universe’s early and late-time acceleration phases. Consequently,
this model can be used to study inflation or as a dark energy candidate.
\end{abstract}
\date{\today}
\maketitle

\tableofcontents
\newpage
\section{Introduction}
\label{sect-1}

The Universe has been observed to be undergoing accelerated expansion in recent cosmological history. The simplest explanation for this is the cosmological constant. However, the issue and the possibility of a dynamic nature have led to two main approaches to its description. The first approach maintains general relativity and introduces the concept of dark energy, which can account for all forms of new, exotic sectors that can be sources of acceleration \cite{Copeland:2006wr, Cai:2009zp, Sharma:2021ayk, Sharma:2021fou}. The second step is to attribute the new degrees of freedom to modifications of the gravitational interaction \cite{CANTATA:2021ktz, Capozziello:2011et, Nojiri:2010wj}, namely to extended theories with general relativity as a limit but generally have a richer structure.
Additionally, there is evidence that most of the Universe's matter content is in Cold Dark Matter (CDM) \cite{Liddle:1993fq, Planck:2018vyg, Abdalla:2022yfr}. Although most cosmologists believe that dark matter should correspond to some particle beyond the Standard Model, the fact that it has not been directly detected in the accelerators led to the investigation of many models in which dark matter can have, partially or entirely, gravitational origin \cite{Nojiri:2006gh, Famaey:2011kh, Sebastiani:2016ras, Addazi:2021xuf}. Modified theories of gravity may arise by extending the Einstein-Hilbert action in a suitable way, such as in $F(R)$ \cite{DeFelice:2010aj, KumarSharma:2022qdf} and $F(G)$ \cite{Nojiri:2005jg, DeFelice:2008wz} gravity, in Lovelock construction \cite{lov1, lov2, lov3, Deruelle:1989fj}, in Horndeski gravity \cite{Horndeski:1974wa}, in generalized galileon theories \cite{DeFelice:2010nf, Deffayet:2011gz}, etc.
Nevertheless, one can construct gravitational modifications starting from  the 
equivalent  torsional
formulation of gravity \cite{Pereira,Maluf:2013gaa} and build theories such as 
$F(T)$ gravity
\cite{Cai:2015emx,Ferraro:2006jd,Linder:2010py}, $F(T,T_{G})$ gravity
\cite{Kofinas:2014owa}, $F(T,B)$ gravity \cite{Bahamonde:2015zma}, etc.
In this framework, one can also introduce scalar fields, i.e. constructing   
scalar-torsion theories \cite{Geng:2011aj}, allowing for non-minimal  
\cite{Geng:2011aj,Geng:2011ka,Gonzalez-Espinoza:2020jss, Paliathanasis:2021nqa, 
Gonzalez-Espinoza:2021qnv,Toporensky:2021poc} or derivative  
 couplings with torsion \cite{Kofinas:2015hla} or more general constructions 
\cite{Geng:2012vtr,Skugoreva:2014ena,Jarv:2015odu,Skugoreva:2016bck,
Hohmann:2018rwf,Hohmann:2018vle,Hohmann:2018ijr,Hohmann:2018dqh,
Emtsova:2019qsl}, then can even be the teleparallel version of Horndeski 
theories 
\cite{Bahamonde:2019shr, Bahamonde:2020cfv, Bahamonde:2021dqn, Bernardo:2021izq}, or allow for a non-minimal scalar-torsion coupling and with boundary term and fermion-torsion coupling. In \cite{Kucukakca:2013mya, Kucukakca:2014vja, Gecim:2017hmn} was performed a detailed study of the teleparallel dark energy in the light of Noether point symmetries. Recently, in \cite{MohseniSadjadi:2015yex, MohseniSadjadi:2016ukp}, the onset of cosmic acceleration from a matter-dominated era ending to a de Sitter phase was studied in the framework of coupled quintessence-torsion model in teleparallel gravity. Late-time acceleration driven by shift-symmetric Galileon in the presence of torsion was investigated in \cite{Banerjee:2018yyi}. Moreover, scalar fields or modified gravity models can be used at galactic scales for the dark matter explanation  \cite{Sharma:2019yix, Sharma:2020vex}.

While four-dimensional spacetime accurately describes the universe, studying theories in higher dimensions is beneficial for several reasons. Spacetime becomes more mathematically tractable when using theories with more than four dimensions ($N>4$), making calculations easier. The field equations for apparent vacuum in five dimensions can be shown to yield Einstein's field equations in 4D with matter because what is commonly referred to as matter can be seen as the result of the embedding in a five-dimensional flat manifold \cite{Wesson:2014raa, Wesson}. Five-dimensional field theory directly extends the four-dimensional spacetime in Einstein's General Relativity. It is considered the minimum dimension in higher-dimensional theories, including ten-dimensional supersymmetry, eleven-dimensional supergravity, and string theory \cite{Wesson}. Other theories in five dimensions include Induced-matter theory, a specific case of Kaluza-Klein theory \cite{Wesson-Ponce, Castillo-Felisola:2016kpe}, and Membrane theory \cite{Randall:1999vf, Arkani-Hamed:1998sfv}, where authors explore a 3-brane in a five-dimensional manifold. Black holes have also been studied in the context of a five-dimensional Einstein-Gauss-Bonnet cosmology \cite{Ghosh:2016ddh}.  The natural generalisation of General Relativity in higher-dimensional spaces is Lovelock's theory of gravity \cite{lov1,lov2,lov3, Deruelle:1989fj}. Lovelock's gravity is a second-order theory that reduces to General Relativity in the case of four-dimensional spacetime.

The Gauss-Bonnet theory is part of Lovelock's gravity, where the Gauss-Bonnet scalar modifies the Einstein-Hilbert Action. The Gauss-Bonnet scalar is a topological invariant in four dimensions, making the theory equivalent to General Relativity \cite{Armaleo:2017lgr}. However, the four-dimensional Einstein-scalar-Gauss-Bonnet gravity has been previously introduced \cite{Nojiri:2005vv, Nojiri:2006je, TerenteDiaz:2023iqk, 
TerenteDiaz:2023kgc, 
Sharma:2023vme, Cognola:2006sp, Nojiri:2007te, Padmanabhan:2013xyr, Chakraborty:2018scm, Fomin:2018typ}, where the scalar field interacts with the Gauss-Bonnet scalar through a non-constant function. In this theory, the mass of the scalar field depends on the Gauss-Bonnet scalar \cite{Kanti:2015pda,Hikmawan:2015rze,Motaharfar:2016dqt,Rashidi:2020wwg}, leading to a non-zero contribution of the Gauss-Bonnet scalar in the gravitational theory. For it to contribute significantly to the four-dimensional model, in \cite{Millano:2023czt,Millano:2023gkt}, the authors considered various coupling functions between the Gauss-Bonnet scalar and the scalar field. It was found that in four dimensions, equilibrium points describe zero acceleration and de Sitter solutions. 

In the present work, we extend the previous studies to a five-dimensional geometry, where the Gauss-Bonnet scalar is not a topological invariant, contributing to the field equations even for a constant coupling function. This work considers a constant and linear coupling function between the Gauss-Bonnet scalar and a quintessence scalar field with its exponential potential.

Dynamical system methods have been widely used in cosmology to obtain relevant information from systems of ordinary differential equations that describe the early and late-time evolution of the models \cite{Leon:2020lnr, Millano:2023vny, Leon:2023idl, Millano:2024rog}. In particular, as mentioned before, in references \cite{Millano:2023czt, Millano:2023gkt}, the authors studied the asymptotic behaviour and dynamics of four-dimensional Einstein-Gauss-Bonnet cosmologies with and without matter with different coupling between the Gauss-Bonnet invariant and the scalar field. 

In cosmology, the deceleration parameter, denoted as $q$, indicates whether an equilibrium point represents an accelerated solution. When combined with the knowledge that an equilibrium point could act as a model's early or late-time attractor, we can determine the potential evolution of the cosmological model from early to late times. A value $q>0$ indicates an eventual collapsing universe, while a value $q<0$ signifies an accelerated universe. Specifically, when $q=-1$, it represents a de Sitter solution. Our five-dimensional model can exhibit both of these behaviours and in addition, it can have equilibrium points that represent scaling solutions, where $q=-1+\frac{3\lambda^2}{4}$. 

This study is self-contained and organized as follows: In section \ref{sect-2}, we briefly review the primary tools used to analyze the dynamical systems that appear throughout this work. 
In section \ref{sect-3}, we present the Action Integral for the five-dimensional Gauss-Bonnet model and derive the corresponding field equations. In section \ref{sect-3-0}, we investigate the dynamics of the five-dimensional cosmological model in a vacuum without the contribution of the Gauss-Bonnet invariant using the standard $H$-normalisation. In section \ref{sect-3-1}
we select a constant coupling function $f$ since, in 5D, the Gauss-Bonnet term is not a topological invariant and contributes to the field equations. The dynamical system analysis for the model with constant coupling function is performed in section \ref{sect-3-1-1}. We also perform numerical integration of the two-dimensional dynamical system and present a possible late-time evolution for the model and the behaviour of the deceleration parameter. Section \ref{app-1} presents an alternative dynamical systems formulation for this constant coupling function model with illustrative topological properties. On the other hand, section \ref{sect-3-2} considers a more general coupling function; that is, $f$ is a linear function. The dynamical system analysis for this choice of coupling function is performed in section \ref{sect-3-2-1}. 

A projection in two dimensions is presented in section \ref{app-2}, where we perform an approximated numerical analysis of the stability of the projected equilibrium points. At the end of this section, one possible late-time evolution of the model is presented using numerical integration and the deceleration parameter. Finally, in section \ref{sect-4}, we present the conclusions of our work and compare our results to other studies in Gauss-Bonnet cosmology.

\section{Dynamical systems techniques}
\label{sect-2}

In this section, we briefly review the standard dynamical system tools that will be used throughout this work. Consider a system of nonlinear ordinary differential equations given by
\begin{equation}
\label{auto}
    \dot{\mathbf{x}}=f(\mathbf{x}),
\end{equation}
where $\mathbf{x}=(x_1,x_2,\ldots, x_n)$ and $\dot{\mathbf{x}}=\frac{d\mathbf{x}}{dt}=\left(\frac{d x_1}{dt},\frac{d x_2}{dt},\ldots \frac{d x_n}{dt}\right)$.  We consider $\textbf{f}=(f_1,f_2,\ldots, f_n)$ to be of class $C^1(E)$ and $E$ is an open subset of $\mathbb{R}^n$.  We observe that system \eqref{auto} is \textit{autonomous} in the sense that there is no explicit dependence on time as opposed to systems like $\dot{\mathbf{x}}=g(\mathbf{x},t)$.  Now we present the following definitions
\begin{definition}
\label{def-1}
    An \textbf{equilibrium point} of system \eqref{auto} is any point $\mathbf{x^*}\in \mathbb{R}^n$ such that $\mathbf{f}(\mathbf{x^*})=0$. 
\end{definition}
\begin{definition}
\label{def-2}
    The \textbf{linearization matrix} of system \eqref{auto} also called Jacobian matrix is \begin{equation}
        \label{Jacobian-1}\mathbf{J}=D\mathbf{f}=\begin{bmatrix}
        \frac{\partial f_1}{\partial x_1}& \frac{\partial f_1}{\partial x_2}& \cdots & \frac{\partial f_1}{\partial x_n}\\ \\
        \frac{\partial f_2}{\partial x_1}& \frac{\partial f_2}{\partial x_2}& \cdots & \frac{\partial f_2}{\partial x_n}\\
        \vdots & \vdots & \ddots& \vdots\\
        \frac{\partial f_n}{\partial x_1}& \frac{\partial f_n}{\partial x_2}& \cdots & \frac{\partial f_n}{\partial x_n}
    \end{bmatrix}
    \end{equation}
\end{definition}
\begin{definition}
\label{def-3}
    An equilibrium point $\mathbf{x^*}$ is called \textbf{hyperbolic} if none of the eigenvalues of $\mathbf{J}(\mathbf{x^*})$ have zero real part. It is called \textbf{non-hyperbolic} if $\mathbf{J}(\mathbf{x^*})$ has eigenvalues with zero real part.
\end{definition}
\begin{definition}
    \label{def-4}
    A hyperbolic equilibrium point $\mathbf{x^*}$ of system \eqref{auto}
    is called \begin{enumerate}
        \item an \textbf{attractor} (stable) if all the eigenvalues of $\mathbf{J}(\mathbf{x^*})$ have negative real part,
        \item a \textbf{source} (unstable) if all the eigenvalues of $\mathbf{J}(\mathbf{x^*})$ have positive real part,
        \item a \textbf{saddle} (unstable) if $\mathbf{J}(\mathbf{x^*})$ has at least one eigenvalue with a positive real part and at least one with a negative real part.
    \end{enumerate}
\end{definition}
\begin{definition}
    \label{def-5}
    A set of non-isolated equilibrium points is \textbf{normally hyperbolic} if the only eigenvalues with zero real part are those whose corresponding eigenvectors are tangent to the set. The sign of the real part of the remaining nonzero eigenvalues determines the stability of the set (or family).
\end{definition}
Definitions \ref{def-1}-\ref{def-4} are taken from \cite{Perko} and definition \ref{def-5} is from \cite{ColeyBook}.
The following result establishes the main tool that will be used to study the dynamical systems in this work.
\begin{theorem}[Hartman-Grobman]\cite{ColeyBook}
\label{teo-1}
    The qualitative structure of the nonlinear system \eqref{auto} near a hyperbolic equilibrium point $\mathbf{x^*}$ is locally topologically equivalent to that of the linearized system \begin{equation}
        \dot{\mathbf{x}}=\mathbf{A}\mathbf{x}
    \end{equation}
    where $\mathbf{A}$ is the $n$ by $n$ matrix $\mathbf{J}(\mathbf{x^*})$. 
\end{theorem}
In what follows, we formulate dynamical systems equivalent to \eqref{auto} that originate from cosmological Gauss-Bonnet models with different coupling functions. Then, we perform a detailed analysis following the results mentioned in this section. 

\section{Higher-dimensions Einstein-Scalar-Gauss-Bonnet field cosmology}
\label{sect-3}
We introduce the five-dimensional spatially flat \textbf{Friedmann-Lemaître-Robertson-Walker} (FLRW) metric tensor $g_{\mu
\nu}$ and line element%
\begin{equation}\label{linelement}
    ds^{2}=-N\left(  t\right)  ^{2}dt^{2}+a\left(  t\right)^{2}  \left(
dr^{2}+r^{2}\left(  d\theta^{2}+\sin^{2}\theta d\varphi^{2}\right)
+dw^{2}\right)  .
\end{equation}
The gravitational Action Integral for the Einstein-Scalar-Gauss-Bonnet theory
of gravity defined as
\begin{equation}
S=\int d^{5}x\sqrt{-g}\left(  \frac{R}{2}-\frac{1}{2}g_{\mu\nu}%
\phi^{;\mu}\phi^{;\nu}-V\left(  \phi\right)  -f\left(  \phi\right)  G\right)
\label{ai.01}%
\end{equation}
where $R$ is the Ricci scalar of the metric tensor $g_{\mu\nu}$, $\phi$ is the
scalar field, $V\left(  \phi\right)  $ the scalar field potential and $G$ is
the Gauss-Bonnet term is expressed as
\begin{equation}
G=R^{2}-4R_{\mu\nu}R^{\mu\nu}+R_{\mu\nu\kappa\lambda}R^{\mu\nu\kappa\lambda}.
\end{equation}
The function $f\left(  \phi\right)  $ is the coupling function between the scalar
field and the Gauss-Bonnet term. 
For the line element \eqref{linelement}, the Ricci scalar and the Gauss-Bonnet terms take the form
\begin{equation}\label{ricciscal}
R=8\frac{\ddot{a}}{aN^{2}}-8\frac{\dot{a}\dot{N}}{aN^{3}}+\frac{12}{N^{2}%
}\left(  \frac{\dot{a}}{a}\right)  ^{2},
\end{equation}%
and
\begin{equation}\label{gaussbonnet}
    G=\frac{24}{N^{2}}\left(  \frac{\dot{a}}{a}\right)  ^{2}\left(  4\left(
\frac{\ddot{a}}{aN^{2}}-\frac{\dot{a}\dot{N}}{aN^{3}}\right)  +\frac{1}{N^{2}%
}\left(  \frac{\dot{a}}{a}\right)  ^{2}\right),  
\end{equation}
or equivalently%
\begin{equation}
    R=4\left(  2\dot{H}+5H^{2}\right)\qquad\text{and}\qquad G=24H^{2}\left(  4\dot{H}+5H^{2}\right)
\end{equation}
when expressed in terms of the Hubble function
\begin{equation}
    H=\frac{1}{N}\frac{\dot{a}}{a}.
\end{equation}
Now let $S=S_{SF}+S_{RG}+S_{GB}$, where
\begin{align*}
&S_{SF}=-\int d^5 x\sqrt{-g}\left(\frac{1}{2}g_{\mu\nu}\phi^{;\mu}\phi^{;\nu}+V\left( \phi\right) \right),\\
&S_{RG}=\frac{1}{2}\int d^5 x\sqrt{-g}R,\\
&S_{GB}=-\int d^5 x\sqrt{-g}f\left(\phi\right)G,
\end{align*}
respectively. The first term can be transformed using the minisuperspace description, as usual, into
\begin{align}
S_{SF}=\int dt\left(\frac{1}{2}\frac{a^{4}}{N}\dot{\phi}^{2}%
-Na^{4}V\left(  \phi\right)  \right).
\end{align}
Now, by inserting \eqref{ricciscal} into the expression for $S_{RG}$ we obtain, after integrating by parts the term containing $\ddot{a}$, the following:
\begin{equation}
    S_{RG}=\int dt\left(  -6\frac{a^{2}}{N}\dot{a}^{2}\right).
\end{equation}
Analogously, by integration by parts, the Gauss-Bonnet term takes the form
\begin{equation}
    S_{GB}=\int dt\left( 8\frac{\dot{a}^3}{N^3}\left(f(\phi)+4a\dot{\phi}\frac{df}{d\phi}\right)\right).
\end{equation}
The point-like Lagrangian is then
\begin{equation}
   \label{general-pl-lagrangian} L(a,\dot a,\phi, \dot{\phi},N)=\frac{1}{2}\frac{a^{4}}{N}\dot{\phi}^{2}%
-Na^{4}V\left(  \phi\right) -6\frac{a^{2}}{N}\dot{a}^{2}+8\frac{\dot{a}^3}{N^3}\left(f(\phi)+4a\dot{\phi}\frac{df}{d\phi}\right).
\end{equation}
The last terms $GB_{5D}= 8\dot{a}^3\left(f(\phi)+4a\dot{\phi}\frac{df}{d\phi}\right)$ are the nontrivial ones due to the coupling between the Gauss-Bonnet term and $\phi$ that remains after integration by parts.

The gravitational field equations are derived from the variation of the point-like Lagrangian \eqref{general-pl-lagrangian} with respect to the dynamical variables ${a, \phi, N}$, while the constraint equation is the Hamiltonian function. The synchronous and separable area gauges are standard temporal gauges used in spherically symmetric cosmological models. In cosmology, especially within the framework of the FLRW metric, the lapse function ($g_{tt}=-N(t)^2$) is often set to one. This choice simplifies the equations without losing generality due to the gauge freedom in general relativity. The FLRW metric assumes a homogeneous and isotropic universe. In such a universe, the lapse function can be chosen to depend only on time, not on spatial coordinates. Setting it to one is a convenient choice aligned with this assumption. As a coordinate choice, it corresponds to a specific coordinate system where the time coordinate ($t$) represents the proper time of comoving observers (observers moving with the Hubble flow), making interpreting the equations more straightforward. However, it is essential to note that this choice is not unique, and different choices of the lapse function can be made depending on the specific problem or the coordinate system used.
Therefore, setting the lapse function to one simplifies the mathematical form of the metric and the Einstein field equations, resulting in the following field equations
\begin{align}
    \label{general_field_2}
    &96 H \dot{H} \dot{\phi} f'(\phi )+144 H^3 \dot{\phi} f'(\phi )+12 H^2 \left(4 \phi
   '^2 f''(\phi )+4 \ddot{\phi} f'(\phi )+4 f(\phi ) \dot{H}-1\right)+48 H^4 f(\phi
   )-6 \dot{H}+2 V(\phi )-  \dot{\phi}^2=0,\\
   \label{general_field_3}
    &96 H^2 \dot{H} f'(\phi )+120 H^4 f'(\phi )+4   H \dot{\phi}+V'(\phi
   )+  \ddot{\phi}=0,
\end{align}
together with the  Friedmann equation 
\begin{equation}
    \label{general_friedmann_1}
    192 H^3 \dot{\phi} f'(\phi )+48 H^4 f(\phi )-12 H^2+2 V(\phi )+  \dot{\phi}^2=0.
\end{equation}
We will assume that the scalar field potential $V(\phi)$ is an exponential one defined by $V(\phi)=V_0 e^{\lambda  \phi }$ where $V_0$ is a non-negative constant. 
 We also define the deceleration parameter as
 \begin{equation}
     \label{def-of-q}
     q=-1-\frac{\dot{H}}{H^2}=\frac{6 H^2 \left(8 f'(\phi ) \left(24 H^4
   f'(\phi )+V'(\phi )\right)+1\right)+144
   H^3 \dot{\phi} f'(\phi )-2 V(\phi )+\dot{\phi}^2}{6 H^2 \left(1-8 H \left(2 f'(\phi
   ) \left(\dot{\phi}-48 H^3 f'(\phi
   )\right)+H f(\phi )\right)\right)}.
   \end{equation}
   
We end this section with a brief discussion on the normalisation variable
   \begin{equation}
 \label{def-of-chi}
     \chi=V(\phi )+  \frac{1}{2}\dot{\phi}^2,
 \end{equation}
that was motivated in \cite{Tot:2022dpr,Leon:2023ywb} and that will be used through our analysis. We will first investigate the implications of $\chi=0$, and subsequently, we will proceed under the assumption that $\chi>0$. 

Recall that the exponential potential is always non-negative by definition, that is, $V(\phi)\geq 0$. Looking at equation \eqref{def-of-chi}, if $\chi=0$ we must have that $V(\phi)=\dot{\phi}=0$ therefore $\phi=k_1\in \mathbb{R}.$
Now Friedmann's equation \eqref{general_friedmann_1} reads
\begin{equation}
\label{friedmann-chi=0}
   -4 H^4 f(\phi)+ H^2=0.
\end{equation}

Based on the coupling functions that will be considered in this work, we have the following three cases:
\begin{enumerate}
\item If there is no coupling at all, that is $f(\phi)=0$, from \eqref{friedmann-chi=0} we obtain
\begin{equation}
H^2=0 ,   
\end{equation} 
which has $H=0$ as its only solution. This means that the model is trivial for $\chi=0.$
    \item If the coupling is constant, that is  $f(\phi)=-\alpha,$ from \eqref{friedmann-chi=0} we obtain the following equation for $H$
  \begin{equation}
 \label{Friedmann-alpha-chi=0}
     4 \alpha  H^4 + H^2= 0,
 \end{equation}
 which yields $H=0$ as its only real solution. Once again, the model is trivial for $\chi=0.$
 \item If the coupling is linear, that is $f(\phi)=-\alpha \phi$, $\alpha \neq 0$, equation \eqref{friedmann-chi=0} now reads
  \begin{equation}
     \alpha  H^3 \dot{\phi}+24 \alpha  H^4 \phi +6 H^2= 0.
 \end{equation}
 Since $\dot{\phi}=0$ we can write $\phi(t)=k_1\in \mathbb{R}.$ Substituting $V(\phi)=0$ and $\phi(t)=k_1$ in the field equations  \eqref{general_field_2} and \eqref{general_field_3} and \eqref{general_friedmann_1}, we obtain the following equations
\begin{align}
& 2 H^2 \left(4 \alpha  k_{1} \left(\dot{H}+H^2\right)+1\right)+ \dot{H}=0, \label{KG-linear-chi=0}\\
& 4 \alpha  H^2 \left(4 \dot{H}+5 H^2\right)=0, \label{Raych-linear-chi=0}\\
& 4 \alpha  k_1 H^4+H^2=0. \label{Friedmann-linear-chi=0}
\end{align}
We will show that the only possibility is $H=0$.
Assuming that $H\neq 0$, $\dot{H}\neq0$ and $k_1\neq 0$, we can solve equation \eqref{Raych-linear-chi=0} to obtain the expression \begin{equation}
  \dot{H}= -\frac{4}{5}H^2.
 \end{equation}
 Replacing this expression in \eqref{KG-linear-chi=0}, the remaining equations are 
 \begin{align}
H^2 \left(8 \alpha k_1 H^2-3\right)=0, \quad H^2 \left(8 \alpha  k_1 H^2+2\right)=0,
\end{align}
which for $H\neq 0$ and $k_1\neq0 $ implies the contradiction  $-3=2$. That yields $H(t)=0$, which makes the model trivial.
\end{enumerate}

Given the implications of $\chi=0$, in the following sections, namely \ref{sect-3-1-1} and \ref{sect-3-2-1}, we will restrict our analysis to the case where $\chi>0$.
\subsection{Five-dimensional scalar field model without the Gauss-Bonnet term}
\label{sect-3-0}
Before focusing on the contribution of the Gauss-Bonnet term to the model by setting two types of coupling functions, in this section, we explore the case where the coupling function $f(\phi)$ in the action \eqref{ai.01} is set to zero. In this case, if we make $f(\phi)=0$, there is no contribution from the Gauss-Bonnet term, so the action now reads
\begin{equation}
S=\int d^{5}x\sqrt{-g}\left(  \frac{R}{2}-\frac{1}{2}g_{\mu\nu}%
\phi^{;\mu}\phi^{;\nu}-V\left(  \phi\right)  \right),
\label{ai-with-f=0}%
\end{equation}
which is the five-dimensional action in a vacuum. Proceeding in the usual way, the point-like Lagrangian is
\begin{equation}
    \label{point-like-lagrangian-f=0}L(a,\dot{a},\phi,\dot{\phi},N)=-\frac{6 a^2 \dot{a}^2}{N}-a^4 N V(\phi )+\frac{a^4 \dot{\phi}^2}{2 N}.
\end{equation}
As mentioned in section \ref{sect-3}, the gravitational field equations are obtained by performing the variation of the point-like Lagrangian \eqref{point-like-lagrangian-f=0} with respect to the dynamical variables ${a, \phi, N}$ and we simplify the equations by setting the lapse function $N$ to $1$ after the variation process. Finally, we obtain the field equations
\begin{align}
    6 \dot{H}+12 H^2-2 V(\phi )+\dot{\phi}^2&=0,\\
    4 H \dot{\phi}+V'(\phi )+\ddot{\phi}&=0,
\end{align}
together with the Friedmann equation as a constraint
\begin{equation}
\label{friedmann-f=0}
       6 H^2=V(\phi )+\frac{1}{2} \dot{\phi}^2.
\end{equation}
To obtain the dynamical system, the dimensionless variables are defined in the usual way, that is, by using the standard $H$-normalisation,
\begin{equation}
    \Omega_{\phi}=\frac{\dot{\phi}}{\sqrt{12} H}\quad \text{and}\quad \Omega_V=\frac{\sqrt{V(\phi )}}{\sqrt{6 H^2}}.
\end{equation}
With these definitions, the dynamical system is

\begin{align}
\frac{d\Omega_{\phi}}{d\tau}&=2 \Omega_{\phi}  \left(-\Omega_V^2+\Omega_{\phi} ^2-1\right)-\sqrt{3} \lambda  \Omega_V^2,\\
   \frac{d\Omega_{V}}{d\tau}&=\Omega_V \left(\sqrt{3} \lambda  \Omega_{\phi} -2 \Omega_V^2+2 \Omega_{\phi}
   ^2+2\right).
\end{align}
Where we defined a new time derivative as $\frac{dg}{d\tau}=\frac{1}{H}\frac{dg}{dt}$. Additionally, in these variables, the Friedmann equation \eqref{friedmann-f=0} is written as
\begin{equation}
\label{friedmann-f=0-newvars}
    \Omega_V^2+\Omega_{\phi} ^2=1,
\end{equation}
since by definition we have that $\Omega_{V}\geq 0$ we can solve equation \eqref{friedmann-f=0-newvars} to reduce the dimension of the system, that is, 
\begin{equation}
\label{def-of-OmegaV}
    \Omega_V=\sqrt{1-\Omega_{\phi} ^2}.
\end{equation}
We can use this expression for $\Omega_V$ to study the dynamics of the following one-dimensional dynamical system \cite{strogatz}
\begin{equation}
\label{one-dimensional-syst}
    \frac{d\Omega_{\phi}}{d\tau}=\left(\Omega_\phi ^2-1\right) \left(\sqrt{3} \lambda +4 \Omega_\phi \right).
\end{equation}
We can also write the deceleration parameter \eqref{def-of-q} in terms of these new variables as $q=1-2 \Omega_V^2+2 \Omega_\phi ^2.$ However, using \eqref{def-of-OmegaV} it simplifies to
\begin{equation}
\label{deceleration-f=0}
     q=4 \Omega_\phi ^2-1.
\end{equation}

The one-by-one Jacobian matrix \eqref{Jacobian-1} for equation \eqref{one-dimensional-syst} is 
\begin{equation}
    \mathbf{J}=2 \Omega_\phi  \left(\sqrt{3} \lambda +4 \Omega_\phi \right)+4 \left(\Omega_\phi ^2-1\right).
\end{equation}
We now perform the stability analysis for equation \eqref{one-dimensional-syst}. The equilibrium points in the coordinate $\Omega_\phi$ are
\begin{enumerate}
    \item $A=1,$ with eigenvalue $\lambda_1=8+2 \sqrt{3} \lambda$  and deceleration parameter $q(A)=3.$ This point is 
    \begin{enumerate}
        \item an attractor for $\lambda <-\frac{4}{\sqrt{3}},$
        \item a source for $\lambda >-\frac{4}{\sqrt{3}},$
        \item non-hyperbolic for $\lambda =-\frac{4}{\sqrt{3}}.$
    \end{enumerate}
    It describes a super-collapsing solution.
    \item $B=-1,$ with eigenvalue $\lambda_1=8-2 \sqrt{3} \lambda$ and deceleration parameter $q(B)=3.$ This point is 
    \begin{enumerate}
        \item an attractor for $\lambda >\frac{4}{\sqrt{3}},$
        \item a source for $\lambda <\frac{4}{\sqrt{3}},$
        \item non-hyperbolic for $\lambda =\frac{4}{\sqrt{3}}.$
    \end{enumerate}
    It describes a super-collapsing solution.
    \item $C=-\frac{\sqrt{3} \lambda }{4},$ with eigenvalue $\lambda_1=\frac{3 \lambda ^2}{4}-4$ and deceleration parameter $q(C)=\frac{3 \lambda ^2}{4}-1.$ Aditionally, it exists for $-\frac{4}{\sqrt{3}}\leq \lambda \leq \frac{4}{\sqrt{3}}.$ This point is 
    \begin{enumerate}
        \item an attractor for  $-\frac{4}{\sqrt{3}}< \lambda < \frac{4}{\sqrt{3}},$
        \item non-hyperbolic for $\lambda=-\frac{4}{\sqrt{3}}$ or  $\lambda=\frac{4}{\sqrt{3}}.$
    \end{enumerate}
    It describes a super-collapsing solution.
\end{enumerate} 

\begin{figure}[h]
        \centering
        \includegraphics[scale=0.6]{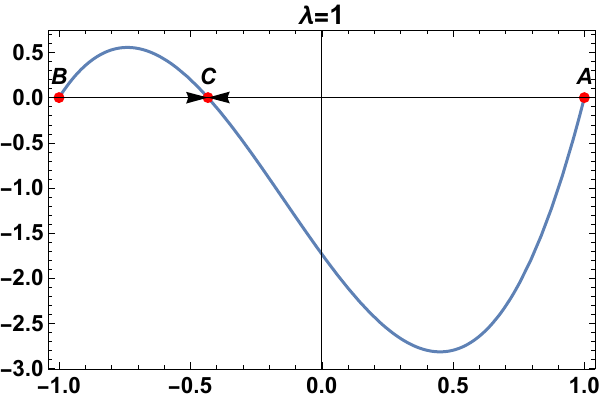}
        \includegraphics[scale=0.6]{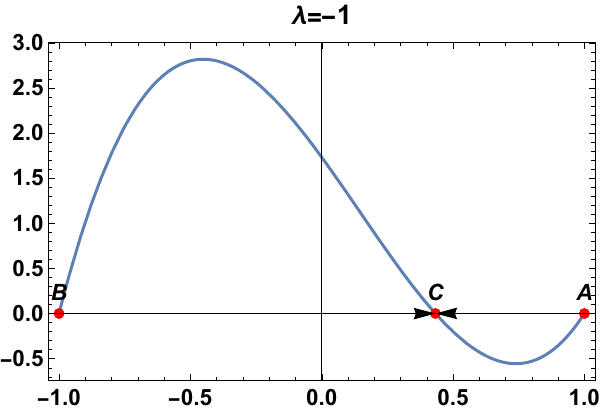}
        \includegraphics[scale=0.6]{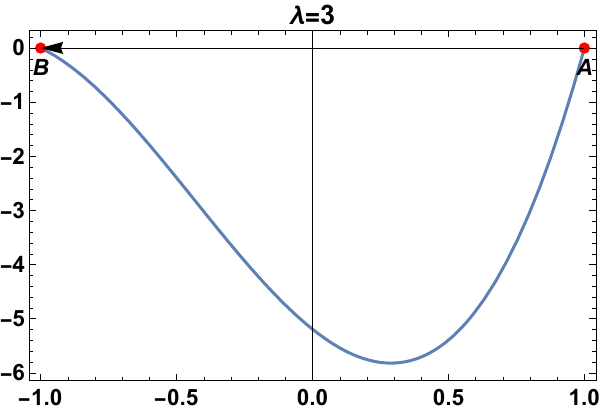}
        \includegraphics[scale=0.6]{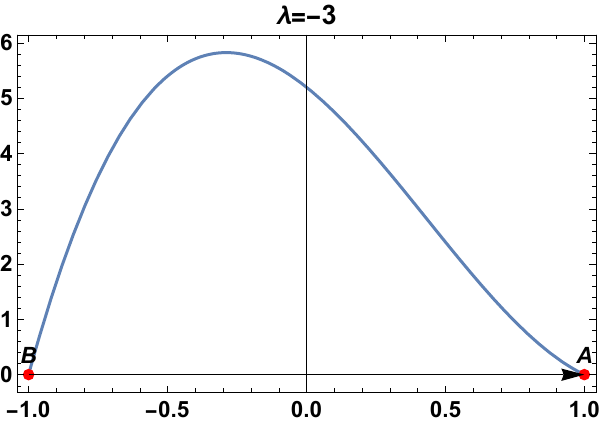}
        \caption{One-dimensional flow for equation \eqref{one-dimensional-syst} for different values of the parameter $\lambda$.}
        \label{fig:one-dimensional-flow}
    \end{figure}

In table \ref{tab:6} we present a summary of the results of this section. In figure \ref{fig:one-dimensional-flow}, we depict four different flows of the one-dimensional system described by equation \eqref{one-dimensional-syst}. We see that a scaling solution $C$ appears in this simple five-dimensional scalar field model without the coupling to the Gauss-Bonnet term.
We can summarise the results of this section in the following results
\begin{theorem}
\label{theo-0}
The late-time attractors for the five-dimensional scalar field model (without coupling to Gauss-Bonnet) are
    \begin{enumerate}
        \item The supper-collapse solution $A=1$ for $\lambda<-\frac{4}{\sqrt{3}}.$
        \item The supper-collapse solution $B=1$ for $\lambda>\frac{4}{\sqrt{3}}.$
        \item The scaling solution $C=-\frac{\sqrt{3}\lambda}{4}$ for $-\frac{4}{\sqrt{3}}<\lambda<\frac{4}{\sqrt{3}}.$
    \end{enumerate}
    To avoid having super-collapsing solutions as late-time attractors we can restrict $\lambda$ to be in the interval
    $-\frac{4}{\sqrt{3}}<\lambda<\frac{4}{\sqrt{3}}.$ In this range of values, the late time attractor for the model is the scaling solution $C.$
\end{theorem}

\begin{table}[h]
    \centering
    \begin{tabular}{|c|c|c|c|c|c|}
    \hline
         Label& $\Omega_\phi$ & $\Omega_V$ & Eigenvalue& Attractor & Interpretation \\ \hline
         $A$& $1$ & $0$ & $8+2\sqrt{3}\lambda$ &Yes, for $\lambda<-\frac{4}{\sqrt{3}}$  & super-collapse \\ \hline
         $B$& $-1$ & $0$ & $8-2\sqrt{3}\lambda$ &Yes, for $\lambda>\frac{4}{\sqrt{3}}$  & super-collapse \\ \hline
         $C$& $-\frac{\sqrt{3} \lambda }{4}$ & $ \sqrt{1-\frac{3 \lambda ^2}{16}}$ & $\lambda_1=\frac{3 \lambda ^2}{4}-4$ &Yes, for $-\frac{4}{\sqrt{3}}< \lambda < \frac{4}{\sqrt{3}}$  & scaling solution \\ \hline
    \end{tabular}
    \caption{\textbf{Summary of the analysis of the one-dimensional system \eqref{one-dimensional-syst}.}}
    \label{tab:6}
\end{table}
\FloatBarrier
    \FloatBarrier
\subsection{Constant coupling function: $f(\phi)=-\alpha$}
\label{sect-3-1}
In \cite{Millano:2023czt,Millano:2023gkt}, the authors studied a four-dimensional Gauss-Bonnet model where the coupling term between the Gauss-Bonnet invariant and the scalar field that remains after integration by parts was 
\begin{equation}
    \label{GB-4D}
    GB_{4D}=8 \dot{a}^3 \dot{\phi}\frac{df}{d\phi}.
\end{equation}
  In this case, the coupling function $f(\phi)$ is necessary for the non-trivial contribution of the Gauss-Bonnet term to the field equations. In the five-dimensional case, we have the following expression for the coupling term
 \begin{equation}
     \label{GB-5D}
     GB_{5D}= 8\dot{a}^3\left(f(\phi)+4a\dot{\phi}\frac{df}{d\phi}\right).
 \end{equation}
 In \eqref{GB-5D} we can set $f(\phi)=-\alpha$ where $\alpha$ is some positive constant and the contribution of the Gauss-Bonnet after integrating by parts is nonzero. For this choice of $f,$ the field equations \eqref{general_field_2}-\eqref{general_field_3} now read
 \begin{align}
     \label{constant_field_1}
     &12 H^2 \left(4 \alpha  \phi  \dot{H}+4 \alpha  \ddot{\phi}+1\right)+96 \alpha  H \dot{H} \dot{\phi}+6 \dot{H}+144
   \alpha  H^3 \dot{\phi}+48 \alpha  H^4 \phi -2 V(\phi
   )+\dot{\phi}^2=0,\\
   &96 \alpha  H^2 \dot{H}+120 \alpha  H^4-4
   H \dot{\phi}-V'(\phi )-\ddot{\phi}=0
 \end{align}
 and the Friedmann equation \eqref{general_friedmann_1} reads
 \begin{equation}
 \label{Friedmann-alpha}
     24 \alpha  H^4 +6 H^2= V(\phi )+  \frac{1}{2}\dot{\phi}^2.
 \end{equation}
The deceleration parameter \eqref{def-of-q} now reads \begin{equation}
     \label{def-of-q-f-constant}
     q=-1+\frac{48 \alpha  H^4+12 H^2-2 V(\phi )+  \dot{\phi}^2}{H^2 \left(48
   \alpha  H^2+6\right)}.
   \end{equation}
 \subsubsection{Dynamical system analysis for constant coupling function}
\label{sect-3-1-1}

 Following \cite{Tot:2022dpr,Leon:2023ywb} we now introduce the following dimensionless variables 
\begin{equation}\label{dimensionlessvars}
     x_1=\frac{2 \sqrt{3} H}{\sqrt{\chi }},\quad x_2=\frac{4 \sqrt{3}
   \sqrt{\alpha } H^2}{\sqrt{\chi }},\quad x_3=\frac{\sqrt{2} \sqrt{V(\phi
)}}{\sqrt{\chi }}\quad\text{and}\quad x_4=\frac{\dot{\phi}}{\sqrt{\chi }},
 \end{equation}
 where the normalisation is done with $\chi>0$ defined in equation \eqref{def-of-chi}
 with inverse transformation given by
 \begin{equation}
     \label{original-vars-in-terms-of-newvars}
     H= \frac{x_1 \sqrt{\chi }}{2 \sqrt{3}},\quad \alpha = \frac{3 x_2^2}{x_1^4 \chi },\quad V(\phi )= \frac{1}{2} x_3^2
   \chi ,\quad \dot{\phi} = x_4 \sqrt{\chi }
 \end{equation}
 By replacing this new set of variables in equation \eqref{Friedmann-alpha}, we obtain the following constraints
 \begin{equation}\label{cont-circum}
     x_1^2+x_2^2=x_3^2+x_4^2=1,
 \end{equation}
 which are the equations for a pair of unit circumferences. Defining a new time derivative $f'=\frac{1}{\sqrt{\chi}}\frac{df}{dt}$,  we obtain the following system

\begin{align}
\label{eq-x1}
    x_1'&=\frac{x_1^2 \left(x_4^2 \left(2 x_1^2+4
   x_2^2-1\right)-x_1^2-x_2^2+x_3^2\right)}{\sqrt{3} \left(x_1^2+2
   x_2^2\right)},\\
   \label{eq-x2}
    x_2'&=\frac{2 x_1 x_2 \left(x_4^2 \left(x_1^2+2
   x_2^2-1\right)-x_1^2-x_2^2+x_3^2\right)}{\sqrt{3} \left(x_1^2+2
   x_2^2\right)}\\
   \label{eq-x3}
    x_3'&=\frac{1}{6} x_3 x_4 \left(3 \lambda +4 \sqrt{3} x_1 x_4\right),\\
    \label{eq-x4}
    x_4'&=\frac{2
   x_1 x_4 \left(x_4^2-1\right)}{\sqrt{3}}.
\end{align}
Defined in the compact space $\{(x_1,x_2,x_3,x_4)\in \mathbb{R}^4: -1\leq x_i\leq 1 \quad \text{for} \quad i=1,4  \quad \text{and} \quad 0\leq x_j\leq 1\quad \text{for} \quad j=2,3 \}$.  We also have that equation \eqref{def-of-q} now reads
\begin{equation}
    \label{def-of-q-newvars}
    q=\frac{x_1^2-2 x_3^2+2 x_4^2}{x_1^2+2 x_2^2}.
\end{equation}
 From the definition of the dimensionless variables \eqref{dimensionlessvars}, we see that $x_2$ and $x_3$ are both positive. We can solve the constraints \eqref{cont-circum} and take the following positive roots
 \begin{equation}
     \label{def-of-x2-x3}
     x_2=\sqrt{1-x_1^2},\quad x_3=\sqrt{1-x_4^2}.
 \end{equation}
 With this, we can then reduce system \eqref{eq-x1}-\eqref{eq-x4} to the following two dimensional system for $x_1$ and $x_4$
 \begin{align}
 \label{eq-x1-reduced}
     x_1'&=\frac{2 x_1^2 \left(x_1^2-1\right) x_4^2}{\sqrt{3} \left(x_1^2-2\right)},\\
     \label{eq-x4-reduced}
     x_4'&=\frac{1}{6}
   \left(x_4^2-1\right) \left(3 \lambda +4 \sqrt{3} x_1 x_4\right).
 \end{align}
Defined in the compact space $\{(x_1,x_2,x_3,x_4)\in \mathbb{R}^4: -1\leq x_i\leq 1 \quad \text{for} \quad i=1,4\}$.  The deceleration parameter \eqref{def-of-q-newvars} now takes the following form
\begin{equation}
    \label{def-of-q-reduced}
    q=-1-\frac{4 x_4^2}{x_1^2-2}.
\end{equation} 

\begin{figure}[ht]
     \centering
     \includegraphics[scale=0.3]{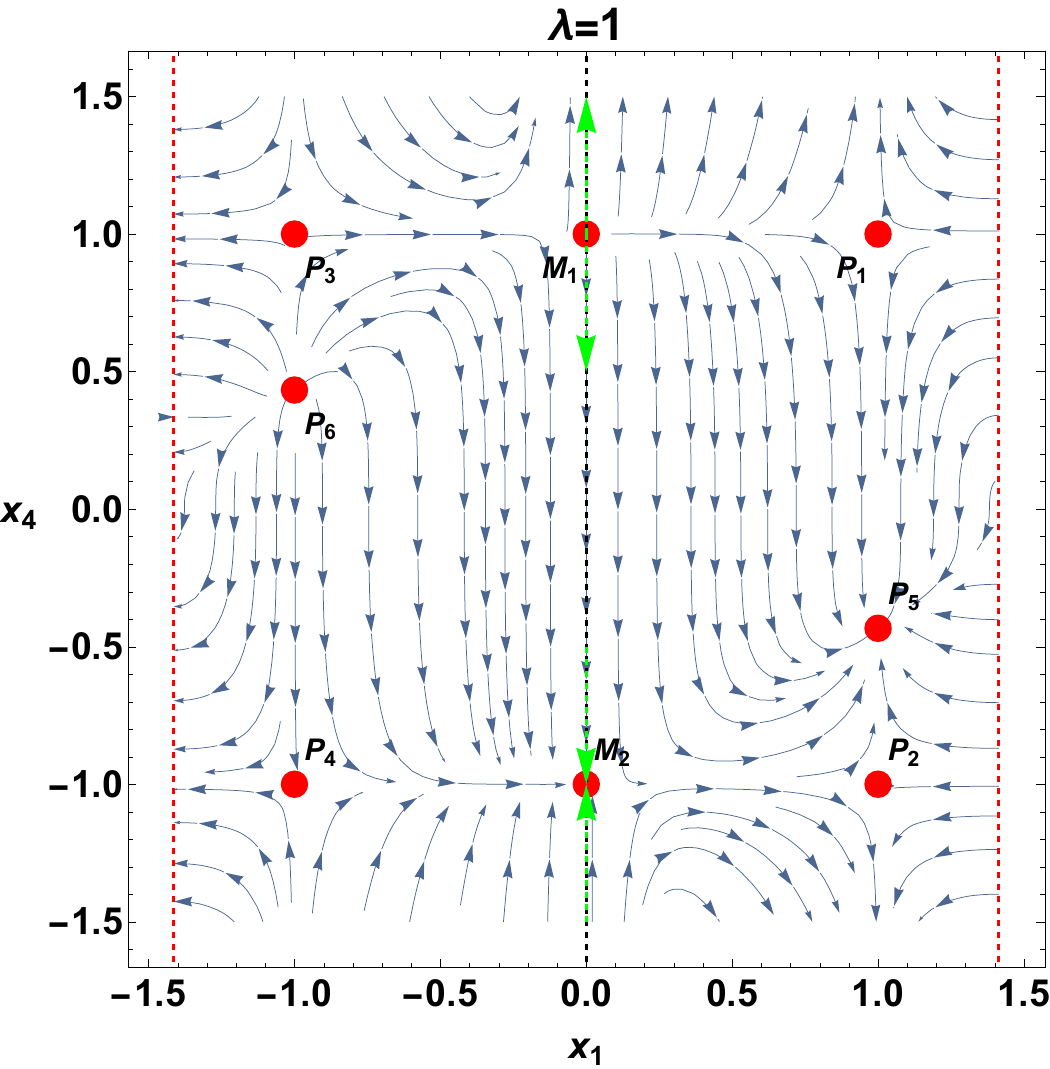}
     \includegraphics[scale=0.3]{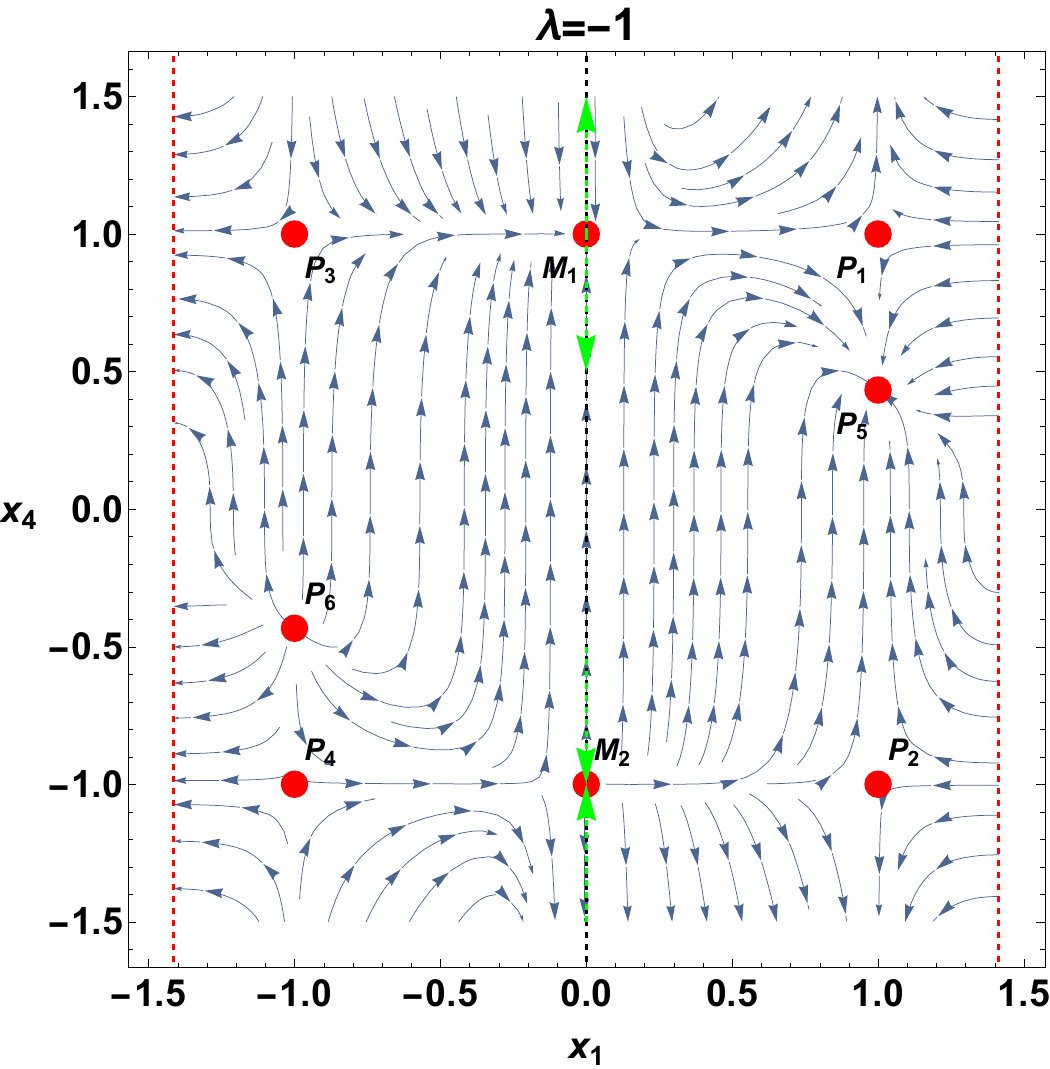}
     \includegraphics[scale=0.3]{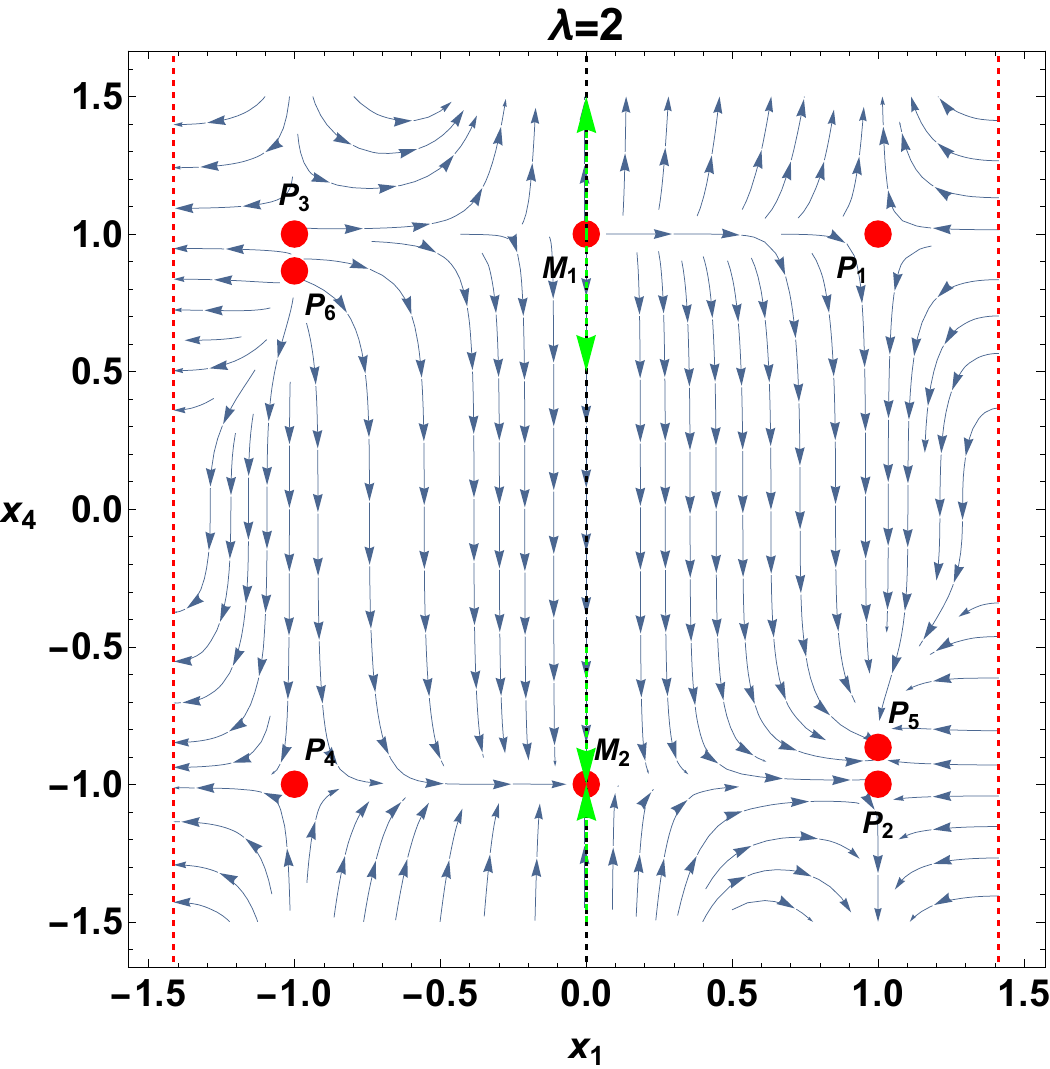}
     \includegraphics[scale=0.3]{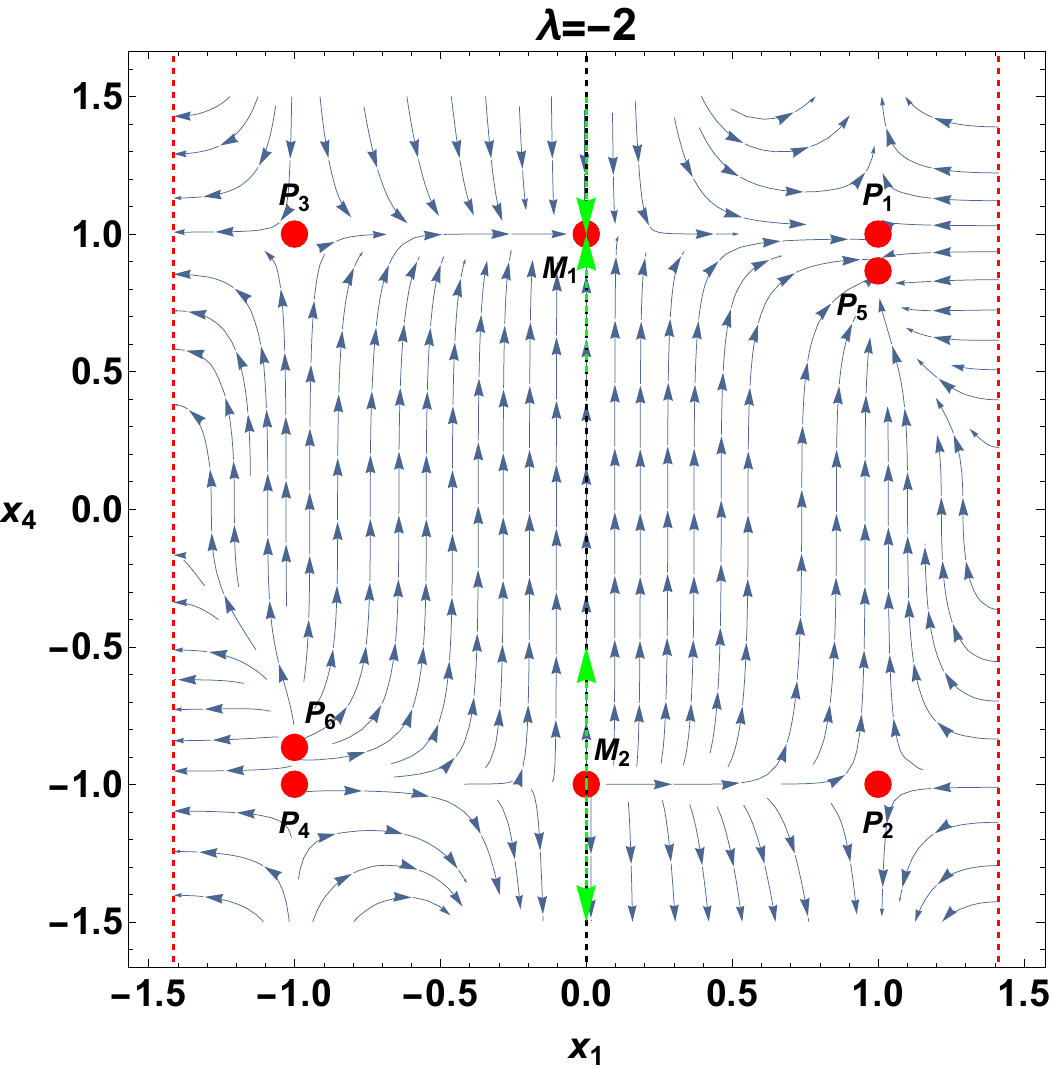}
     \includegraphics[scale=0.3]{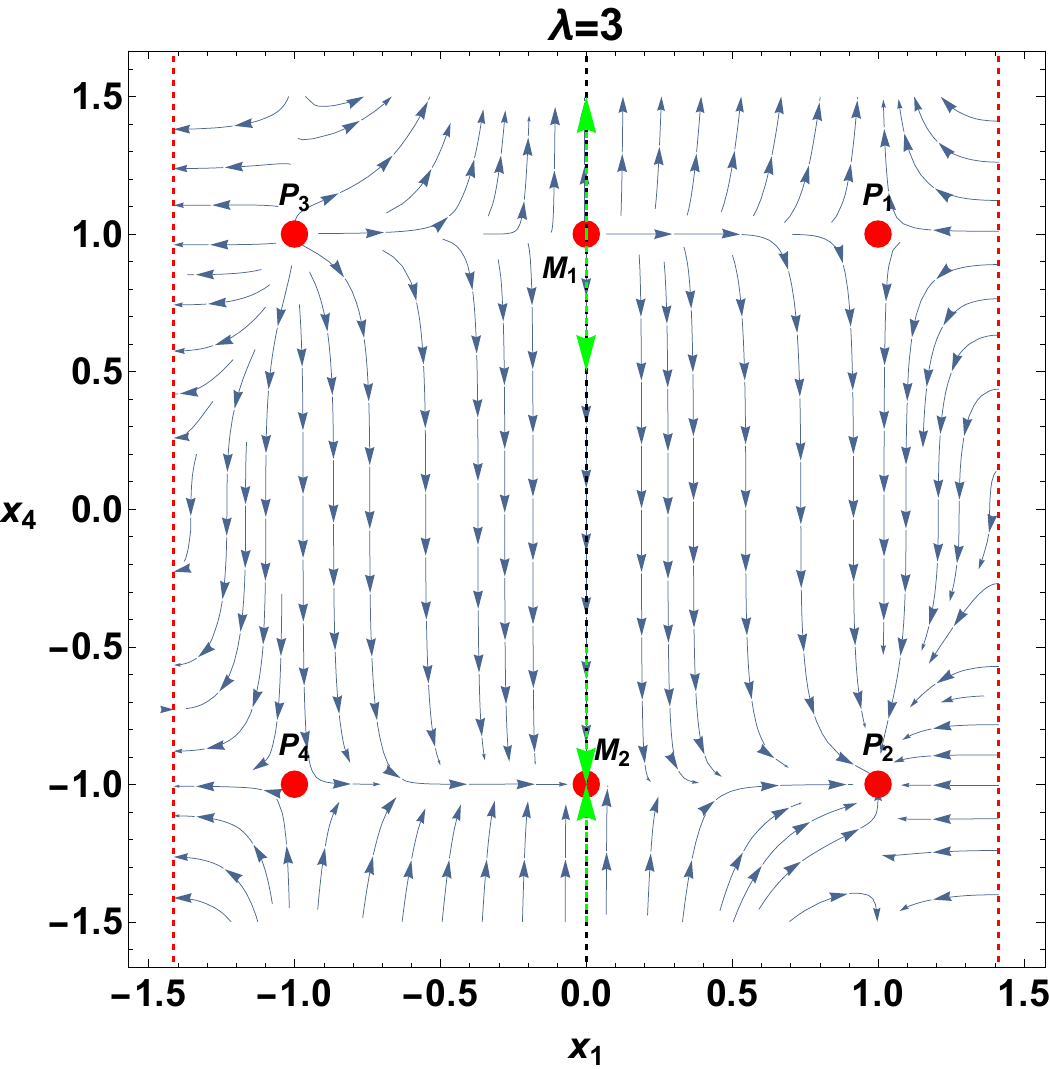}
     \includegraphics[scale=0.3]{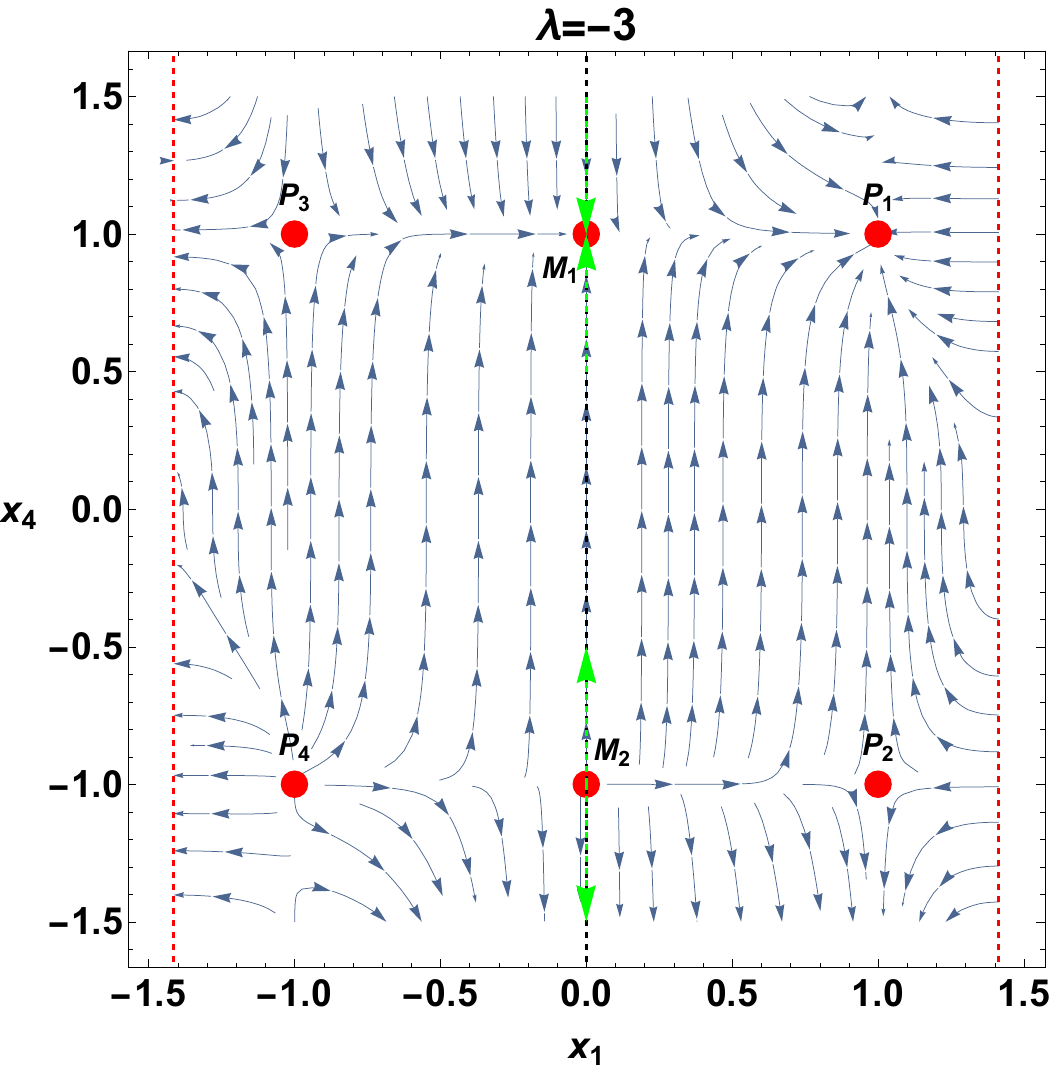}
     \caption{Phase-space plots for system \eqref{eq-x1-reduced}-\eqref{eq-x4-reduced} for the values $\lambda=1,\pm 2, \pm 3$.  The black dotted line is the manifold for points $M_{1,2}$.  The red dotted lines correspond to the vertical line $ x_=\pm \sqrt{2}$. These lines are the singularities previously mentioned in which the direction of the flow changes.}
     \label{fig:1a}
 \end{figure}

\begin{figure}[h!]
    \centering
    \includegraphics[scale=0.35]{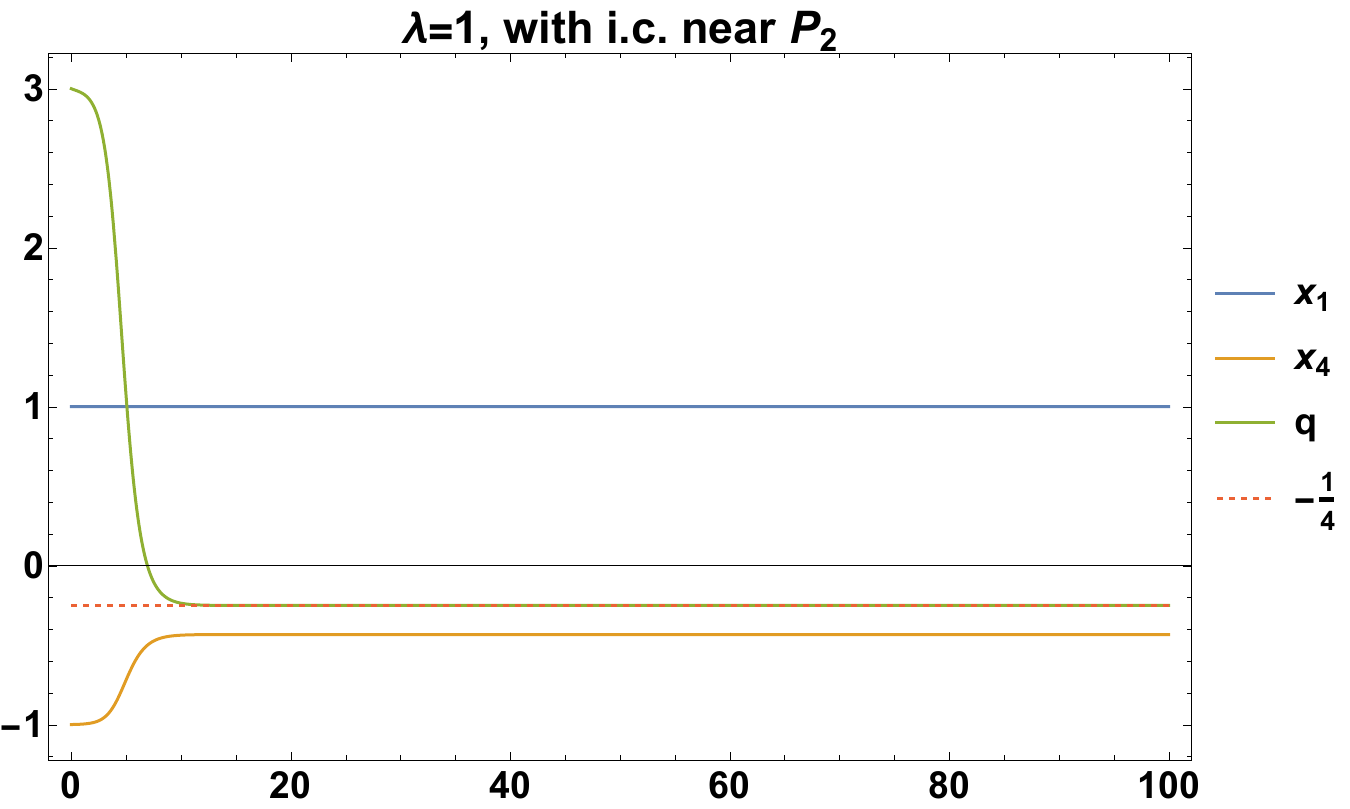}
    \includegraphics[scale=0.35]{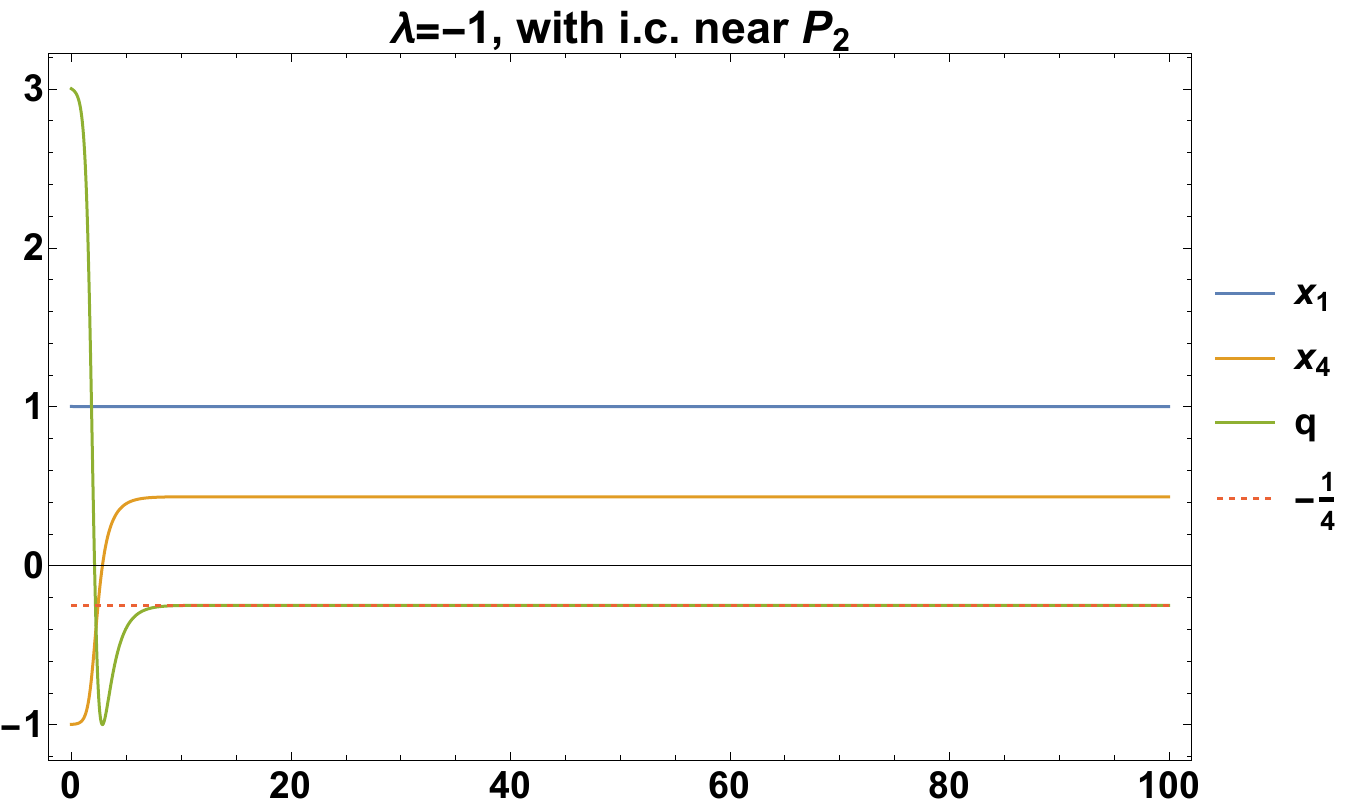}
    \caption{Numerical solution for system \eqref{eq-x1-reduced}-\eqref{eq-x4-reduced} for $\lambda=\pm 1$ and initial conditions near the point $P_2$ with a displacement of $\xi=10^{-3}$. In both cases, $P_2$ is a saddle point, which means that the system initially near $P_2$ will converge into the nearest attractor, in this case, $P_5$ with coordinates $x_1=1$ and $x_4=-\sqrt{3}/4$. Also, the deceleration parameter is depicted as a green line in the plots. For both values of $\lambda$, we see that after an initially decelerated stage, the solution accelerates into the scaling solution for $q(P_5)\to-1/4$. However, for $\lambda=-1$, we see that there is an intermediate accelerated behaviour after the initial super-collapsing stage and before the scaling solution regime since $q$ reaches the de Sitter value $-1$. }
    \label{fig:6}
\end{figure}

Note that in both the first equation of system \eqref{eq-x1-reduced}-\eqref{eq-x4-reduced} and in equation\eqref{def-of-q-reduced}, the value $x_1=\sqrt{2}$ is a singularity value.
In what follows, we calculate the equilibrium points of system \eqref{eq-x1-reduced}-\eqref{eq-x4-reduced}, then compute the Jacobian matrix \eqref{Jacobian-1} of the system and evaluate it in each of the previously mentioned equilibrium points. We will classify each point according to the signs of the real part of their respective eigenvalues. The points (and their stability) are
 \begin{enumerate}
     \item $P_1=(1,1)$,  with eigenvalues $\left\{-\frac{4}{\sqrt{3}},\lambda +\frac{4}{\sqrt{3}}\right\}$.  This point is an attractor for $\lambda <-\frac{4}{\sqrt{3}}$,  a saddle for $\lambda >-\frac{4}{\sqrt{3}}$ and non-hyperbolic for $\lambda =-\frac{4}{\sqrt{3}}$.  We also have $q(P_1)=3$, which describes a super-collapse solution.
     \item $P_2=(1,-1)$,  with eigenvalues $\left\{-\frac{4}{\sqrt{3}},\frac{4}{\sqrt{3}}-\lambda \right\}$.  This point is an attractor for $\lambda >\frac{4}{\sqrt{3}}$,  a saddle for $\lambda <\frac{4}{\sqrt{3}}$ and non hyperbolic for $\lambda =\frac{4}{\sqrt{3}}$.  As before, $q(P_2)=3$, and we also have a super-collapse scenario.
     \item $P_3=(-1,1)$,  with eigenvalues $\left\{\frac{4}{\sqrt{3}},\lambda -\frac{4}{\sqrt{3}}\right\}$.  This point is a source for $\lambda >\frac{4}{\sqrt{3}}$,  a saddle for $\lambda <\frac{4}{\sqrt{3}}$ and non-hyperbolic for $\lambda =\frac{4}{\sqrt{3}}$.  As the previous two points, $q(P_3)=3$. 
     \item $P_4=(-1,-1)$,  with eigenvalues $\left\{\frac{4}{\sqrt{3}},-\lambda -\frac{4}{\sqrt{3}}\right\}$.  This point is a source for $\lambda <-\frac{4}{\sqrt{3}}$,  a saddle for $\lambda >-\frac{4}{\sqrt{3}}$ and non-hyperbolic for $\lambda =-\frac{4}{\sqrt{3}}$.  Once again $q(P_4)=3$. 
     \item $P_5=(1,-\frac{\sqrt{3} \lambda }{4})$,  with eigenvalues $\left\{-\frac{1}{4} \sqrt{3} \lambda ^2,\frac{3 \lambda ^2-16}{8 \sqrt{3}}\right\}$.  This point exists for $-\frac{4}{\sqrt{3}}\leq \lambda \leq \frac{4}{\sqrt{3}}$ and is an attractor for $-\frac{4}{\sqrt{3}}<\lambda <0$ or $0<\lambda <\frac{4}{\sqrt{3}}$,  a saddle for $\lambda <-\frac{4}{\sqrt{3}}$ or $\lambda >\frac{4}{\sqrt{3}}$ and non-hyperbolic for $\lambda =0$ or $ \lambda =\pm \frac{4}{\sqrt{3}}$.  In this case we have $q(P_5)=-1+\frac{3 \lambda ^2}{4}$, which is a scaling solution. This means that the solution describes acceleration in the interval $-\frac{2}{\sqrt{3}}<\lambda <\frac{2}{\sqrt{3}}$ and particularly it is a de Sitter solution for $\lambda=0$. 
     \item $P_6=(-1,\frac{\sqrt{3} \lambda }{4})$,  with eigenvalues $\left\{\frac{\sqrt{3} \lambda ^2}{4},\frac{16-3 \lambda ^2}{8 \sqrt{3}}\right\}$.  This point exists for $-\frac{4}{\sqrt{3}}\leq \lambda \leq \frac{4}{\sqrt{3}}$ and is a source for $-\frac{4}{\sqrt{3}}<\lambda <0$ or  $0<\lambda <\frac{4}{\sqrt{3}}$,  a saddle for $\lambda <-\frac{4}{\sqrt{3}}$ or $\lambda >\frac{4}{\sqrt{3}}$  and non-hyperbolic for $\lambda =0$ or $ \lambda =\pm \frac{4}{\sqrt{3}}$.  In this case we have $q(P_6)=-1+\frac{3 \lambda ^2}{4}$ which is a scaling solution with the same interpretation as the previous one.
     \item $M_1=(0,1)$,  with eigenvalues $\{\lambda,0\}$.  This point is globally a saddle. However, along a 1-dimensional manifold (a line) is stable for $\lambda<0$ and is unstable along the same manifold for $\lambda>0$.  We verify that $q(M_1)=1$,  describing a decelerated solution.
     \item $M_2=(0,-1)$,   with eigenvalues $\{-\lambda,0\}$.  This point is globally a saddle. However, it has a  1-dimensional  unstable manifold (a line) for $\lambda<0$ and a  1-dimensional stable manifold for $\lambda>0$.  Once again, we see that $q(M_2)=1$. 
 \end{enumerate}
 We observe that in the previous analysis, the values $\lambda=0, \pm \frac{4}{\sqrt{3}}$ are bifurcation values for the stability of the system. In figure \ref{fig:1a}, we present the flow of the system for different values of the parameter $\lambda$.  
 The results of this section are summarised in table \ref{tab:1}. In figure \ref{fig:6} we present some numerical solutions for system \eqref{eq-x1-reduced}-\eqref{eq-x4-reduced} for $\lambda=\pm 1$ and initial conditions near the point $P_2$ with a displacement of $\xi=10^{-3}$. In both cases, $P_2$ is a saddle point, which means that the system initially near $P_2$ will converge into the nearest attractor, in this case, $P_5$ with coordinates $x_1=1$ and $x_4=-\sqrt{3}/4$. Also, the deceleration parameter is depicted as a green line in the plots. For both values of $\lambda$, we see that after an initially decelerated stage, the solution accelerates into the scaling solution for $q(P_5)\to-1/4$. However, for $\lambda=-1$, we see that there is an intermediate accelerated behaviour after the initial super-collapsing stage and before the scaling solution regime since $q$ reaches the de Sitter value $-1$. Finally, we can write the findings of this section in the following way
 
\begin{theorem}
\label{teo-2}
    The late-time attractors for the five-dimensional Gauss-Bonnet model with constant coupling function are
    \begin{enumerate}
        \item The super-collapse solution $P_1=(1,1)$ for $\lambda<-\frac{4}{\sqrt{3}}$. 
        \item The super-collapse solution $P_2=(1,-1)$ for $\lambda>\frac{4}{\sqrt{3}}$. 
        \item The scaling solution $P_5=(1,-\frac{\sqrt{3}\lambda}{4})$ for $-\frac{4}{\sqrt{3}}<\lambda<0$ or $0<\lambda<\frac{4}{\sqrt{3}}$. 
    \end{enumerate}
\end{theorem}
To avoid super-collapse late-time attractors we can restrict the parameter $\lambda$ in the range $-\frac{4}{\sqrt{3}}<\lambda<0$ or $0<\lambda<\frac{4}{\sqrt{3}}$. Hence, according to the third statement in Theorem \ref{teo-2}, we have the following 
\begin{corollary}
\label{corol-1}
    The late-time attractor for the five-dimensional Gauss-Bonnet model with constant coupling function in the range $-\frac{4}{\sqrt{3}}<\lambda<0$ or $0<\lambda<\frac{4}{\sqrt{3}}$ is the scaling solution $P_5=(1,-\frac{\sqrt{3}\lambda}{4})$. 
\end{corollary}
Additionally, in section \ref{app-1}, we present a different formulation of the dynamical system for constant coupling function using polar coordinates since the constraints on the dimensionless variables describe two circumferences. The reader is encouraged to read the section to see an illustrative topological example of how to display the flow of a two-dimensional dynamical system on the surface of a torus.
\begin{table}[ht]
\begin{tabular}{|c|c|c|c|c|c|}
\hline
Label & $(x_1,x_4)$ & $(x_2,x_3)$ & Attractor? & Acceleration?& Interpretation \\ \hline
$P_1$ & $(1,1)$ & $(0,0)$ & Yes* & No, $q=3$ & Super-collapse\\ \hline
$P_2$ & $(1,-1)$ & $(0,0)$ & Yes* & No, $q=3$ & Super-collapse \\ \hline
$P_3$ & $(-1,1)$ & $(0,0)$ & No & No, $q=3$& Super-collapse \\ \hline
$P_4$ & $(-1,-1)$ & $(0,0)$ & No & No, $q=3$ & Super-collapse \\ \hline
$P_5$ & $(1,-\frac{\sqrt{3}\lambda}{4})$ & $\left(0,\sqrt{1-\frac{3 \lambda ^2}{16}}\right)$ & Yes &  Yes, $q=-1+\frac{3\lambda ^2}{4}$ & Scaling solution\\ \hline
$P_6$ & $(-1,\frac{\sqrt{3}\lambda}{4})$ & $\left(0,\sqrt{1-\frac{3 \lambda ^2}{16}}\right)$ & No & Yes, $q=-1+\frac{3\lambda ^2}{4}$ & Scaling solution\\ \hline
$M_1$ & $(0,1)$ & $(1,0)$ & No & No, $q=1$ & decelerated\\ \hline
$M_2$ & $(0,-1)$ & $(1,0)$ & No & No, $q=1$ & decelerated \\ \hline
\end{tabular}
\caption{Summary of the analysis of system \eqref{eq-x1-reduced}-\eqref{eq-x4-reduced}. The $x_3$ coordinate gives the interval of existence for $P_{5,6}$.  The asterisk indicates that we can restrict the parameter $\lambda$ if we do not wish to have a super-collapse solution as a late-time attractor (see Theorem \ref{teo-2} and the comment after it.}
\label{tab:1}
\end{table}

 \subsubsection{Alternative formulation for the constant coupling function model}
 \label{app-1}
 This section presents an alternative dynamical system formulation for the constant coupling function of section \ref{sect-3-1-1}. Since the constraints \eqref{cont-circum} describe two circumferences, we choose a reparametrization in polar coordinates as follows
 \begin{equation}
 \label{reparametrization}
     x_1=\cos (\theta_1),\quad x_2=\sin (\theta_1),\quad x_3=\cos (\theta_2),\quad x_4=\sin (\theta_2).
 \end{equation}

 With \eqref{reparametrization}, we define the new variables for the dynamical systems analysis 

 \begin{equation}
 \label{angles}
     \theta_1=\arctan \left(\frac{x_2}{x_1}\right)=\arctan(2 \sqrt{\alpha } H)\quad \text{and} \quad \theta_2=\arctan \left(\frac{x_4}{x_3}\right)=\arctan \left(\frac{\dot{\phi}}{\sqrt{2} \sqrt{V(\phi )}}\right).
 \end{equation}
 With inverse transformation given by
 \begin{equation}
     \label{inverse-transformation}
     H= \frac{\tan (\theta_1)}{2 \sqrt{\alpha }},\quad \dot{\phi}= \sqrt{2} \tan (\theta_2) \sqrt{V(\phi)}
 \end{equation}
 Using the definition \eqref{angles} and setting $\tau=\ln (a)$ we obtain the following two-dimensional system of first-order differential equations

 \begin{align}
     \label{dy-sys-1}\frac{d\theta_1}{d\tau}&=\frac{4 \sin (\theta_1) \cos ^2(\theta_1) \sin ^2(\theta_2)}{\sqrt{3} (\cos (2 \theta_1)-3)},\\
     \label{dy-sys-2}
     \frac{d\theta_2}{d\tau}&=-\frac{1}{6} \cos (\theta_2) \left(4 \sqrt{3} \cos (\theta_1) \sin (\theta_2)+3 \lambda \right).
 \end{align}

 From equations \eqref{inverse-transformation}, we remark that for $H$ to be continuous and simply so that the inverse transformations are defined, we must restrict the angles $\theta_{1,2}$ to $ (-\frac{\pi}{2},\frac{\pi}{2})$.  The equilibrium points for \eqref{dy-sys-1}-\eqref{dy-sys-2} are periodic, but we will consider the following ranges $\theta_1\in [-\pi,\pi]$ and $\theta_2\in [-\pi,\pi]$,  this choice will allow us to transfer from one branch of the solutions to the other one. Considering $c_1, c_2\in \mathbb{Z}$,  the equilibrium points for system \eqref{dy-sys-1}-\eqref{dy-sys-2} are
 \begin{enumerate}
     \item $T_1=(2 \pi  c_1, 2 \pi  c_2+\frac{\pi }{2})$,
   \item $T_2=(2 \pi  c_1, 2 \pi c_2-\frac{\pi }{2})$,
   \item $T_3=(2 \pi 
   c_1-\frac{\pi }{2},2 \pi  c_2+\frac{\pi }{2})$,
   \item $T_4=(2 \pi  c_1-\frac{\pi }{2},2\pi  c_2-\frac{\pi }{2})$,
   \item $T_5=(2 \pi  c_1+\frac{\pi }{2},2 \pi  c_2-\frac{\pi}{2})$,
   \item $T_6=(2
   \pi  c_1+\frac{\pi }{2},2 \pi  c_2+\frac{\pi }{2})$,
   \item $T_7=(2 \pi  c_1+\pi ,2 \pi c_2+\frac{\pi }{2})$,
   \item $T_8=(2 \pi  c_1+\pi,2 \pi  c_2-\frac{\pi }{2})$,
   \item $T_9=(2 \pi  c_1,\tan ^{-1}\left(\sqrt{48-9 \lambda ^2},-3 \lambda \right)+2 \pi  c_2)$,
   \item $T_{10}=(2 \pi  c_1,\tan
   ^{-1}\left(-\sqrt{48-9 \lambda ^2},-3 \lambda \right)+2 \pi c_2)$,
   \item $T_{11}=(2
   \pi  c_1+\pi,\tan ^{-1}\left(\sqrt{48-9 \lambda ^2},3 \lambda \right)+2 \pi c_2)$,
   \item $T_{12}=(2 \pi  c_1+\pi, \tan ^{-1}\left(-\sqrt{48-9 \lambda ^2},3 \lambda \right)+2 \pi  c_2)$.  
 \end{enumerate}
 We have used the notation $\tan ^{-1}(x,y)=\tan ^{-1}\left(\frac{y}{x}\right)$.  The first eight points lie within the previously mentioned range for $c_1=c_2=0$.  Additionally, points seven and eight can have $c_1=-1$.  Points nine and ten also require $c_1=0$ and additionally $\frac{-\tan ^{-1}\left(\pm \sqrt{48-9 \lambda ^2},-3 \lambda \right)-\pi }{2 \pi }\leq c_2\leq \frac{\pi -\tan ^{-1}\left(\pm\sqrt{48-9 \lambda ^2},-3
   \lambda \right)}{2 \pi }$.  The last two points are in the desired range for $c_1=0$ or $c_1=-1$ and $\frac{-\tan ^{-1}\left(\pm \sqrt{48-9 \lambda ^2},-3 \lambda \right)-\pi }{2 \pi }\leq c_2\leq \frac{\pi -\tan ^{-1}\left(\pm\sqrt{48-9 \lambda ^2},-3
   \lambda \right)}{2 \pi }$. 
To illustrate the relationship between these points and the ones from section \ref{sect-3-1-1}, taking into consideration the values of $\lambda$ used to plot the dynamics, we will consider only $c_1=c_2=0$ and $c_1=-1$,  $c_2=0$. 

\begin{figure}[h!]
    \centering
    \includegraphics[scale=0.33]{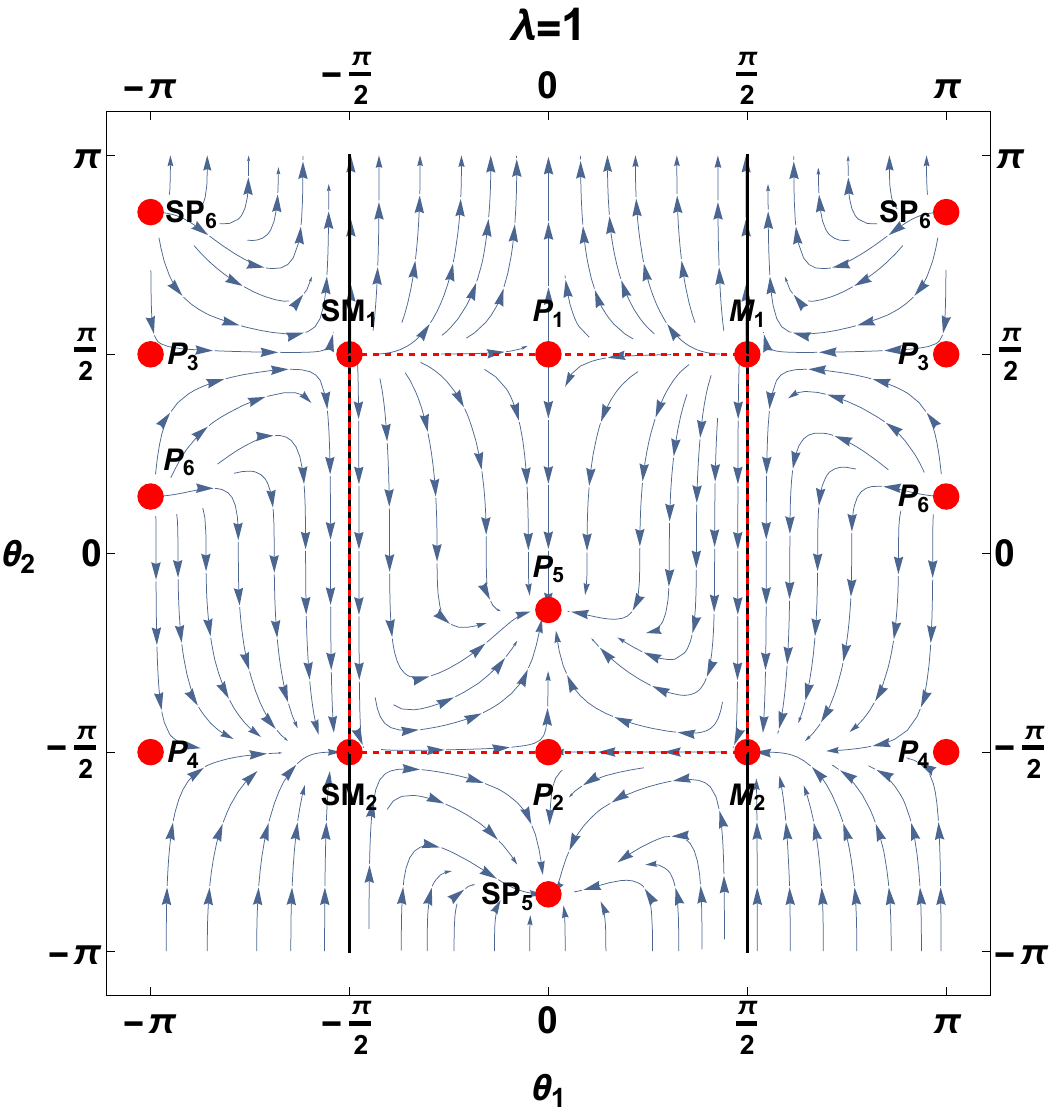}
    \includegraphics[scale=0.33]{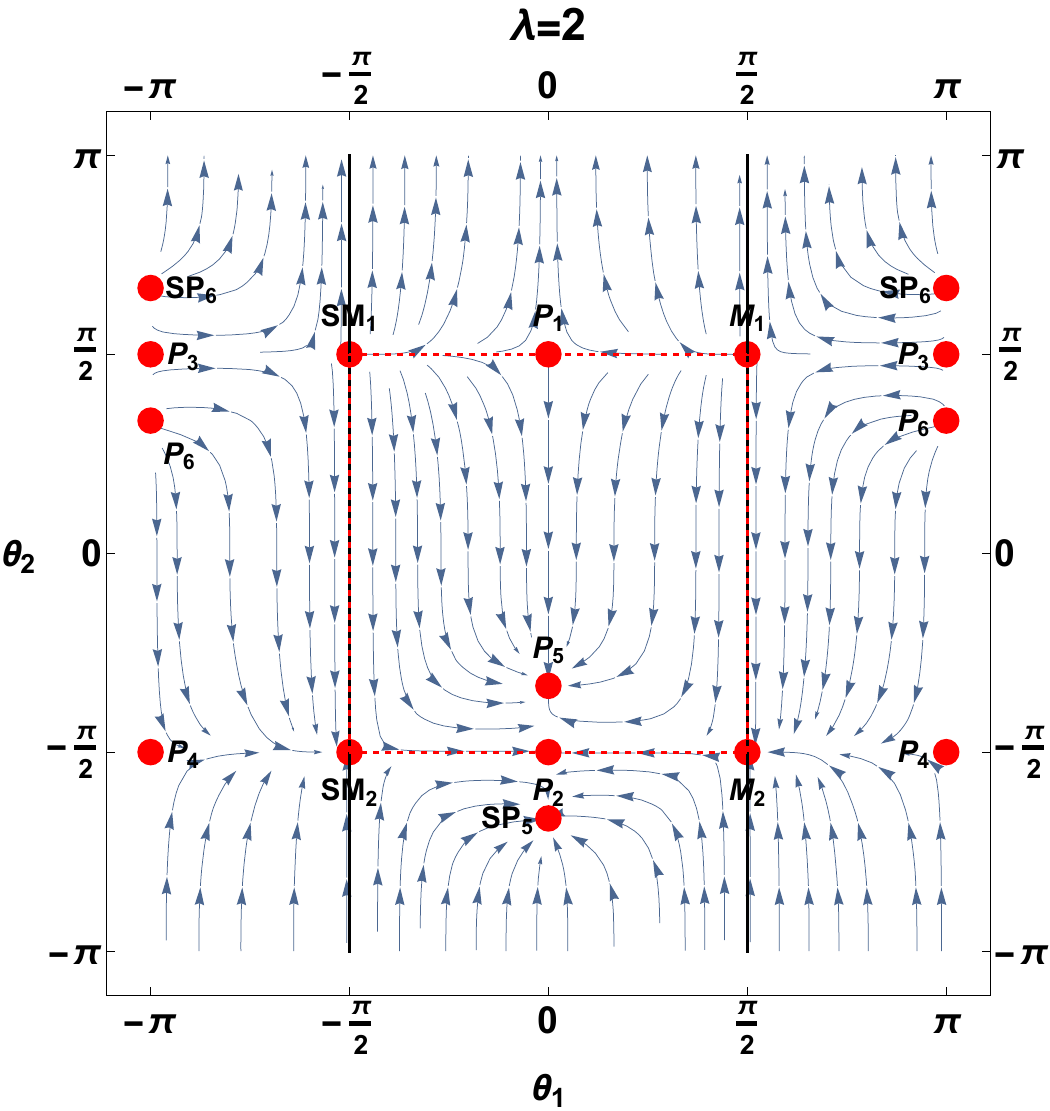}
    \includegraphics[scale=0.33]{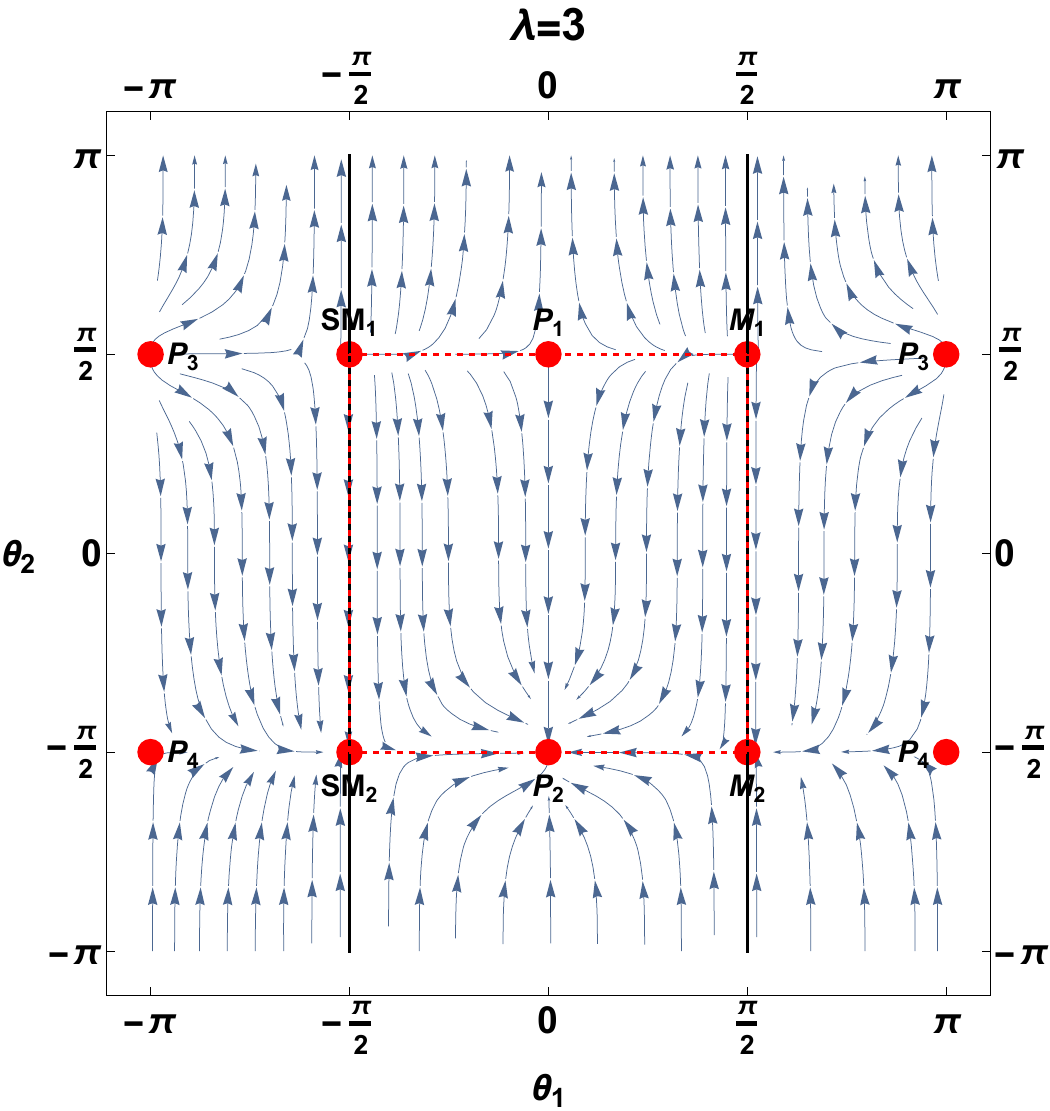}
    \caption{Phase plane analysis for system \eqref{dy-sys-1}-\eqref{dy-sys-2} for the values $\lambda= 1, 2, 3$.  The physical region under this formulation is the inside of the orange square. We labelled the points according to the corresponding points in section \ref{sect-3-1-1}. The points with different labels like $SM_1$ represent symmetric points that were not present In the original analysis because of the constraints on the $x_i$ variables. The black vertical lines correspond to the values $\pm \frac{\pi}{2}$ where the inverse transformations are not defined.}
    \label{fig:1}
\end{figure}

\begin{figure}
    \centering
    \includegraphics[scale=0.33]{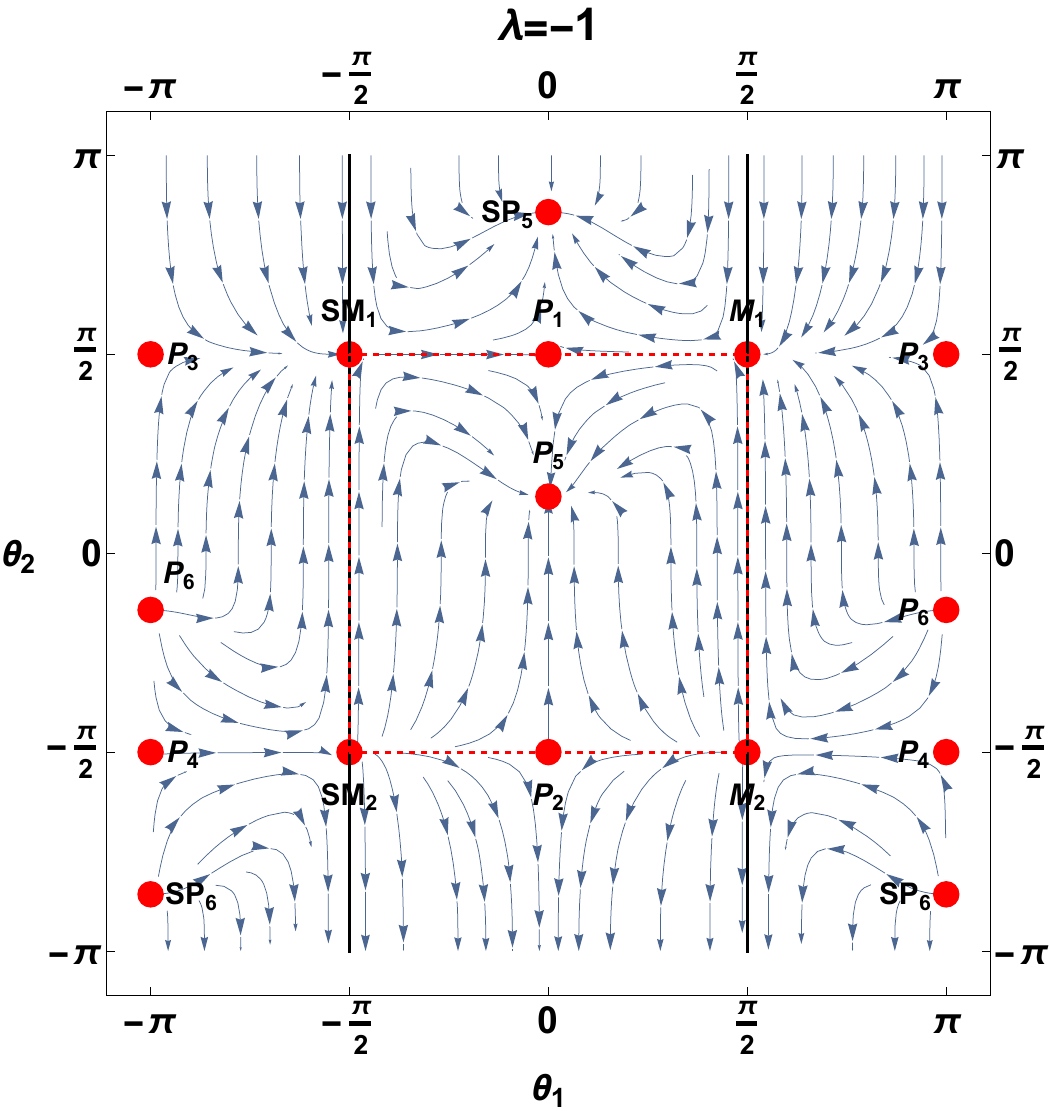}
    \includegraphics[scale=0.33]{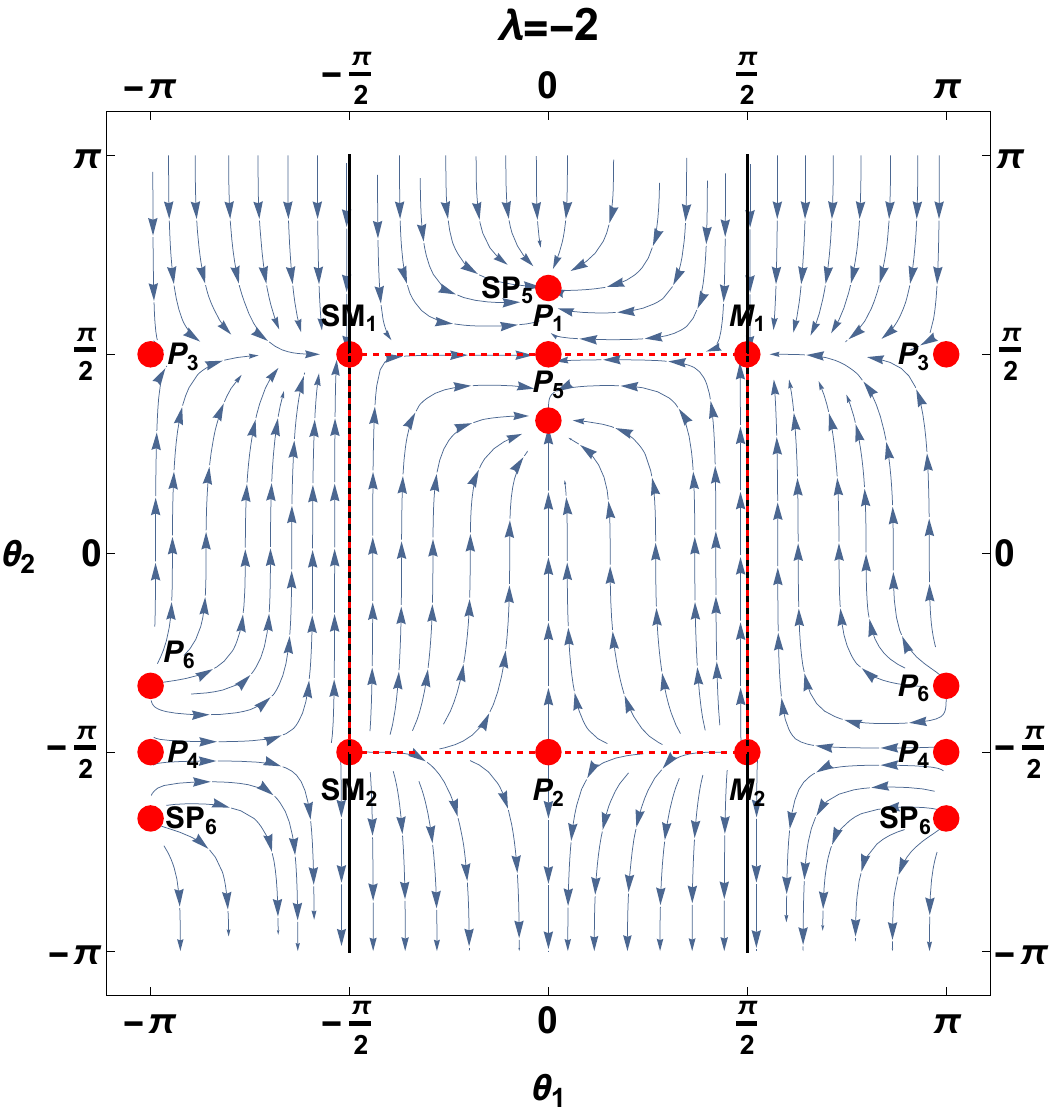}
    \includegraphics[scale=0.33]{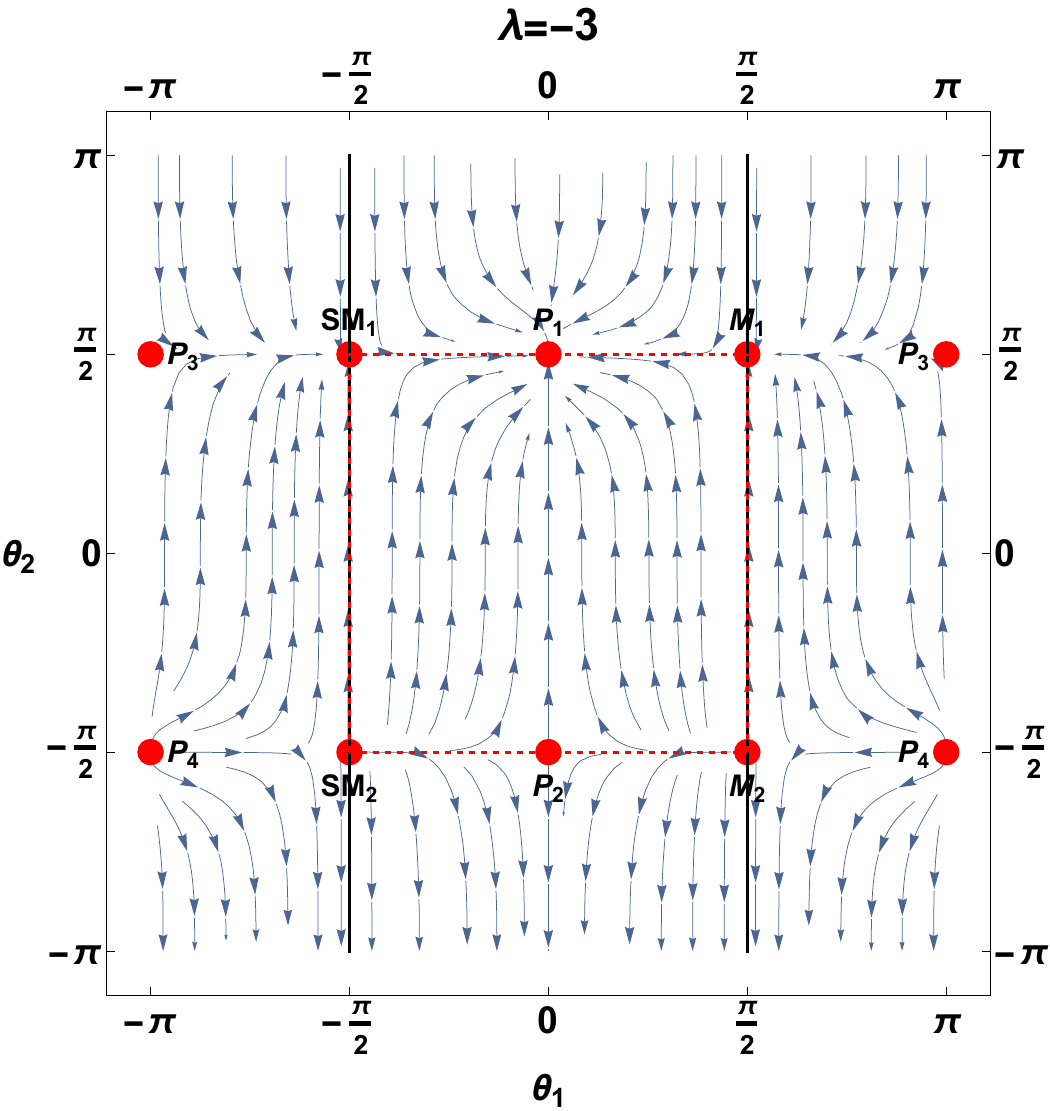}
    \caption{Phase plane analysis for system \eqref{dy-sys-1}-\eqref{dy-sys-2} for the values $\lambda=-1, -2, -3$.  The physical region under this formulation is the inside of the orange square. We labelled the points according to the corresponding points in section \ref{sect-3-1-1}. The points with different labels like $SM_1$ represent symmetric points that were not present In the original analysis because of the constraints on the $x_i$ variables. The black vertical lines correspond to the values $\pm \frac{\pi}{2}$ where the inverse transformations are not defined.}
    \label{fig:1aa}
\end{figure}

In Table \ref{tab:3}, we present the selected equilibrium points and their corresponding counterparts of section \ref{sect-3-1-1}. We chose some values for the constants $c_{1,2}$ that appear because of the periodicity of the trigonometric functions. These choices ensure that the equilibrium points lie in the desired range. For the plots presented in the following figures, we use the labels of the points from section \ref{sect-3-1-1}. 
In figures \ref{fig:1} and \ref{fig:1aa}, we depict the two-dimensional phase portraits for system \eqref{dy-sys-1}-\eqref{dy-sys-2} for different values of the parameter $\lambda$.  Recall that under this alternative formulation, the physical region is highlighted inside the dashed-orange square $(-\frac{\pi}{2},\frac{\pi}{2}),\times (-\frac{\pi}{2},\frac{\pi}{2})$.  On the other hand, the vertical black lines correspond to the values where the tangent function in equation \eqref{inverse-transformation} is not defined.

\begin{figure}[h!]
    \centering
    \includegraphics[scale=0.35]{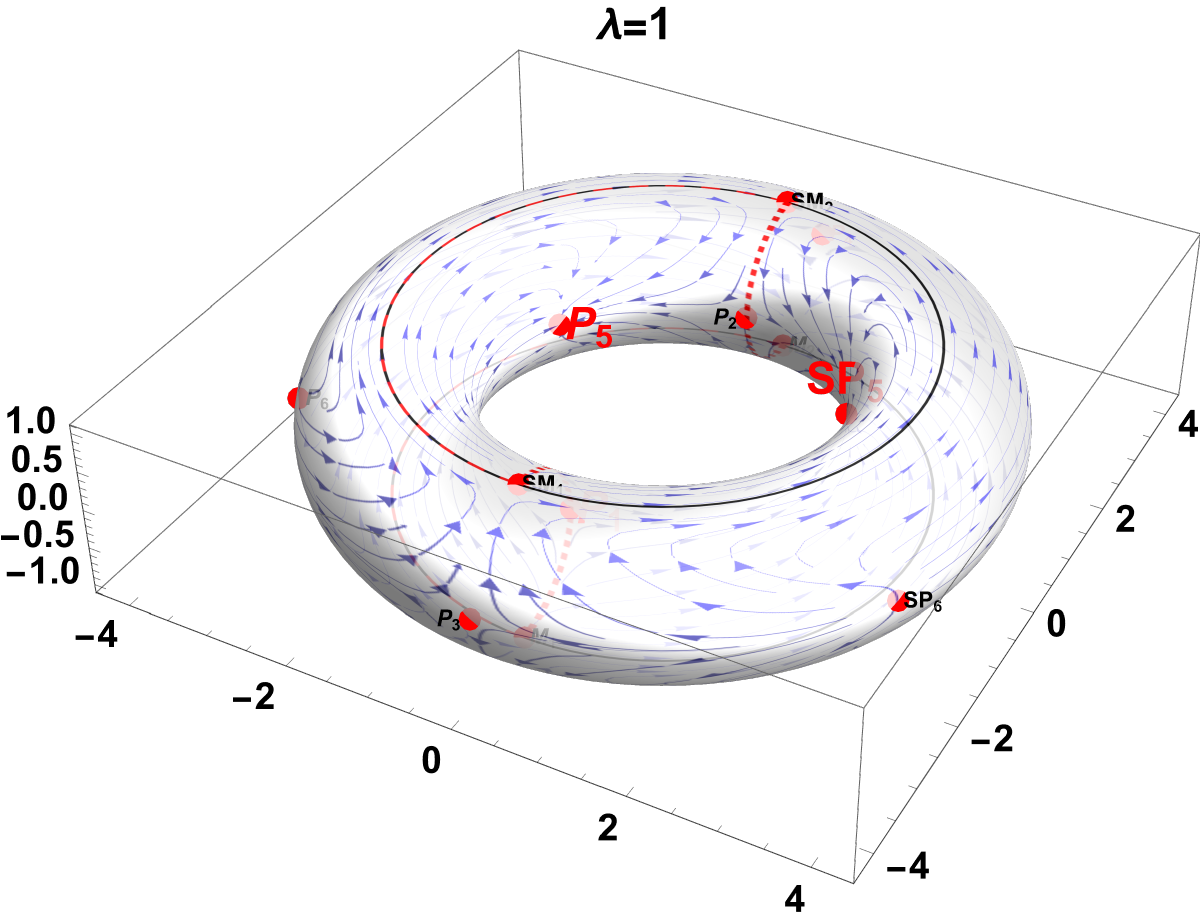}
    \includegraphics[scale=0.35]{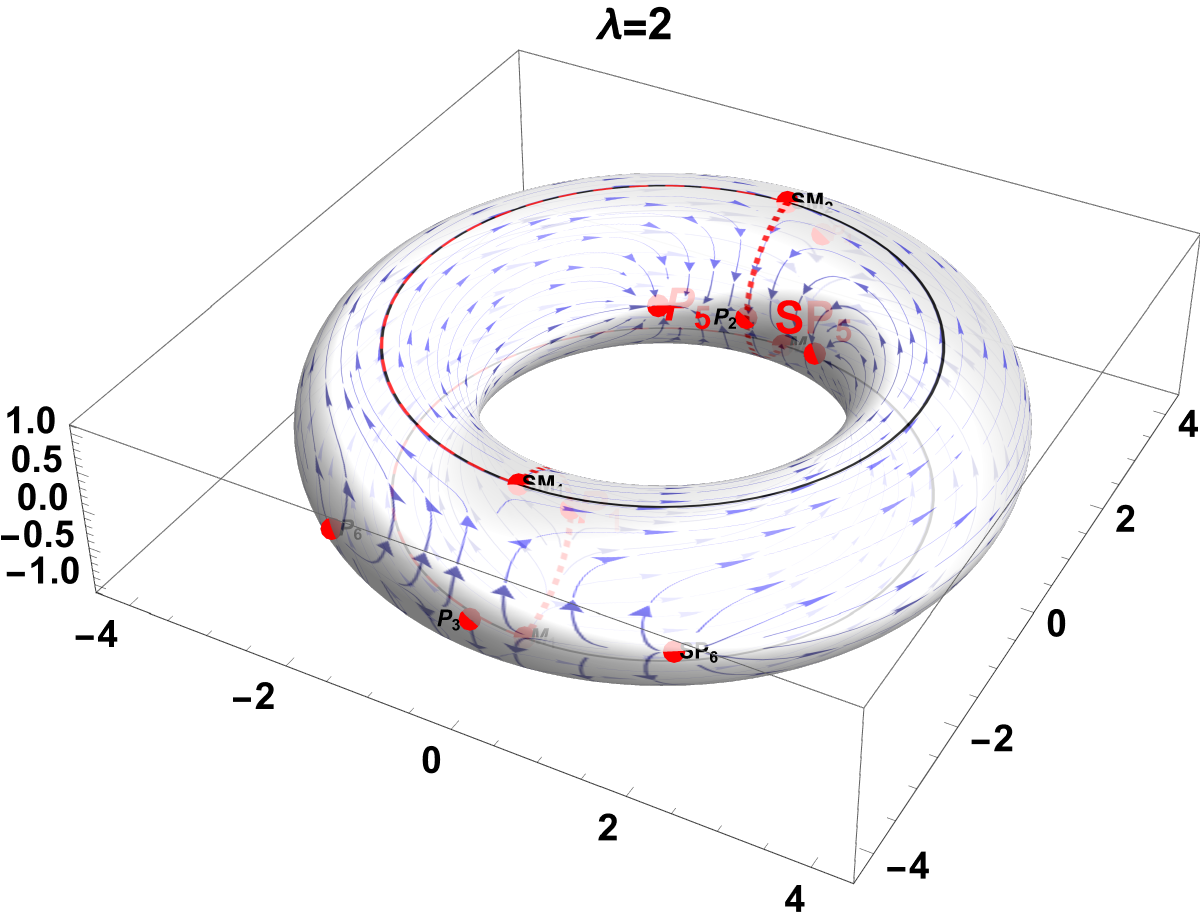}
    \includegraphics[scale=0.35]{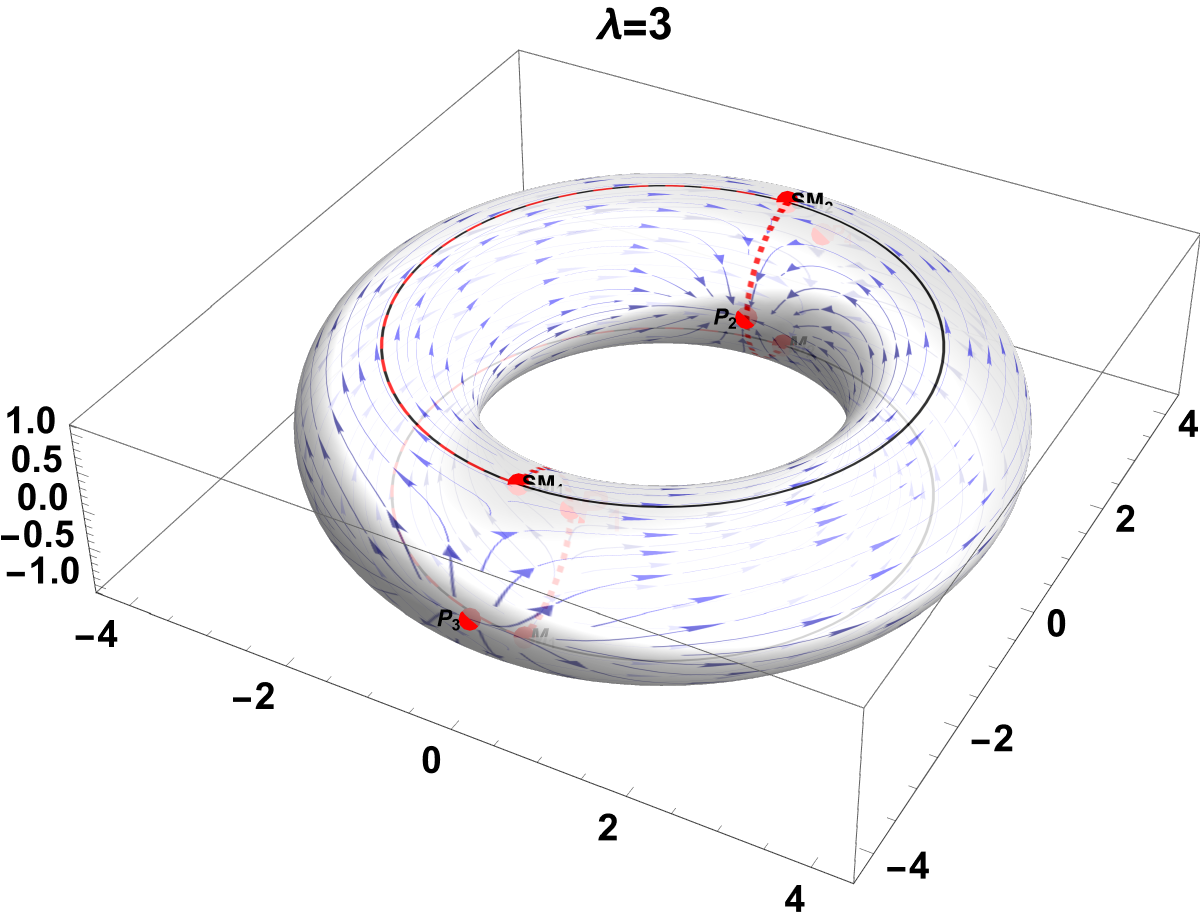}
    \caption{Dynamics of system \eqref{dy-sys-1}-\eqref{dy-sys-2} in the surface of a torus using the following parametrization $x=(R+r\cos(\theta_1-\pi))\cos(\theta_2-\pi),\quad x=(R+r\cos(\theta_1-\pi))\sin(\theta_2-\pi)$,  and $z=r\sin(\theta_1-\pi)$ for $r=1$ and $R=3$.  We have considered $\theta_1$ and $\theta_2$ as the poloidal and toroidal directions. As the 2-dimensional phase plots of Fig. \ref{fig:1} wrap around the torus, we can transfer from one branch to another. The values $\lambda=1,2,3$ were considered noting that in figure \ref{fig:1} the dynamic for the negative values is symmetrical.}
    \label{fig:2}
\end{figure}

\begin{figure}[h!]
    \centering
    \includegraphics[scale=0.35]{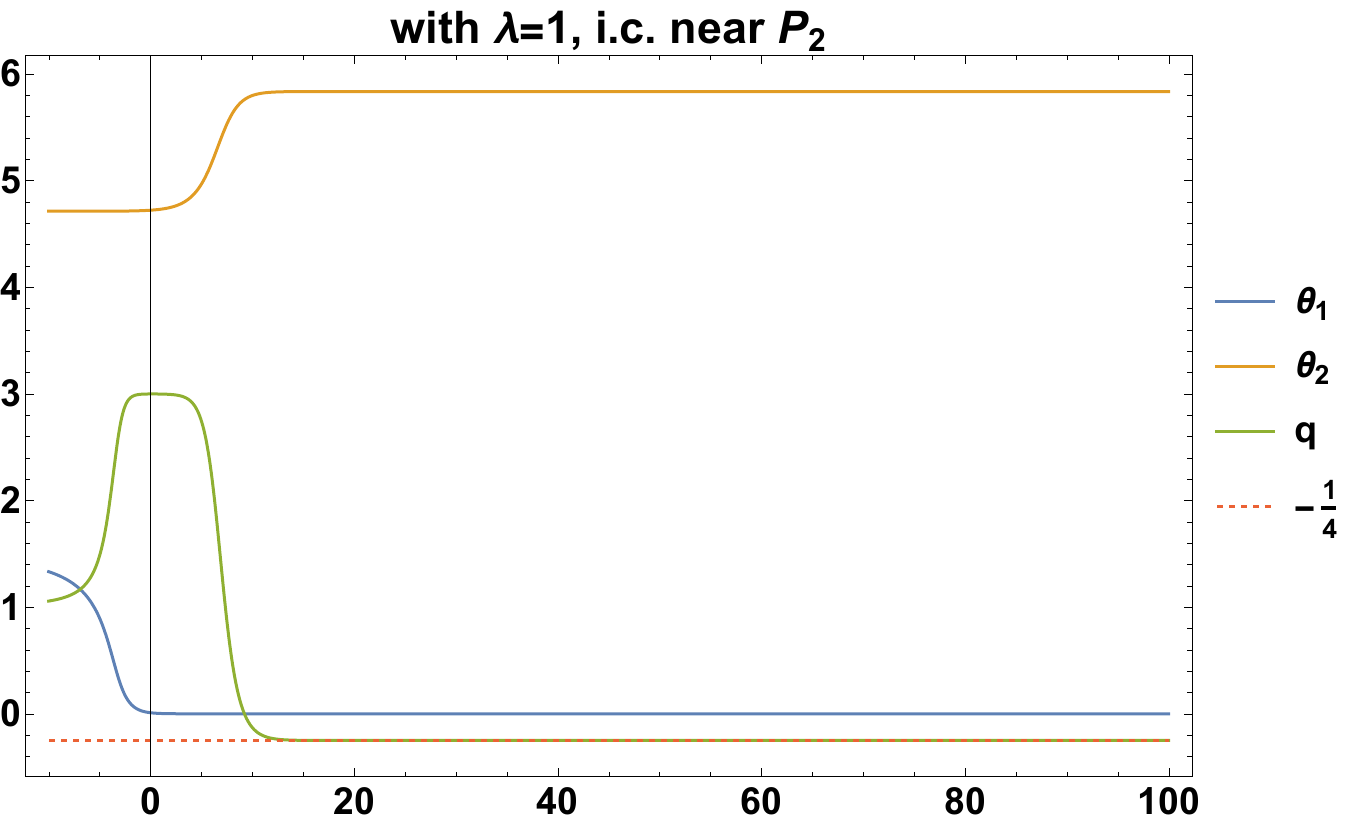}
    \includegraphics[scale=0.35]{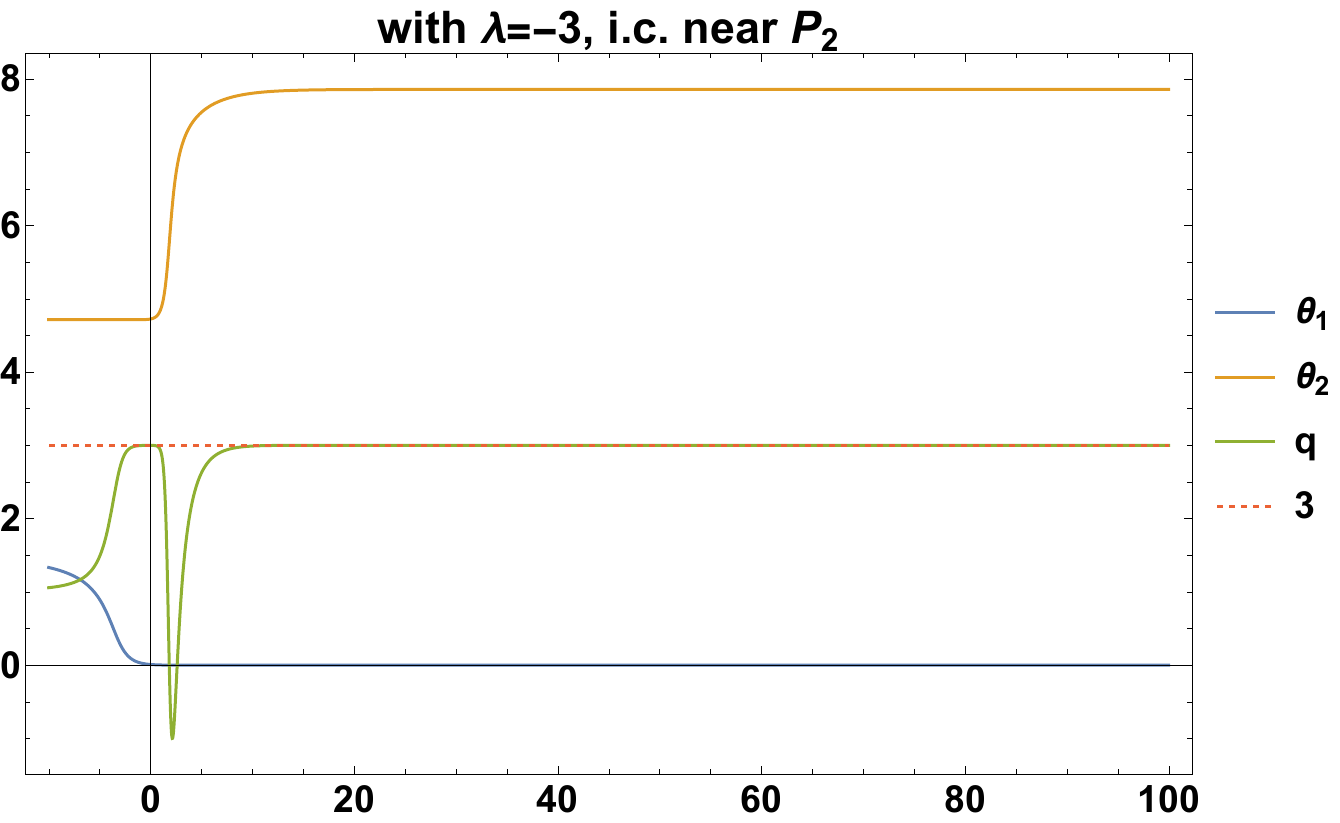}
    \caption{Numerical solutions of system \eqref{dy-sys-1}-\eqref{dy-sys-2} for different values of the parameter $\lambda=1,-3$ and initial conditions near $P_2$ displacement of $\xi=10^{-3}$.  Also, the deceleration parameter is depicted as a green line in the plots. For $\lambda=1$, we see that after an initial decelerated stage, the solution accelerates to $q(P_5)\rightarrow -\frac{1}{4}$,  the scaling solution. For $\lambda=- 3$, the solution is decelerated since $q\rightarrow 3$ but has a brief accelerated stage when $q<0$. }
    \label{fig:3}
\end{figure}

In figure \ref{fig:2} the dynamics of system \eqref{dy-sys-1}-\eqref{dy-sys-2} is depicted as ``wrapped'' around a torus given by the parametrization  $x=(R+r\cos(\theta_1-\pi))\cos(\theta_2-\pi),\quad y=(R+r\cos(\theta_1-\pi))\sin(\theta_2-\pi)$,  and $z=r\sin(\theta_1-\pi)$ for $r=1$ and $R=3$.  We are motivated to present our results this way because the original variables $x_i$ lie within two circumferences defined by the constraint \eqref{cont-circum}. One is represented by the angle $\theta_1$ and the other by $\theta_2$, the poloidal and toroidal directions of the torus.
In figure \ref{fig:3} we present some numerical solutions of system \eqref{dy-sys-1}-\eqref{dy-sys-2} for two values of the parameter $\lambda=1,-3$ and initial conditions near $P_2$ displacement of $\xi=10^{-3}$.  A possible model evolution can be determined by analysing the behaviour of the deceleration parameter, depicted as a green line. For $\lambda=1$, we see that after an initial decelerated stage, the solution accelerates to $q(P_5)\rightarrow -\frac{1}{4}$,  the scaling solution. The solution is decelerated for $\lambda=- 3$ since $q\rightarrow 3$ but has a brief accelerated stage when $q<0$. 
    
 \begin{table}[ht]
\begin{tabular}{|c|c|c|c|}
\hline
Label & $(\theta_1,\theta_2)$ & $c_1, c_2$ & Sect. \ref{sect-3-1-1} point \\ \hline
$T_1$ & $(0,\frac{\pi}{2})$ & $0,0$ & $P_1$ \\ \hline
$T_2$ & $(0,-\frac{\pi}{2})$ & $0,0$ &  $P_2$ \\ \hline
$T_3$ & $\left(\frac{\pi}{2},\frac{\pi}{2}\right)$ & $0,0$ & $M_1$ \\ \hline
$T_4$ & $\left(-\frac{\pi}{2},\frac{\pi}{2}\right)$ & $0,0$ & Symmetric to $M_1$ \\ \hline
$T_5$ & $\left(\frac{\pi}{2},-\frac{\pi}{2}\right)$ & $0,0$ & $M_2$\\ \hline
$T_6$ & $\left(-\frac{\pi}{2},-\frac{\pi}{2}\right)$ & $0,0$ & Symmetric to $M_2$ \\ \hline
$T_7$ & $\left(\pi,\frac{\pi}{2}\right)$ & $0,0$  & $P_3$ \\ \hline
$T_{7}*$ & $\left(-\pi,\frac{\pi}{2}\right)$ & $-1,0$  & $P_3$\\ \hline
$T_8$ & $\left(\pi,-\frac{\pi}{2}\right)$ & $0,0$  & $P_4$ \\ \hline
$T_{8}*$ & $\left(-\pi,-\frac{\pi}{2}\right)$ & $-1,0$  & $P_4$\\ \hline
$T_9$ & $\left(0,\tan
   ^{-1}\left(\sqrt{48-9 \lambda ^2},-3 \lambda \right)\right)$ & $0,0$ & $P_5$ \\ \hline
$T_{10}$ & $\left(0,\tan
   ^{-1}\left(-\sqrt{48-9 \lambda ^2},-3 \lambda \right)\right)$ & $0,0$ & Symmetric to $P_5$ \\ \hline
$T_{11}$ & $\left(\pi, \tan ^{-1}\left(\sqrt{48-9 \lambda ^2},3 \lambda \right)\right)$ & $0,0$ & $P_6$\\ \hline
$T_{11}*$ & $\left(-\pi, \tan ^{-1}\left(\sqrt{48-9 \lambda ^2},3 \lambda \right)\right)$ & $-1,0$ &  $P_6$\\ \hline
$T_{12}$ & $\left(\pi, \tan ^{-1}\left(-\sqrt{48-9 \lambda ^2},3 \lambda \right)\right)$ & $0,0$ & Symmetric to $P_6$\\ \hline
$T_{12}*$ & $\left(-\pi, \tan ^{-1}\left(-\sqrt{48-9 \lambda ^2},3 \lambda \right)\right)$ & $-1,0$ &  Symmetric to $P_6$\\ \hline
\end{tabular}
\caption{Summary of the equilibrium points of system \eqref{dy-sys-1}-\eqref{dy-sys-2}. The corresponding points from section \ref{sect-3-1-1} are derived by applying the definitions in equation \eqref{reparametrization}. The asterisk next to some points means a different choice of $c_1$ was made from the original list. For every point, the stability analysis and the value of the deceleration parameter are the same as in section \ref{sect-3-1-1}.}
\label{tab:3}
\end{table}

 \subsection{Linear coupling function: $f(\phi)=-\alpha\phi$}
 \label{sect-3-2}
In this section, we define the coupling between the scalar function and the Gauss-Bonnet term as a linear function $f(\phi)=-\alpha\phi$. Hence, the field equations are
 \begin{align}
\label{linear_field_1}
12 H^2 \left(4
   \alpha  \phi  \dot{H}+4 \alpha  \ddot{\phi}+1\right)+96 \alpha  H \dot{H} \dot{\phi}+6 \dot{H}+144 \alpha  H^3 \dot{\phi}+48 \alpha  H^4 \phi -2 V(\phi )+\dot{\phi}^2=& 0,\\
   \label{linear_field_2}
   96 \alpha  H^2 \dot{H}+120 \alpha  H^4-4 H \dot{\phi}-V'(\phi )-\ddot{\phi}=&0.
 \end{align}
 We can also write Friedmann's equation \eqref{general_friedmann_1} as 
 \begin{equation}
         \label{linear_friedmann}96 \alpha  H^3 \dot{\phi}+24 \alpha  H^4 \phi +6 H^2= V(\phi )+\frac{1}{2} \dot{\phi}^2,
 \end{equation}
 and the deceleration parameter \eqref{def-of-q} as
 \begin{equation}
 \label{linear_deceleration}
     q=\frac{-144 \alpha  H^3 \dot{\phi}+6 H^2 \left(192 \alpha ^2
   H^4-8 \alpha  \lambda  V(\phi )+1\right)-2 V(\phi
   )+\dot{\phi}^2}{6 H^2 \left(8 \alpha  H \left(96 \alpha
    H^3+H \phi +2 \dot{\phi}\right)+1\right)}.
 \end{equation}
\subsubsection{Dynamical system analysis for linear $f$}
\label{sect-3-2-1}
Once again we use the normalisation variable $\chi$ defined in \eqref{def-of-chi}  and define the following dimensionless variables
 \begin{equation}
 \label{linear_dimensionless_var}
     x_1=\frac{\sqrt{6} H}{\sqrt{\chi
   }},\quad x_2=\frac{2 \sqrt{6} \sqrt{\alpha  H^4
   \phi }}{\sqrt{\chi
   }},\quad x_3=\frac{\sqrt{V(\phi )}}{\sqrt{\chi
   }},\quad x_4=\frac{\dot{\phi}}{\sqrt{2} \sqrt{\chi
   }},\quad x_5=\frac{4 \sqrt{6} \sqrt{\alpha 
   H^3}}{\sqrt[4]{\chi }},
 \end{equation}
 with inverse transformation given by
 \begin{equation}
     H= \frac{x_1 \sqrt{\chi
   }}{\sqrt{6}},\quad \phi = \frac{4 \sqrt{6}
   x_2^2}{x_1 x_5^2},
   \quad V(\phi )=
   x_3^2 \chi,
   \quad \dot{\phi}= \sqrt{2} x_4
   \sqrt{\chi },
   \quad \alpha = \frac{\sqrt{\frac{3}{2}}
   x_5^2}{8 x_1^3 \chi }.
 \end{equation}
 By replacing these variables in \eqref{linear_friedmann}, we obtain the following restrictions
 \begin{equation}
 \label{rest_linear}
     x_1^2+x_2^2+\sqrt{2} x_4
   x_5^2=1,\quad x_3^2+x_4^2=1,
 \end{equation}
 we also have the following ranges for the new variables $0\leq x_2$,  $0\leq x_3\leq 1$,  $-1\leq x_4\leq 1$,  $0\leq x_5$. 
With this, we can write the following five-dimensional dynamical system
 \begin{align}
     \label{linear_eq_x1}
     x_1'&=\frac{x_1 \left(x_1^3 \left(32 x_4^2+3
   \sqrt{2} x_4 x_5^2-16\right)+x_1
   \left(-2 \sqrt{2} x_4 x_5^2
   \left(x_2^2+4 x_3^2-2\right)+2
   x_4^2 \left(32
   x_2^2+x_5^4-8\right)\right)\right)}{4 \sqrt{6} \left(2
   x_1^2+4 x_2^2+2 \sqrt{2} x_4
   x_5^2+x_5^4\right)}\\ \nonumber &+\frac{x_1 \left(-16 x_2^2+16
   x_3^2+40 \sqrt{2} x_4^3 x_5^2-5
   x_5^4\right)+4 \sqrt{3} \lambda  x_3^2
   x_5^2 \left(\sqrt{2}-x_4
   x_5^2\right)}{4 \sqrt{6} \left(2
   x_1^2+4 x_2^2+2 \sqrt{2} x_4
   x_5^2+x_5^4\right)}\\
      \label{linear_eq_x2}
     x_2'&=\frac{2 x_1^3 \left(16 \sqrt{2} x_2^2
   \left(x_4^2-1\right)+\left(3
   x_2^2+1\right) x_4
   x_5^2\right)-4 \sqrt{3} \lambda  x_2^2
   x_3^2 x_5^2 \left(\sqrt{2} x_4
   x_5^2-4\right)}{8 \sqrt{3} x_2 \left(2
   x_1^2+4 x_2^2+2 \sqrt{2} x_4
   x_5^2+x_5^4\right)}\\ 
   \nonumber &+\frac{x_1 \left(-4 x_2^2
   x_4 x_5^2 \left(x_2^2+4
   x_3^2-20 x_4^2-5\right)+32 \sqrt{2}
   x_2^2 \left(\left(2 x_2^2-1\right)
   x_4^2-x_2^2+x_3^2\right)+2 \sqrt{2}
   x_5^4 \left(x_2^2
   \left(x_4^2-5\right)+x_4^2\right)+x_4 x_5^6\right)}{8 \sqrt{3} x_2 \left(2
   x_1^2+4 x_2^2+2 \sqrt{2} x_4
   x_5^2+x_5^4\right)}\\ 
   \label{linear_eq_x3}
     x_3'&=\frac{x_3 x_4 \left(x_1^3 \left(32
   \sqrt{3} x_4+3 \sqrt{6} x_5^2\right)+24
   \lambda  x_1^2+2 \sqrt{3} x_1
   \left(x_2^2 \left(32 x_4-\sqrt{2}
   x_5^2\right)-4 \sqrt{2} x_5^2
   \left(x_3^2-5 x_4^2\right)+x_4
   x_5^4\right)\right)}{12
   \sqrt{2} \left(2 x_1^2+4 x_2^2+2 \sqrt{2}
   x_4 x_5^2+x_5^4\right)}\\
   \nonumber &+\frac{x_3 x_4 \left(12 \lambda  \left(4
   x_2^2-x_3^2 x_5^4+2 \sqrt{2}
   x_4 x_5^2+x_5^4\right)\right)}{12
   \sqrt{2} \left(2 x_1^2+4 x_2^2+2 \sqrt{2}
   x_4 x_5^2+x_5^4\right)}\\
    \label{linear_eq_x4}
     x_4'&=\frac{\sqrt{3} x_1^3 \left(x_4^2-1\right)
   \left(32 x_4+3 \sqrt{2} x_5^2\right)-12
   \lambda  x_3^2 \left(4 x_2^2+x_4
   x_5^2 \left(x_4 x_5^2+2
   \sqrt{2}\right)\right)}{12 \sqrt{2} \left(2
   x_1^2+4 x_2^2+2 \sqrt{2} x_4
   x_5^2+x_5^4\right)}\\
   \nonumber &+\frac{-24
   \lambda  x_1^2 x_3^2+2 \sqrt{3} x_1
   \left(x_4^2-1\right) \left(x_2^2 \left(32
   x_4-\sqrt{2} x_5^2\right)-4 \sqrt{2}
   x_5^2 \left(x_3^2-5
   x_4^2\right)+x_4 x_5^4\right)}{12 \sqrt{2} \left(2
   x_1^2+4 x_2^2+2 \sqrt{2} x_4
   x_5^2+x_5^4\right)}\\
         \label{linear_eq_x5}
     x_5'&=\frac{16
   x_1 x_5 \left(x_4^2 \left(2 x_1^2+4
   x_2^2-3\right)-3
   \left(x_1^2+x_2^2-x_3^2\right)\right)+x_5^5 \left(x_1 \left(2 x_4^2-15\right)-4 \sqrt{3}
   \lambda  x_3^2 x_4\right)}{8 \sqrt{6} \left(2
   x_1^2+4 x_2^2+2 \sqrt{2} x_4
   x_5^2+x_5^4\right)}\\
   \nonumber &+\frac{16
   x_1 x_5 \left(x_4^2 \left(2 x_1^2+4
   x_2^2-3\right)-3
   \left(x_1^2+x_2^2-x_3^2\right)\right)+x_5^5 \left(x_1 \left(2 x_4^2-15\right)-4 \sqrt{3}
   \lambda  x_3^2 x_4\right)}{8 \sqrt{6} \left(2
   x_1^2+4 x_2^2+2 \sqrt{2} x_4
   x_5^2+x_5^4\right)}.
   \end{align}
 In these variables, the deceleration parameter \eqref{linear_deceleration} reads
 \begin{equation}
     \label{deceleration_5d}
     q=\frac{8 x_1^3+x_1 \left(-16 x_3^2+16
   x_4^2-12 \sqrt{2} x_4
   x_5^2+x_5^4\right)-4 \sqrt{6} \lambda  x_3^2
   x_5^2}{4 x_1 \left(2 x_1^2+4 x_2^2+2
   \sqrt{2} x_4 x_5^2+x_5^4\right)}
 \end{equation}

 We can use the restrictions \eqref{rest_linear} together with the fact that $x_2\geq0$ and $x_3\geq0$ to obtain the following positive roots
 \begin{equation}
 \label{def_of_x2_x3}
     x_2= \sqrt{-x_1^2-\sqrt{2} x_4
   x_5^2+1},\quad x_3= \sqrt{1-x_4^2}
 \end{equation}
provided $x_1^2 + \sqrt{2} x_4  x_5^2 \leq 1$.
   
 With these definitions, we can obtain the following reduced three-dimensional dynamical system
 \begin{small}
  \begin{align}
     \label{reduced_linear_eq_x1}
     x_1'&=\frac{x_1 \left(\sqrt{2} x_5^2 \left(-5 x_1^3
   x_4+2 x_1 x_4 \left(8 x_4^2-5\right)+4
   \sqrt{3} \lambda  \left(x_4^2-1\right)\right)+32
   x_1 \left(x_1^2-1\right) x_4^2+x_5^4
   \left(x_1 \left(5-6 x_4^2\right)-4 \sqrt{3} \lambda
    x_4 \left(x_4^2-1\right)\right)\right)}{4 \sqrt{6}
   \left(2 x_1^2+2 \sqrt{2} x_4
   x_5^2-x_5^4-4\right)},\\
     \label{reduced_linear_eq_x4}
     x_4'&=\frac{\left(x_4^2-1\right) \left(x_1^3 \left(32
   \sqrt{3} x_4-5 \sqrt{6} x_5^2\right)+24 \lambda 
   x_1^2+2 \sqrt{3} x_1 \left(\sqrt{2} \left(8
   x_4^2+5\right) x_5^2-3 x_4 x_5^4-32
   x_4\right)+12 \lambda  \left(-x_4^2 x_5^4+2
   \sqrt{2} x_4 x_5^2-4\right)\right)}{12 \sqrt{2}
   \left(2 x_1^2+2 \sqrt{2} x_4
   x_5^2-x_5^4-4\right)},\\
     \label{reduced_linear_eq_x5}
     x_5'&=\frac{\sqrt{2} x_5^3 \left(5 x_1^3 x_4+2
   x_1 x_4 \left(25-8 x_4^2\right)-12 \sqrt{3}
   \lambda  \left(x_4^2-1\right)\right)-32 x_1
   \left(x_1^2+1\right) x_4^2 x_5+x_5^5
   \left(3 x_1 \left(2 x_4^2-5\right)+4 \sqrt{3}
   \lambda  x_4 \left(x_4^2-1\right)\right)}{8
   \sqrt{6} \left(-2 x_1^2-2 \sqrt{2} x_4
   x_5^2+x_5^4+4\right)}.
 \end{align}
 \end{small}

\begin{figure}
    \centering
    \includegraphics[scale=0.35]{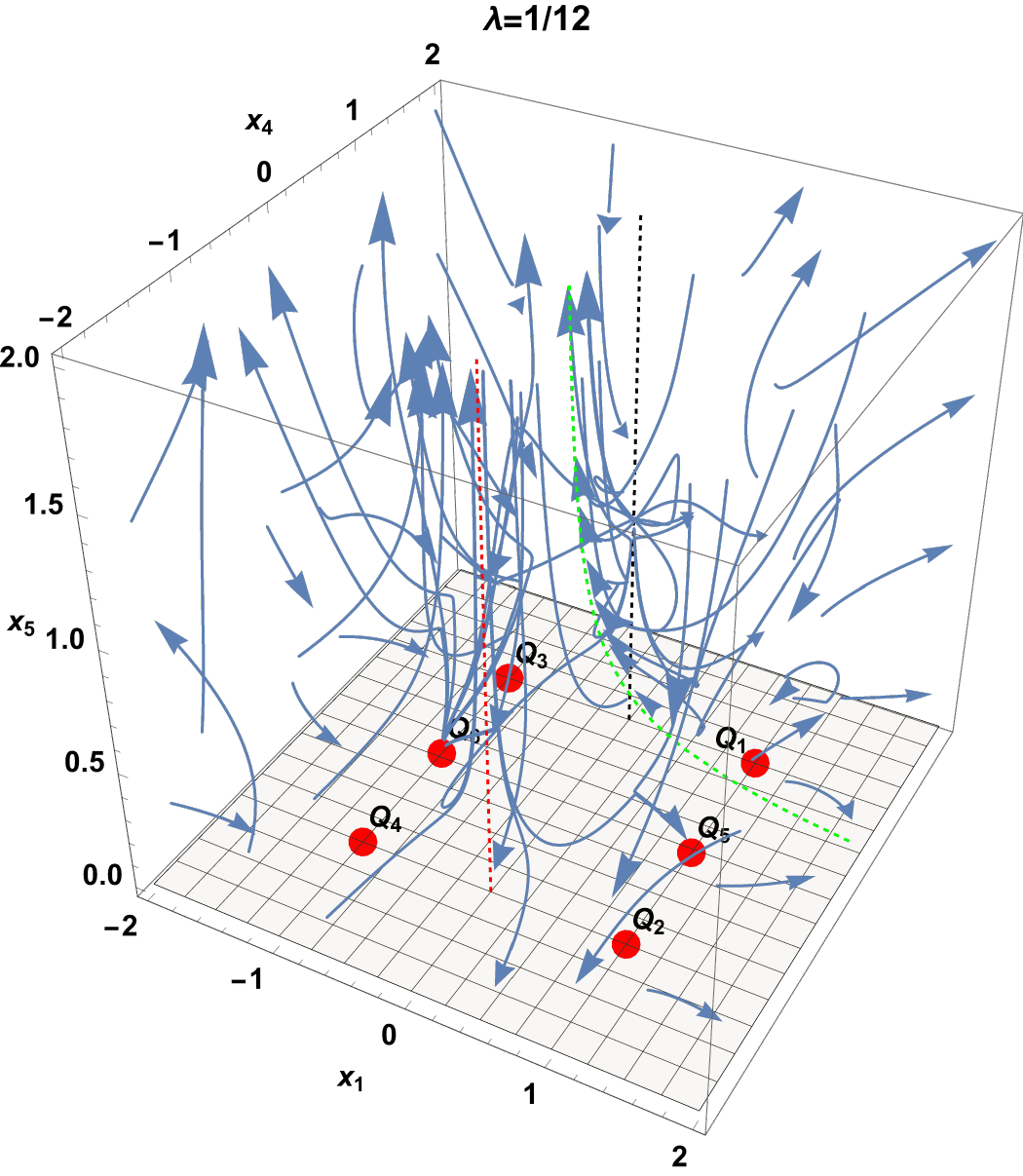}
    \includegraphics[scale=0.35]{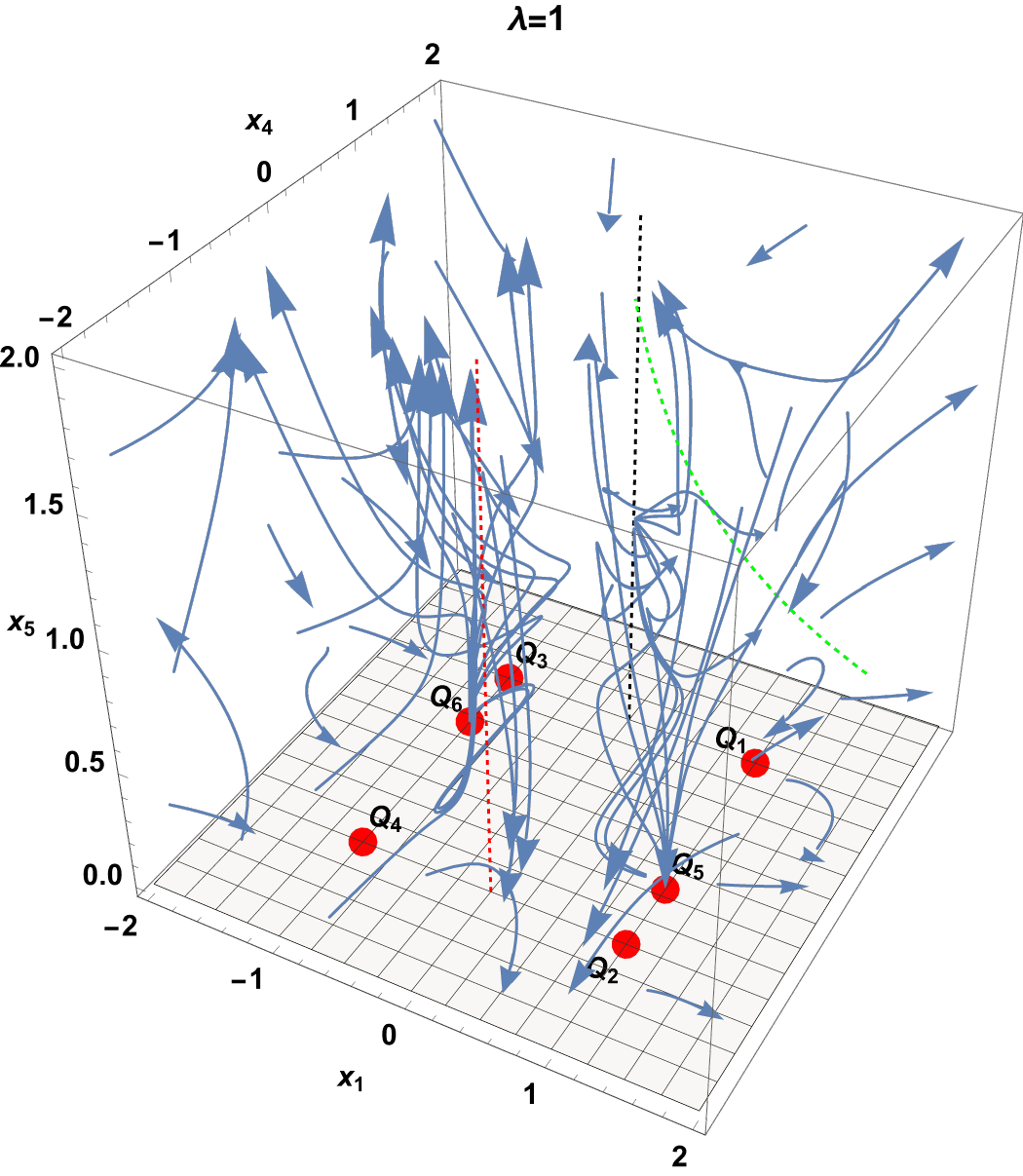}
    \includegraphics[scale=0.35]{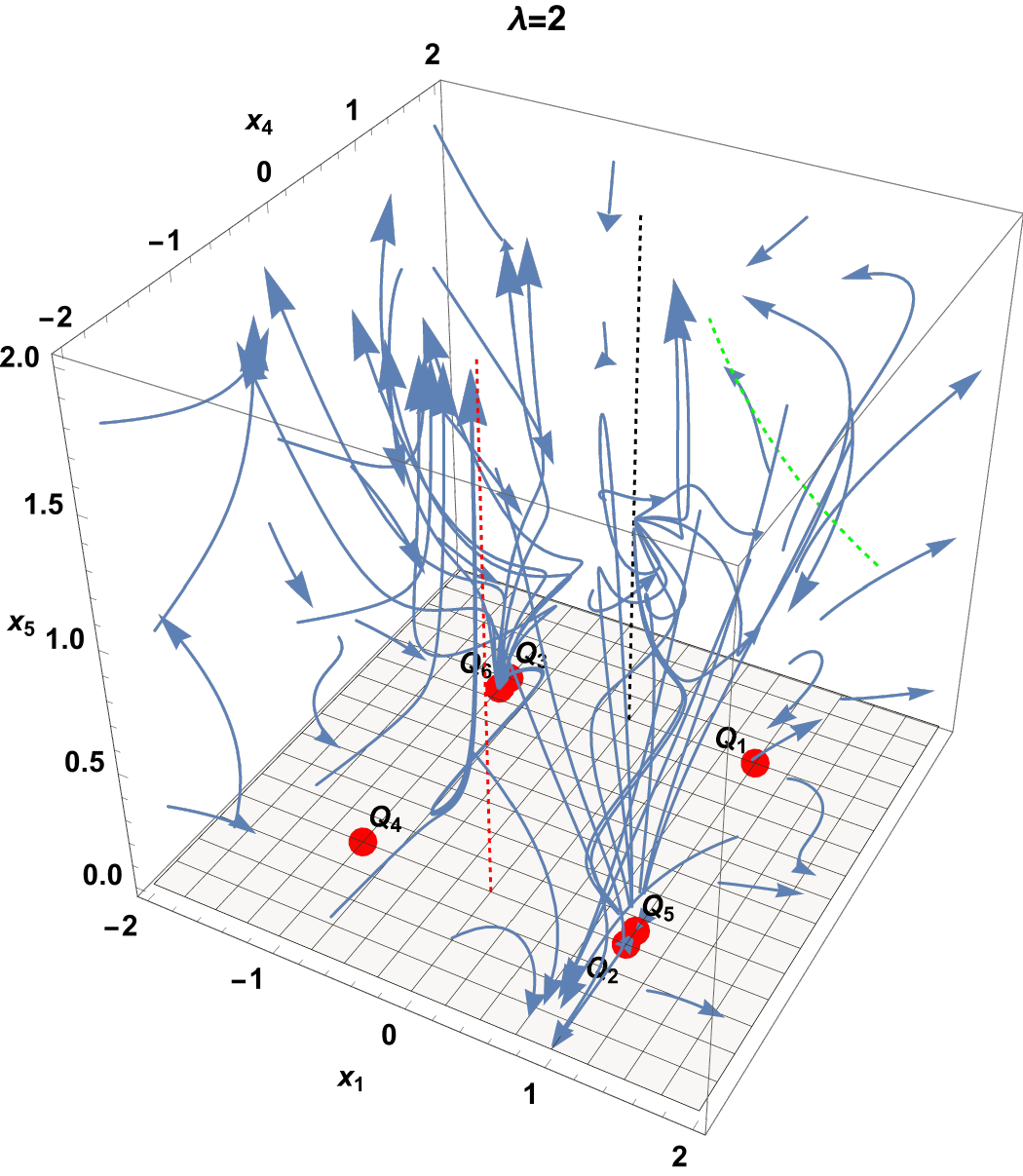}
    \includegraphics[scale=0.35]{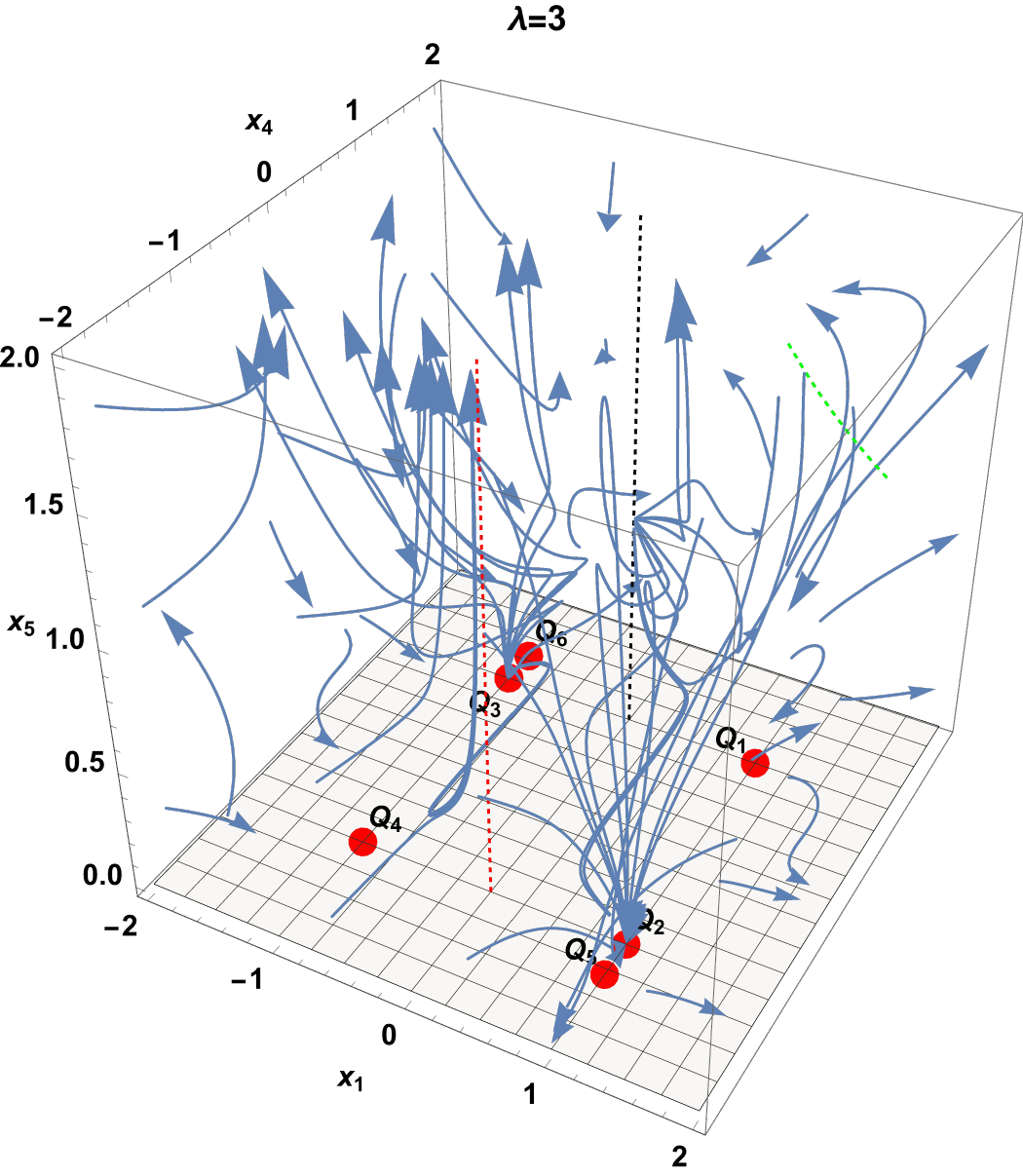}
    \caption{Dynamics of the three-dimensional system \eqref{reduced_linear_eq_x1}-\eqref{reduced_linear_eq_x5} for positive values of $\lambda$.  The green-dashed curve corresponds to the family of equilibrium points $F_1$, which is the attractor for $\lambda=\frac{1}{12}$.  The black-dashed line represents the family $L_1 $, and the red-dashed line represents the family $L_2$. }
    \label{fig:3d-1}
\end{figure}

\begin{figure}
    \centering
    \includegraphics[scale=0.35]{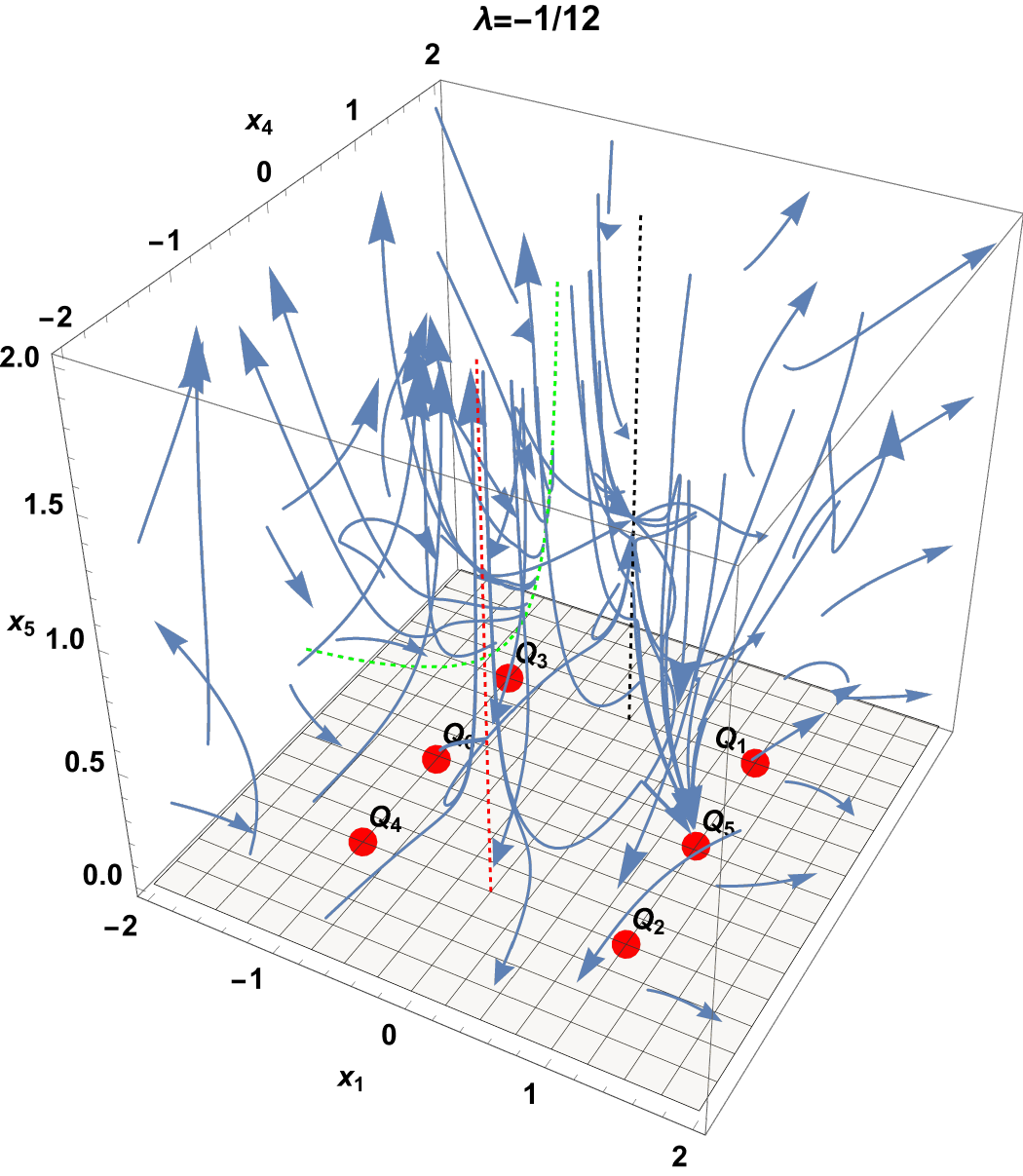}
    \includegraphics[scale=0.35]{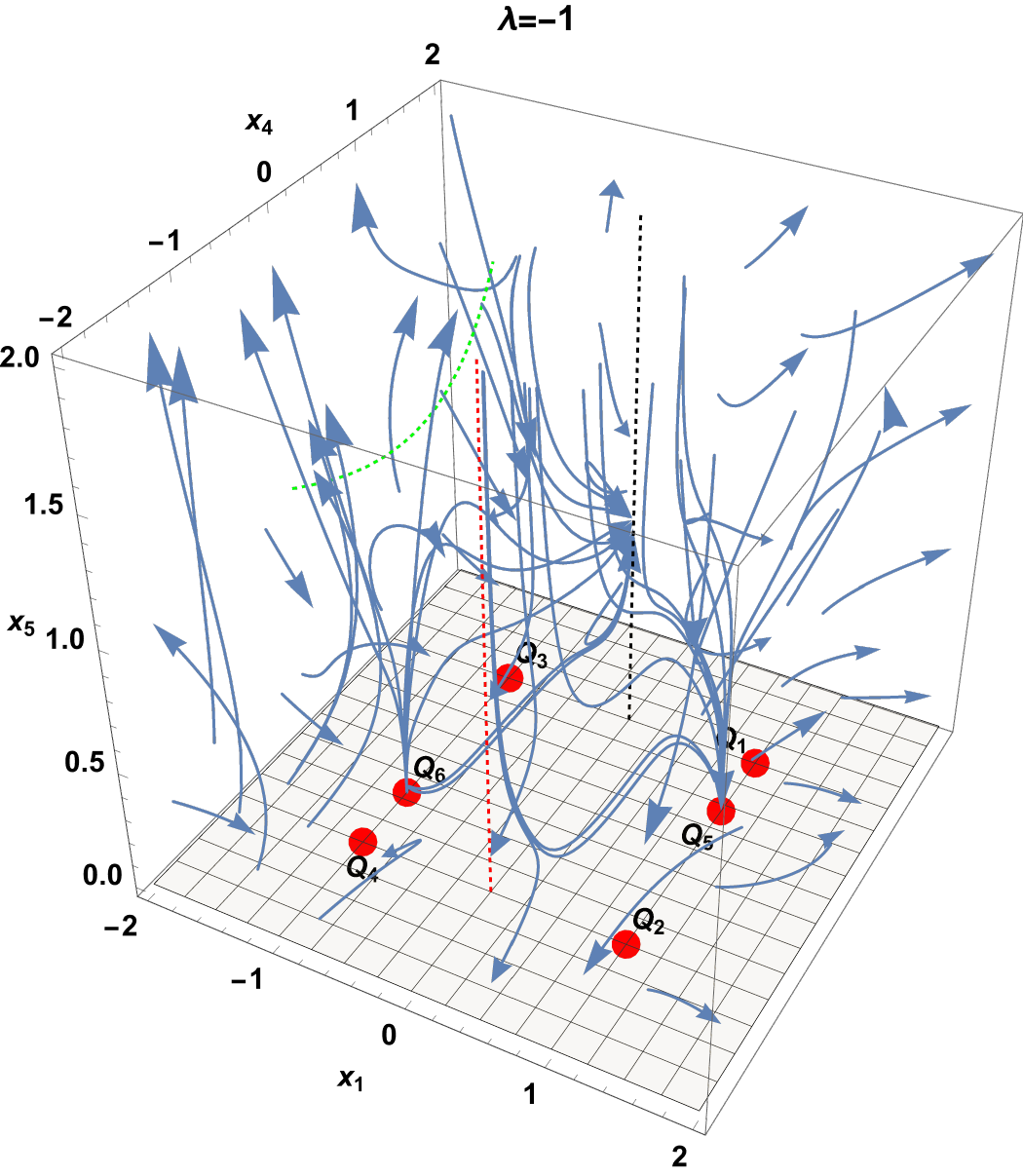}
    \includegraphics[scale=0.35]{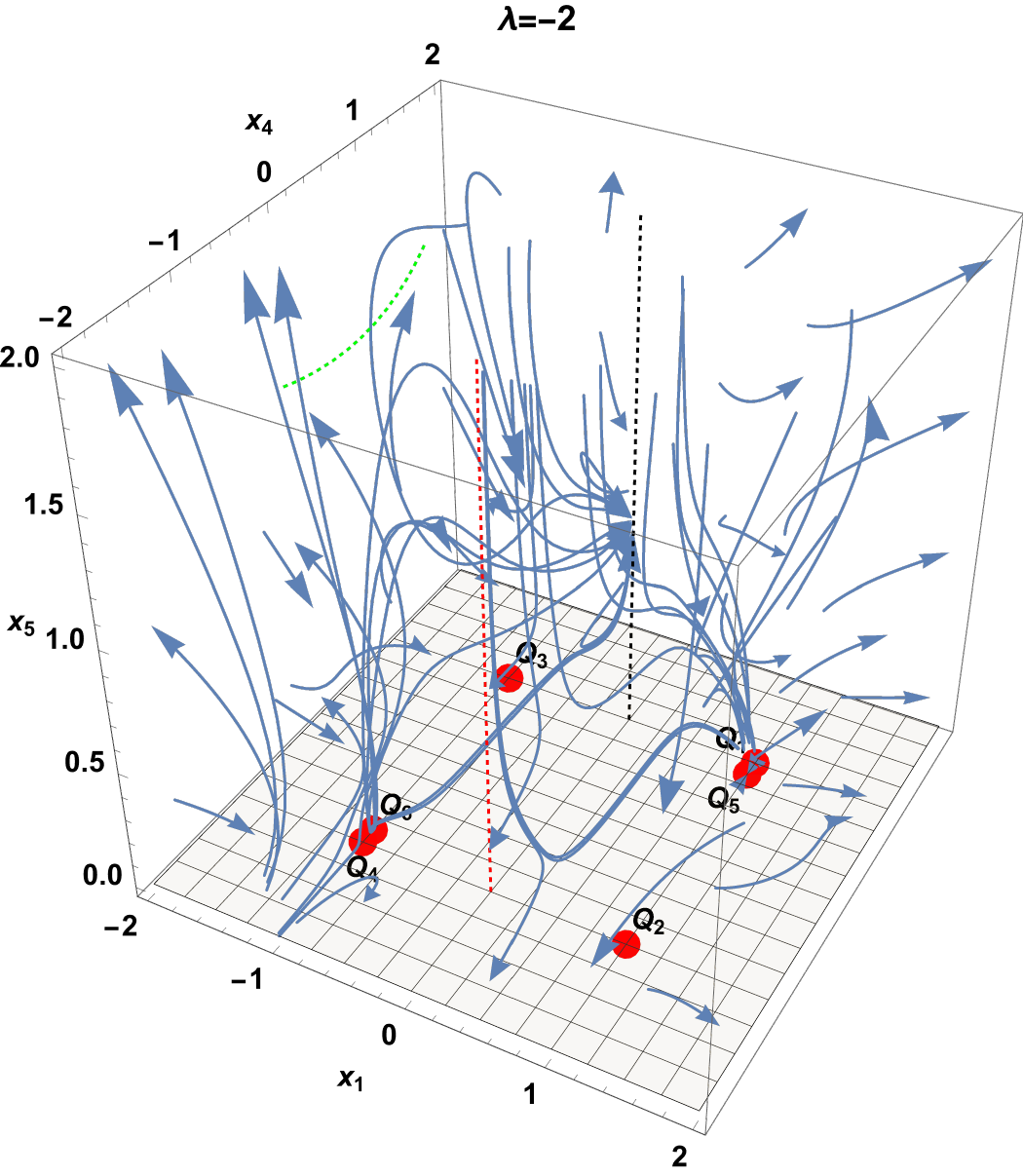}
    \includegraphics[scale=0.35]{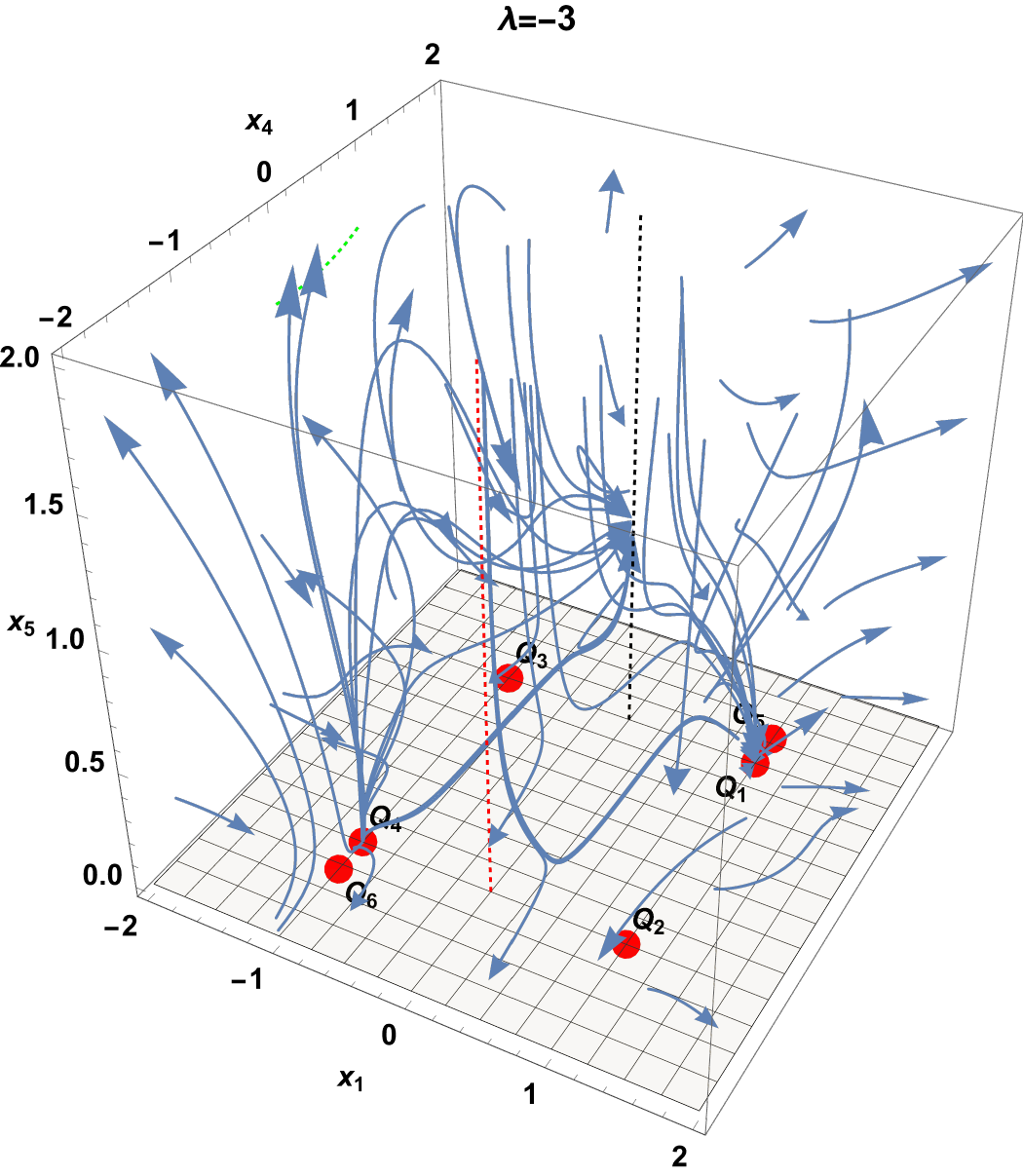}
    \caption{Dynamics of the three-dimensional system \eqref{reduced_linear_eq_x1}-\eqref{reduced_linear_eq_x5} for negative values of $\lambda$.  The green-dashed curve corresponds to the family of equilibrium points $F_1$.  The black-dashed line represents the family $L_1 $, and the red-dashed line represents the family $L_2$.  The behaviour is symmetric to that depicted in figure \ref{fig:3d-1}.}
    \label{fig:3d-2}
\end{figure}

\begin{figure}
    \centering
    \includegraphics[scale=0.35]{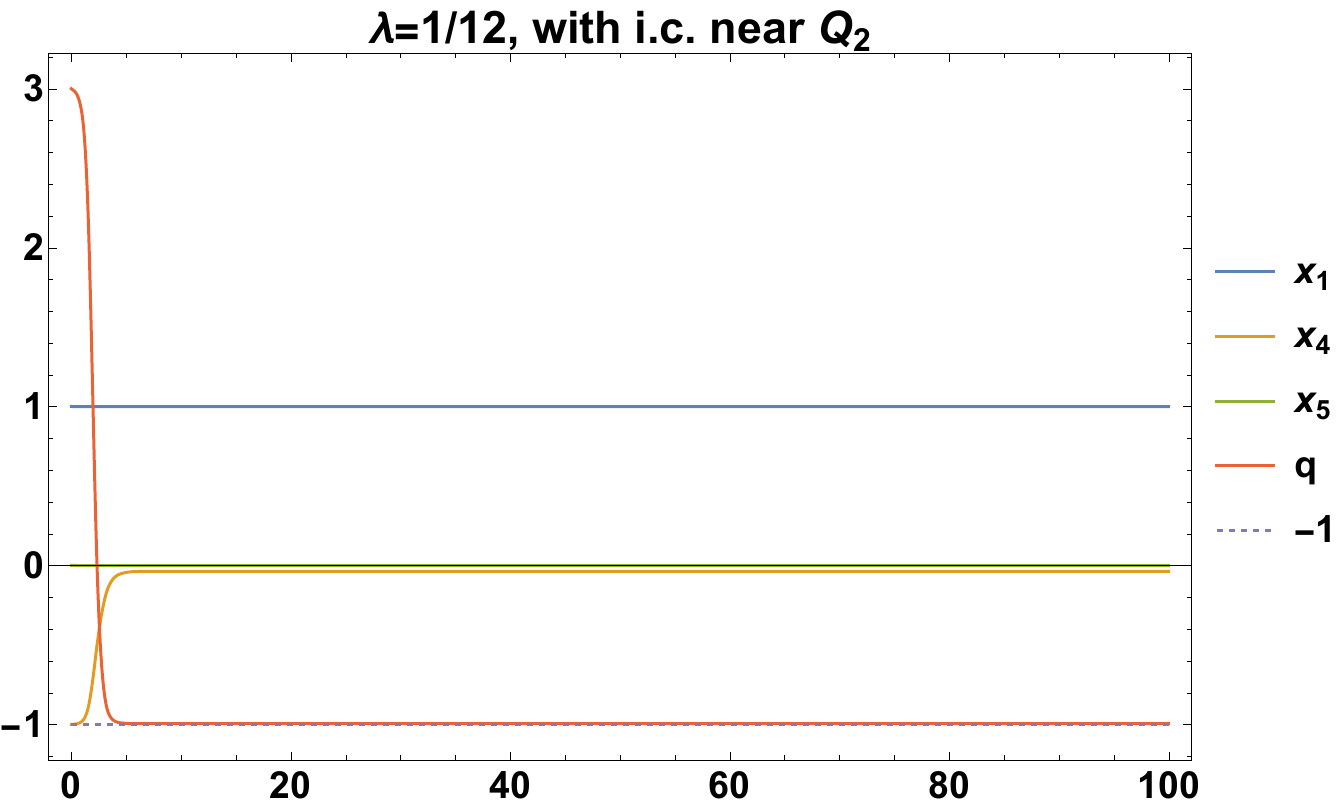}
    \includegraphics[scale=0.35]{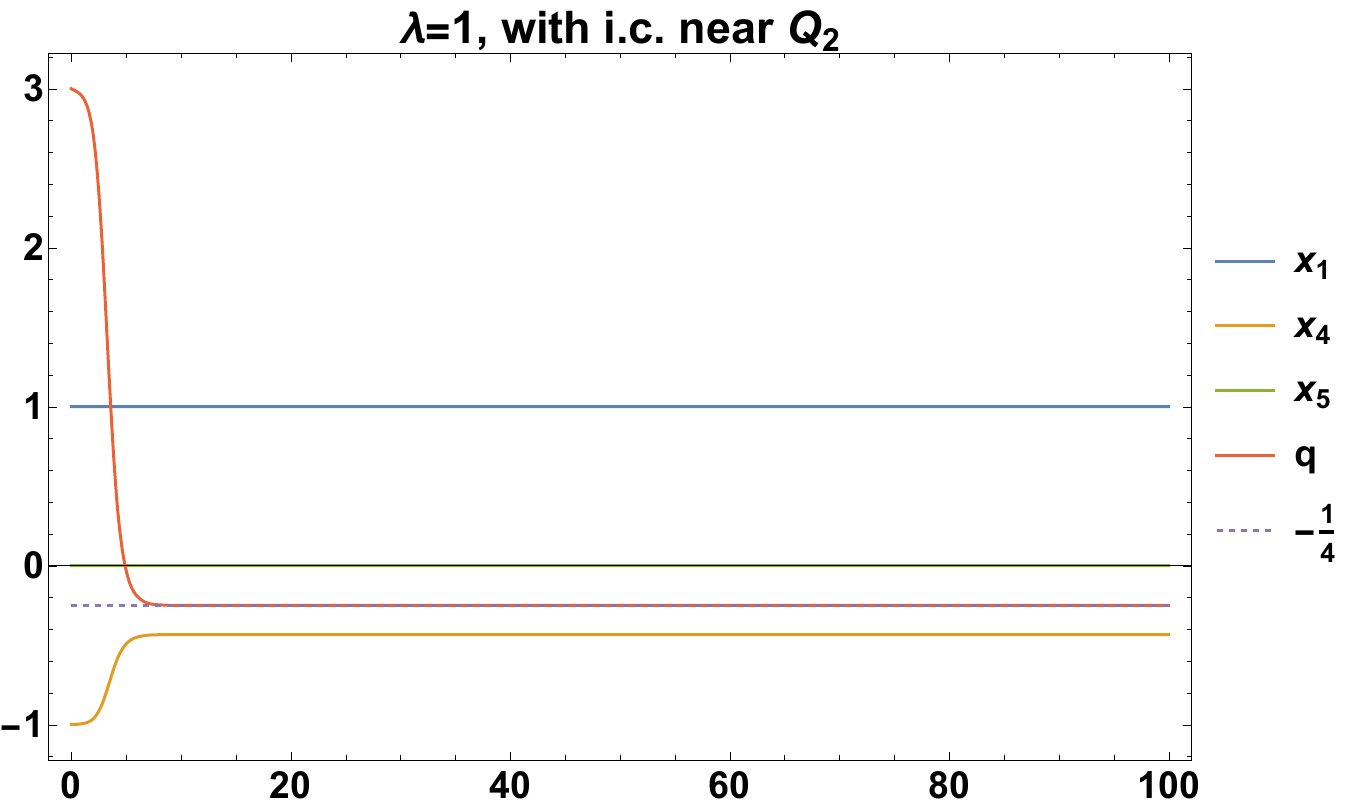}
    \caption{Possible late-time evolution for system \eqref{reduced_linear_eq_x1} -\eqref{reduced_linear_eq_x5} for different values of $\lambda$ with initial conditions near $Q_2$ with a displacement of $\xi=10^{-3}$.   Also, the deceleration parameter is depicted as a red line in the plots. For both values of $\lambda$, we see that after an initially super-collapsing stage, the solution accelerates into one of the following two cases: towards the family $F_1$ since $q\rightarrow-1;$ towards the scaling solution $Q_5$ for which $q\rightarrow -\frac{1}{4}$.}
    \label{fig:num-3d}
\end{figure}

 Using \eqref{linear_deceleration} and \eqref{def_of_x2_x3} deceleration parameter is 
 \begin{equation}
     \label{linear_deceleration_3dimensions}
     q=-\frac{8 x_1^3+x_1 \left(32 x_4^2-12 \sqrt{2}
   x_4 x_5^2+x_5^4-16\right)+4 \sqrt{6} \lambda 
   \left(x_4^2-1\right) x_5^2}{4 x_1 \left(2
   x_1^2+2 \sqrt{2} x_4
   x_5^2-x_5^4-4\right)}.
 \end{equation}
 From equations \eqref{deceleration_5d} and \eqref{linear_deceleration_3dimensions}, we see that the deceleration parameter is not defined for $x_1=0$,  but we can still study its limit as $x_1\rightarrow 0$ for the equilibrium points where $x_1=0$. 
 The equilibrium points for system \eqref{reduced_linear_eq_x1}-\eqref{reduced_linear_eq_x5} are
 \begin{enumerate}
     \item The normally hyperbolic family $F_1=(\frac{4 \sqrt{6} \lambda }{x_{5_c}^2},0,x_{5_c})$.  This family exists for $x_{5_c}>0, -\frac{5 \sqrt{x_{5_c}^4}}{4 \sqrt{6}}<\lambda <\frac{5
   \sqrt{x_{5_c}^4}}{4 \sqrt{6}}$.  The eigenvalues are $\{0,\frac{1536 \lambda ^3-200 \lambda 
   \left(x_{5_c}^4+4\right) x_{5_c}^4-2 \sqrt{\lambda ^2
   \left(25 x_{5_c}^4 \left(x_{5_c}^4+4\right)-192 \lambda
   ^2\right) \left(-3072 \lambda ^2+275 x_{5_c}^8+400 \left(3
   \lambda ^2+4\right) x_{5_c}^4\right)}}{125 x_{5_c}^6
   \left(x_{5_c}^4+4\right)-960 \lambda ^2 x_{5_c}^2},\\ \frac{2
   \left(768 \lambda ^3-100 \lambda  \left(x_{5_c}^4+4\right)
   x_{5_c}^4+\sqrt{\lambda ^2 \left(25 x_{5_c}^4
   \left(x_{5_c}^4+4\right)-192 \lambda ^2\right) \left(-3072
   \lambda ^2+275 x_{5_c}^8+400 \left(3 \lambda ^2+4\right)
   x_{5_c}^4\right)}\right)}{125 x_{5_c}^6
   \left(x_{5_c}^4+4\right)-960 \lambda ^2 x_5^2}\}$,  see Fig. \ref{fig:4} for a representation of the real part of the nonzero eigenvalues. The nonzero eigenvalues determine the stability of the family. The family is 
   \begin{enumerate}
       \item an attractor for $\lambda >0, \sqrt{15} x_{5_c}^2>12 \lambda$, 
       \item a source for $-\frac{1}{4} \sqrt{\frac{5}{3}} x_{5_c}^2<\lambda <0$, 
       \item a saddle for $-\frac{5 x_{5_c}^2}{4 \sqrt{6}}<\lambda <-\frac{1}{4}
   \sqrt{\frac{5}{3}} x_{5_c}^2$ or $ \frac{1}{4}
   \sqrt{\frac{5}{3}} x_{5_c}^2<\lambda <\frac{5 x_{5_c}^2}{4
   \sqrt{6}}$, 
       \item non-hyperbolic for $\lambda =0$ or $ 12 \lambda +\sqrt{15} x_{5_c}^2=0$ or $ \sqrt{15}
   x_{5_c}^2=12 \lambda$. 
   \begin{figure}[h!]
       \centering
       \includegraphics[scale=0.7]{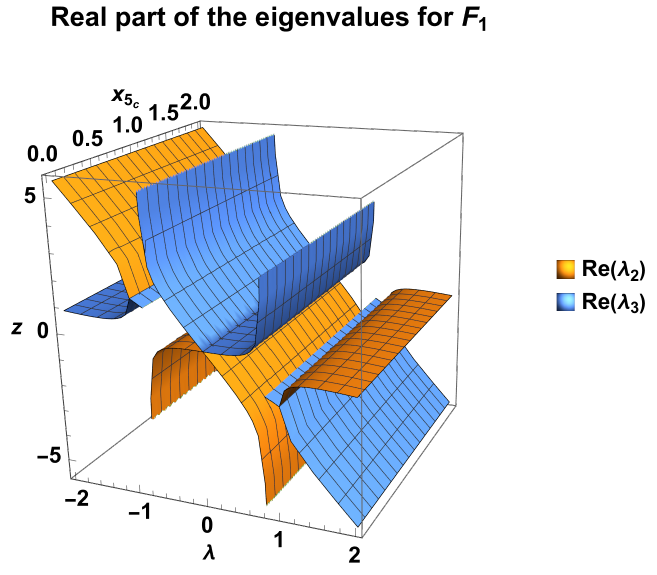}
       \caption{Real part of the eigenvalues for the normally hyperbolic family of points $F_1$.  Here we see $\lambda_2$ and $\lambda_3$ the nonzero eigenvalues.}
       \label{fig:4}
   \end{figure}
   We also verify that $q(F_1)=-1$, which means that the family describes a de Sitter solution.
   \end{enumerate}
     \item The normally hyperbolic family $L_1=(0,1,x_{5_c})$ has eigenvalues $\{0,0,\sqrt{2}\lambda\}$.  This family exists for $0\leq x_{5_c}\leq \frac{1}{\sqrt[4]{2}}$ and is unstable for $\lambda>0$ and stable for $\lambda<0$.  Since $x_1=0$,  we study $q$ as $x_1\rightarrow 0$,  this means that $\lim_{x_1\rightarrow 0} q(L_1)=\frac{x_{5_c}^4-12 \sqrt{2} x_{5_c}^2+16}{4 x_{5_c}^4-8
   \sqrt{2} x_{5_c}^2+16}$.  This means that $L_1$ cannot describe an accelerated solution since $0\leq \frac{x_{5_c}^4-12 \sqrt{2} x_{5_c}^2+16}{4 x_{5_c}^4-8
   \sqrt{2} x_{5_c}^2+16}\leq 1$.
     \item  The normally hyperbolic family $L_2=(0,-1,x_{5_c})$ has eigenvalues $\{0,0,-\sqrt{2}\lambda\}$.  This family exists for $x_{5_c}\geq0$ and is unstable for $\lambda<0$ and stable for $\lambda>0$.  Since $x_1=0$,  we study $q$ as $x_1\rightarrow 0$,  this means that $\lim_{x_1\rightarrow 0} q(L_1)=\frac{-x_{5_c}^4+12 \sqrt{2} x_{5_c}^2-16}{-4 x_{5_c}^4+8
   \sqrt{2} x_{5_c}^2-16}$.  This means that $L_2$ describes an accelerated solution for $\sqrt{2 \left(3 \sqrt{2}-\sqrt{14}\right)}<x_{5_c}<\sqrt{2
   \left(3 \sqrt{2}+\sqrt{14}\right)}$
     \item $Q_1=(1,1,0)$,  with eigenvalues $\left\{-4 \sqrt{\frac{2}{3}},-2 \sqrt{\frac{2}{3}},\frac{1}{3}
   \sqrt{2} \left(3 \lambda +4 \sqrt{3}\right)\right\}$.  This point is 
   \begin{enumerate}
       \item an attractor for for $\lambda <-\frac{4}{\sqrt{3}}$, 
       \item a saddle for $\lambda >-\frac{4}{\sqrt{3}}$, 
       \item non-hyperbolic for $\lambda =-\frac{4}{\sqrt{3}}$. 
   \end{enumerate}
   We also verify that $Q_1$ describes a super-collapse solution because $q(Q_1)=3$. 
     \item $Q_2=(1,-1,0)$,  with eigenvalues $\left\{-4 \sqrt{\frac{2}{3}},-2 \sqrt{\frac{2}{3}},\frac{1}{3}
   \sqrt{2} \left(4 \sqrt{3}-3 \lambda \right)\right\}$.  This point is 
   \begin{enumerate}
       \item an attractor for $\lambda >\frac{4}{\sqrt{3}}$, 
       \item a saddle for $\lambda <\frac{4}{\sqrt{3}}$, 
       \item non-hyperbolic for $\lambda =\frac{4}{\sqrt{3}}$. 
   \end{enumerate}
   As before, $Q_2$ describes a super-collapse solution because $q(Q_2)=3$. 
     \item $Q_3=(-1,1,0)$,  with eigenvalues $\left\{4 \sqrt{\frac{2}{3}},2 \sqrt{\frac{2}{3}},\frac{1}{3}
   \sqrt{2} \left(3 \lambda -4 \sqrt{3}\right)\right\}$.  This point is
   \begin{enumerate}
       \item a source for $\lambda >\frac{4}{\sqrt{3}}$, 
       \item a saddle for $\lambda <\frac{4}{\sqrt{3}}$, 
       \item non-hyperbolic for $\lambda =\frac{4}{\sqrt{3}}$. 
   \end{enumerate}
   We verify that $Q_3$ describes the same type of super-collapse solution as $Q_{1,2}$ because $q(Q_3)=3$. 
     \item $Q_4=(-1,-1,0)$,  with eigenvalues $\left\{4 \sqrt{\frac{2}{3}},2 \sqrt{\frac{2}{3}},\frac{1}{3}
   \sqrt{2} \left(-3 \lambda -4 \sqrt{3}\right)\right\}$.  This point is
   \begin{enumerate}
       \item a source for $\lambda <-\frac{4}{\sqrt{3}}$, 
       \item a saddle for $\lambda >-\frac{4}{\sqrt{3}}$, 
       \item non-hyperbolic for $\lambda =-\frac{4}{\sqrt{3}}$. 
   \end{enumerate}
   As with $Q_3$,  we observe that $q(Q_4)=3$. 
     \item $Q_5=(1,-\frac{\sqrt{3} \lambda }{4},0)$,  with eigenvalues $\left\{-\frac{1}{2} \sqrt{\frac{3}{2}} \lambda ^2,-\frac{1}{4}
   \sqrt{\frac{3}{2}} \lambda ^2,\frac{3 \lambda ^2-16}{4
   \sqrt{6}}\right\}$.  This point exists for $-\frac{4}{\sqrt{3}}\leq \lambda \leq \frac{4}{\sqrt{3}}$ and is 
   \begin{enumerate}
       \item an attractor for $-\frac{4}{\sqrt{3}}<\lambda <0$ or $ 0<\lambda <\frac{4}{\sqrt{3}}$, 
       \item non-hyperbolic for $\lambda =0$ or $ \lambda =-\frac{4}{\sqrt{3}}$ or $ \lambda
   =\frac{4}{\sqrt{3}}$. 
   \end{enumerate}
   The deceleration parameter is $q(Q_5)=-1+\frac{3 \lambda ^2}{4}$, which is a scaling solution. This means that $Q_5$ describes an accelerated solution for $-\frac{2}{\sqrt{3}}<\lambda <\frac{2}{\sqrt{3}}$.  It describes a decelerated solution for $\lambda <-\frac{2}{\sqrt{3}}$ or $ \lambda >\frac{2}{\sqrt{3}}$.  Finally it describes a de Sitter solution for $\lambda=0$. 
     \item $Q_6=(-1,\frac{\sqrt{3} \lambda }{4},0)$,  with eigenvalues $\left\{\frac{1}{4} \sqrt{\frac{3}{2}} \lambda ^2,\frac{1}{2}
   \sqrt{\frac{3}{2}} \lambda ^2,-\frac{3 \lambda ^2-16}{4
   \sqrt{6}}\right\}$.  This point exists for $-\frac{4}{\sqrt{3}}\leq \lambda \leq \frac{4}{\sqrt{3}}$ and is 
   \begin{enumerate}
       \item a source for $-\frac{4}{\sqrt{3}}<\lambda <0$ or $ 0<\lambda <\frac{4}{\sqrt{3}}$, 
       \item non-hyperbolic for $\lambda =0$ or $ \lambda =-\frac{4}{\sqrt{3}}$ or $ \lambda
   =\frac{4}{\sqrt{3}}$. 
   \end{enumerate}
   This point describes the same type of solutions as $Q_5$ since $q(Q_6)=-1+\frac{3 \lambda ^2}{4}$. 
 \end{enumerate}
 In Table \ref{tab:2}, we present a summary of the results of this section.
 Figures \ref{fig:3d-1} and \ref{fig:3d-2} depict different phase-plots for system \eqref{reduced_linear_eq_x1}-\eqref{reduced_linear_eq_x5} for different values of the parameter $\lambda$.  We used $\lambda=\frac{1}{12}$ in order to confirm the family of equilibrium points $F_1$ is attractive. In figure \ref{fig:num-3d} we present a possible late-time evolution for system \eqref{reduced_linear_eq_x1} -\eqref{reduced_linear_eq_x5} for different values of $\lambda$ with initial conditions near $Q_2$ with a displacement of $\xi=10^{-3}$.   Also, the deceleration parameter is depicted as a red line in the plots. For both values of $\lambda$, we see that after an initially super-collapsing stage, the solution accelerates into one of the following two cases: towards the family $F_1$ since $q\rightarrow-1;$ towards the scaling solution $Q_5$ for which $q\rightarrow -\frac{1}{4}$.
 As in the previous section, we can write our findings in the following results
 \begin{theorem}
     \label{teo-3}
     The late-time attractors for the five-dimensional Gauss-Bonnet model with linear coupling function are
     \begin{enumerate}
         \item The normally hyperbolic family $F_1=(\frac{4 \sqrt{6} \lambda }{x_{5_c}^2},0,x_{5_c})$ for $\lambda >0, \sqrt{15} x_{5_c}^2>12 \lambda$. 
         \item The normally hyperbolic family $L_1=(0,1,x_{5_c})$ for $\lambda<0$. 
         \item The normally hyperbolic family $L_2=(0,-1,x_{5_c})$ for $\lambda>0$. 
         \item The super-collapse solution $Q_1=(1,1,0)$ for $\lambda<-\frac{4}{\sqrt{3}}$. 
         \item The super-collapse solution $Q_2=(1,-1,0)$ for $\lambda>\frac{4}{\sqrt{3}}$. 
         \item The scaling solution $Q_5=(1,-\frac{\sqrt{3}\lambda}{4},0)$ for $-\frac{4}{\sqrt{3}}<\lambda<0$ or $0<\lambda<\frac{4}{\sqrt{3}}$. 
     \end{enumerate}
 \end{theorem}
 As before, to avoid having super-collapsing solutions as late-time attractors for this model, we can restrict the numerical range for the parameter $\lambda$ in the range $-\frac{4}{\sqrt{3}}<\lambda<0$ or $0<\lambda<\frac{4}{\sqrt{3}}$. That modifies the region of parameters for some of the equilibrium points. We formulate this in the following result 
 \begin{corollary}
     \label{corol-2}
      The late-time attractors for the five-dimensional Gauss-Bonnet model with linear coupling function in the range $-\frac{4}{\sqrt{3}}<\lambda<0$ or $0<\lambda<\frac{4}{\sqrt{3}}$ are 
           \begin{enumerate}
         \item The normally hyperbolic family $F_1=(\frac{4 \sqrt{6} \lambda }{x_{5_c}^2},0,x_{5_c})$ for $0<\lambda<\frac{4}{\sqrt{3}} , \sqrt{15} x_{5_c}^2>12 \lambda.$ 
         \item The normally hyperbolic family $L_1=(0,1,x_{5_c})$ for $-\frac{4}{\sqrt{3}}<\lambda<0.$  
         \item The normally hyperbolic family $L_2=(0,-1,x_{5_c})$ for $0<\lambda<\frac{4}{\sqrt{3}}.$ 
         \item The scaling solution $Q_5=(1,-\frac{\sqrt{3}\lambda}{4},0)$ for $-\frac{4}{\sqrt{3}}<\lambda<0$ or $0<\lambda<\frac{4}{\sqrt{3}}$. 
     \end{enumerate}
 \end{corollary}
Notice that the late-time attractor for the model mentioned in the last item of corollary \ref{corol-2} corresponds to the sixth result in Theorem \ref{teo-3}.
 Additionally, in section \ref{app-2}, we study a projection of the three-dimensional dynamical system for linear coupling function to better visualise the dynamics for a fixed value of $x_{5_c}$ since the families $F_1, L_1, L_2$ depend on this free parameter.
 \begin{table}[ht]
\begin{tabular}{|c|c|c|c|c|c|}
\hline
Label & $(x_1,x_4,x_5)$ & $(x_2,x_3)$ & Attractor? & Acceleration? &Interpretation\\ \hline
$F_1$ & $\left(\frac{4\sqrt{6}\lambda}{5x_{5_c}^{2}},0,x_{5_c}\right)$ & $\left(\sqrt{1-\frac{96\lambda^2}{25x_{5_c}^{4}}},1\right)$ & Yes & Yes, $q=-1$ & de Sitter\\ \hline
$L_1$ & $\left(0,1,x_{5_c}\right)$ & $\left(\sqrt{1-\sqrt{2}x_{5_c}},0\right)$ & Yes & No, see text & decelerated\\ \hline
$L_2$ & $(0,-1,x_{5_c})$ & $\left(\sqrt{1+\sqrt{2}x_{5_c}},0\right)$ & Yes & Yes, see text &accelerated\\ \hline
$Q_1$ & $(1,1,0)$ & $(0,0)$ & Yes* & No, $q=3$ & super-collapse\\ \hline
$Q_2$ & $(1,-1,0)$ & $(0,0)$ & Yes* &  No, $q=3$ & super-collapse\\ \hline
$Q_3$ & $(-1,1,0)$ & $(0,0)$ & No & No, $q=3$ &super-collapse\\ \hline
$Q_4$ & $(-1,-1,0)$ & $(0,0)$ & No & No, $q=3$ &super-collapse\\ \hline
$Q_5$ & $\left(1,-\frac{\sqrt{3}}{4}\lambda,0\right)$ & $(0,\sqrt{1-\frac{3}{16}\lambda^2})$ & Yes & Yes, $q=-1+\frac{3\lambda^2}{4}$ &scaling solution\\ \hline
$Q_6$ & $\left(1,\frac{\sqrt{3}}{4}\lambda,0\right)$ & $(0,\sqrt{1-\frac{3}{16}\lambda^2})$ & No & Yes, $q=-1+\frac{3\lambda^2}{4}$ &scaling solution\\ \hline
\end{tabular}
\caption{Summary of the analysis of system \eqref{reduced_linear_eq_x1}-\eqref{reduced_linear_eq_x5}. The $x_2$ coordinate gives the interval of existence for the families $F_1$ and $L_{1,2}$. The $x_3$ coordinate gives the interval of existence for $Q_{5,6}$. The asterisk in $Q_{1,2}$ means that if we constrain the parameter $\lambda$,  the super-collapse points would not be attractors for the model.}
\label{tab:2}
\end{table}

\subsubsection{Two-dimensional projection of the dynamical system for linear coupling function}
\label{app-2}
In the previous section, we obtained three families of equilibrium points that depend on the constant $x_{5}$. Now we wish to better observe the dynamics for a fixed value of $x_{5_c}$ so we consider a projection to $x_5=\mu$ in system \eqref{reduced_linear_eq_x1}-\eqref{reduced_linear_eq_x5} for $\mu\geq 0$.  

Setting $x_5=\mu$ in system \eqref{reduced_linear_eq_x1}-\eqref{reduced_linear_eq_x5} and keeping the $x_1$ and $x_4$ equations we obtain
\begin{align}
\label{projection-1}
    x_1'&=\frac{x_1 \left(\sqrt{2} \mu ^2 \left(-5 x_1^3 x_4+2 x_1 x_4 \left(8 x_4^2-5\right)-\frac{x_4^2-1}{\sqrt{3}}\right)+32 x_1
   \left(x_1^2-1\right) x_4^2+\mu ^4 \left(x_1 \left(5-6 x_4^2\right)+\frac{x_4 \left(x_4^2-1\right)}{\sqrt{3}}\right)\right)}{4 \sqrt{6} \left(-\mu
   ^4+2 x_1^2+2 \sqrt{2} \mu ^2 x_4-4\right)},\\
   \label{projection-2}
   x_4'&=\frac{\left(x_4^2-1\right) \left(\sqrt{3} x_1 \left(-5 \sqrt{2} \mu ^2 \left(x_1^2-2\right)+x_4 \left(-6 \mu ^4+32 x_1^2-64\right)+16 \sqrt{2} \mu ^2
   x_4^2\right)-2 x_1^2+\mu ^4 x_4^2-2 \sqrt{2} \mu ^2 x_4+4\right)}{12 \sqrt{2} \left(-\mu ^4+2 x_1^2+2 \sqrt{2} \mu ^2 x_4-4\right)}.
\end{align}
The points on the curve $x_4=-\frac{-\mu ^4+2 x_1^2-4}{2 \sqrt{2} \mu ^2}$ are those for which the denominator of system \eqref{projection-1}-\eqref{projection-2} vanishes. This singularity curve is a parabola for which the direction of the flow changes.

The deceleration parameter for this system depends on $\mu$ and is written as

\begin{equation}
\label{q-app}
    q=-\frac{8 x_1^3+x_1 \left(\mu ^4+32 x_4^2-12 \sqrt{2} \mu ^2 x_4-16\right)+4 \sqrt{6} \lambda  \mu ^2 \left(x_4^2-1\right)}{4 x_1 \left(-\mu ^4+2
   x_1^2+2 \sqrt{2} \mu ^2 x_4-4\right)}.
\end{equation}

As in section \ref{sect-3-2-1}, the deceleration parameter is not defined for equilibrium points with $x_1=0$, so we will study its limit as $x_1\rightarrow 0$.  In the following list, an asterisk means that the equilibrium point was originally part of one of the families from section \ref{sect-3-2-1}. The equilibrium points for system \eqref{projection-1}-\eqref{projection-2} are
\begin{enumerate}
    \item $F_1^*=\left(\frac{4\sqrt{6}\lambda}{5\mu^{2}},0\right)$,  as in section \ref{sect-3-2-1} we see that $q(F_1^*)=-1$,  this means it describes a de Sitter solution.
    \item $L_1^*=(0,1)$,  for this point, we observe that $\lim_{x_1\rightarrow 0} q(L_1^*)=\frac{\mu ^4-12 \sqrt{2} \mu ^2+16}{4 \mu ^4-8 \sqrt{2} \mu ^2+16}$ and this describes acceleration for $\sqrt{2 \left(3 \sqrt{2}-\sqrt{14}\right)}<\mu <\sqrt{2 \left(3 \sqrt{2}+\sqrt{14}\right)}$ which is approximately $1.00098<\mu <3.99607$. 
    \item $L_2^*=(0,-1)$,  for this point we have $\lim_{x_1\rightarrow 0} q(L_2^*)=\frac{5 \sqrt{2} \mu ^2+6}{2 \mu ^4+4 \sqrt{2} \mu ^2+8}+\frac{1}{4}$ it cannot describe acceleration for any value of $\mu$. 
    \item $A=\left(\sqrt{\frac{\mu ^4+6 \sqrt{2} \mu ^2+32}{5 \sqrt{2} \mu ^2+32}},1\right)$,  this point verifies $q(A)=\frac{-45 \mu ^4-100 \sqrt{2} \mu ^2+352}{10 \sqrt{2} \mu ^6+20 \mu ^4-112 \sqrt{2} \mu ^2+128}+\frac{1}{4}$.  It describes acceleration for the approximated ranges $1.02547<\mu <1.27391$ or $1.39621<\mu <3.78861$. 
    \item $B=\left(-\sqrt{\frac{\mu ^4+6 \sqrt{2} \mu ^2+32}{5 \sqrt{2} \mu ^2+32}},1\right)$.  Similarly to $A$,  we see that $q(B)=\frac{-45 \mu ^4-100 \sqrt{2} \mu ^2+352}{10 \sqrt{2} \mu ^6+20 \mu ^4-112 \sqrt{2} \mu ^2+128}+\frac{1}{4}$ meaning it can describe acceleration for the same intervals. 
    \item $C=\left(\sqrt{\frac{\mu ^4+6 \sqrt{2} \mu ^2+32}{5 \sqrt{2} \mu ^2+32}},-1\right)$,  for this point we have $q(C)=\frac{11}{2 \sqrt{2} \mu ^2+4}+\frac{1}{4}$,  therefore it cannot describe acceleration.
    \item $D=\left(-\sqrt{\frac{\mu ^4+6 \sqrt{2} \mu ^2+32}{5 \sqrt{2} \mu ^2+32}},-1\right)$,  we see that $q(D)=\frac{11}{2 \sqrt{2} \mu ^2+4}+\frac{1}{4}$,  as $C$,  it cannot describe acceleration. 
    \item $E=\left(\frac{\sqrt{\left(32-5 \mu ^4\right)^2 \left(384 \lambda ^2+\mu ^4\right)}+5 \mu ^6-32 \mu ^2}{8 \sqrt{6} \lambda  \left(5 \mu ^4-32\right)},-\frac{\sqrt{\left(32-5 \mu
   ^4\right)^2 \left(384 \lambda ^2+\mu ^4\right)}-32 \left(6 \lambda ^2-11\right) \mu ^2+5 \mu ^6}{32 \sqrt{2} \left(\left(3 \lambda ^2-1\right) \mu ^4-32\right)}\right)$.  This point exists for $0\leq \mu <2 \sqrt[4]{\frac{2}{5}}$ or $ \mu >2 \sqrt[4]{\frac{2}{5}}$ and it verifies that $q(E)=-\frac{\sqrt{\left(32-5 \mu ^4\right)^2 \left(384 \lambda ^2+\mu ^4\right)}+32 \left(11-6 \lambda ^2\right) \mu ^2+5 \mu ^6}{32 \sqrt{2} \left(\left(3 \lambda ^2-1\right) \mu
   ^4-32\right)}$.  In this case, the deceleration parameter depends on $\lambda$ and $\mu$,  however, it describes acceleration for 
   \begin{enumerate}
       \item $\mu >2 \sqrt[4]{\frac{2}{5}}$, $-\frac{1}{4} \sqrt{5} \mu ^2<\lambda <-\sqrt{\frac{32}{3 \mu ^4}+\frac{1}{3}}$, 
   \item $\mu >2 \sqrt[4]{\frac{2}{5}}$, $  \sqrt{\frac{32}{3 \mu ^4}+\frac{1}{3}}<\lambda <\frac{\sqrt{5}
   \mu ^2}{4}$. 
   \end{enumerate}
    \item $F=\left(-\frac{\sqrt{\frac{3}{2}} \left(5 \mu ^6+\sqrt{\left(32-5 \mu ^4\right)^2 \left(\mu ^4+\frac{8}{3}\right)}-32 \mu ^2\right)}{2 \left(5 \mu ^4-32\right)},-\frac{5 \mu
   ^6+\sqrt{\left(32-5 \mu ^4\right)^2 \left(\mu ^4+\frac{8}{3}\right)}+\frac{1052 \mu ^2}{3}}{32 \sqrt{2} \left(-\frac{47 \mu ^4}{48}-32\right)}\right)$.  The existence conditions for this point are the same ones as point $E$ but the deceleration parameter is $q(F)=\frac{-6 \mu ^2 \sqrt{384 \lambda ^2+\mu ^4}+3 \lambda ^2 \left(\mu ^2 \sqrt{384 \lambda ^2+\mu ^4}+3 \mu ^4-64\right)+2 \mu ^4+256}{8 \left(3 \lambda ^2-1\right) \mu ^4-256}$.  This point describes acceleration for \begin{enumerate}
       \item $0\leq \mu <2 \sqrt[4]{\frac{2}{5}}$, $ -\frac{\sqrt{\mu ^4-\sqrt{\mu ^4 \left(\mu ^4+512\right)}+128}}{4 \sqrt{6}}<\lambda
   <\sqrt{\frac{\mu ^4}{96}-\frac{1}{96} \sqrt{\mu ^4 \left(\mu ^4+512\right)}+\frac{4}{3}}$, 
       \item $ \mu >2 \sqrt[4]{\frac{2}{5}} $, $  -\frac{\sqrt{\mu ^4-\sqrt{\mu ^4 \left(\mu ^4+512\right)}+128}}{4 \sqrt{6}}<\lambda
   <\sqrt{\frac{\mu ^4}{96}-\frac{1}{96} \sqrt{\mu ^4 \left(\mu ^4+512\right)}+\frac{4}{3}}$. 
   \end{enumerate}
\end{enumerate}

\begin{figure}[h!]
    \centering
    \includegraphics[scale=0.4]{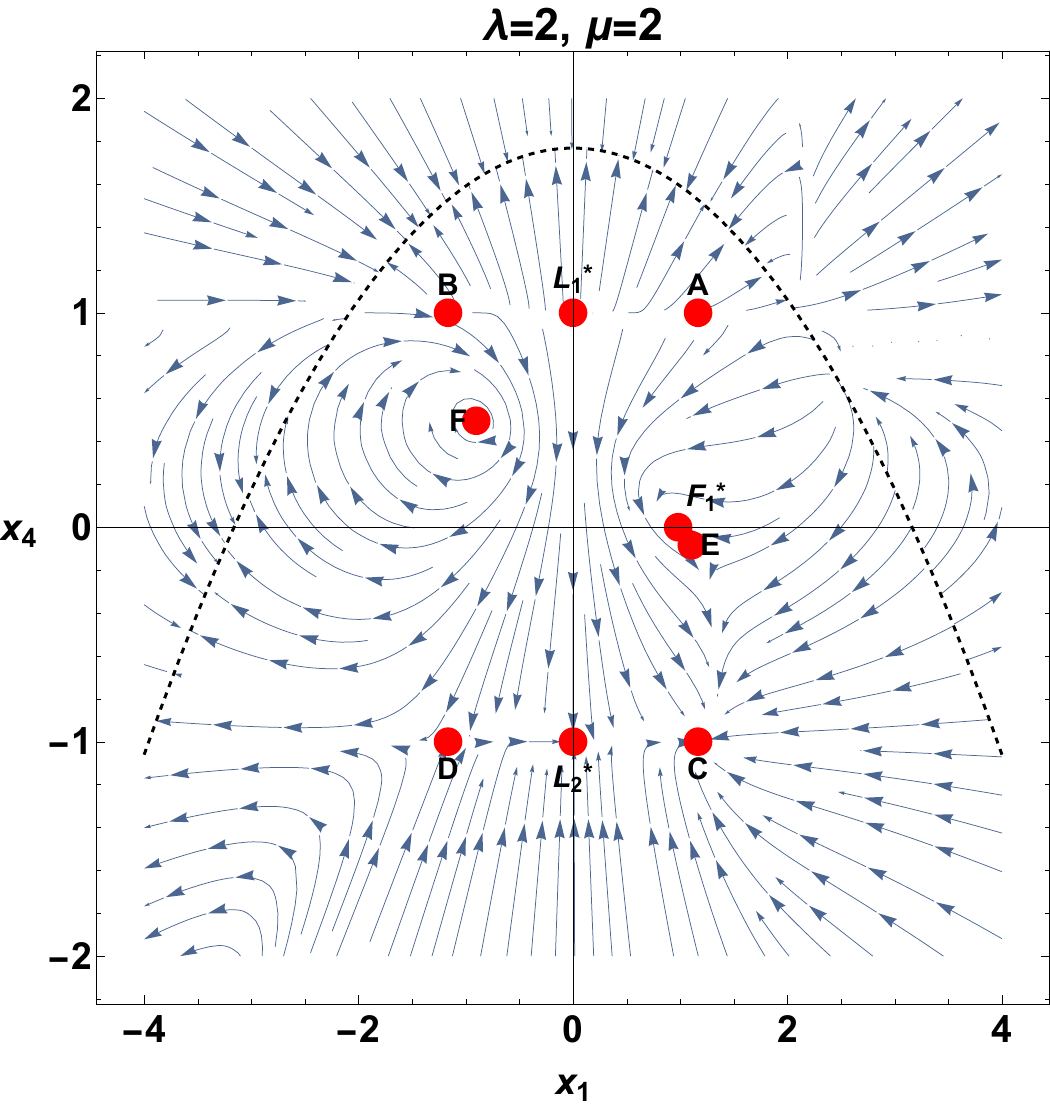}
    \includegraphics[scale=0.4]{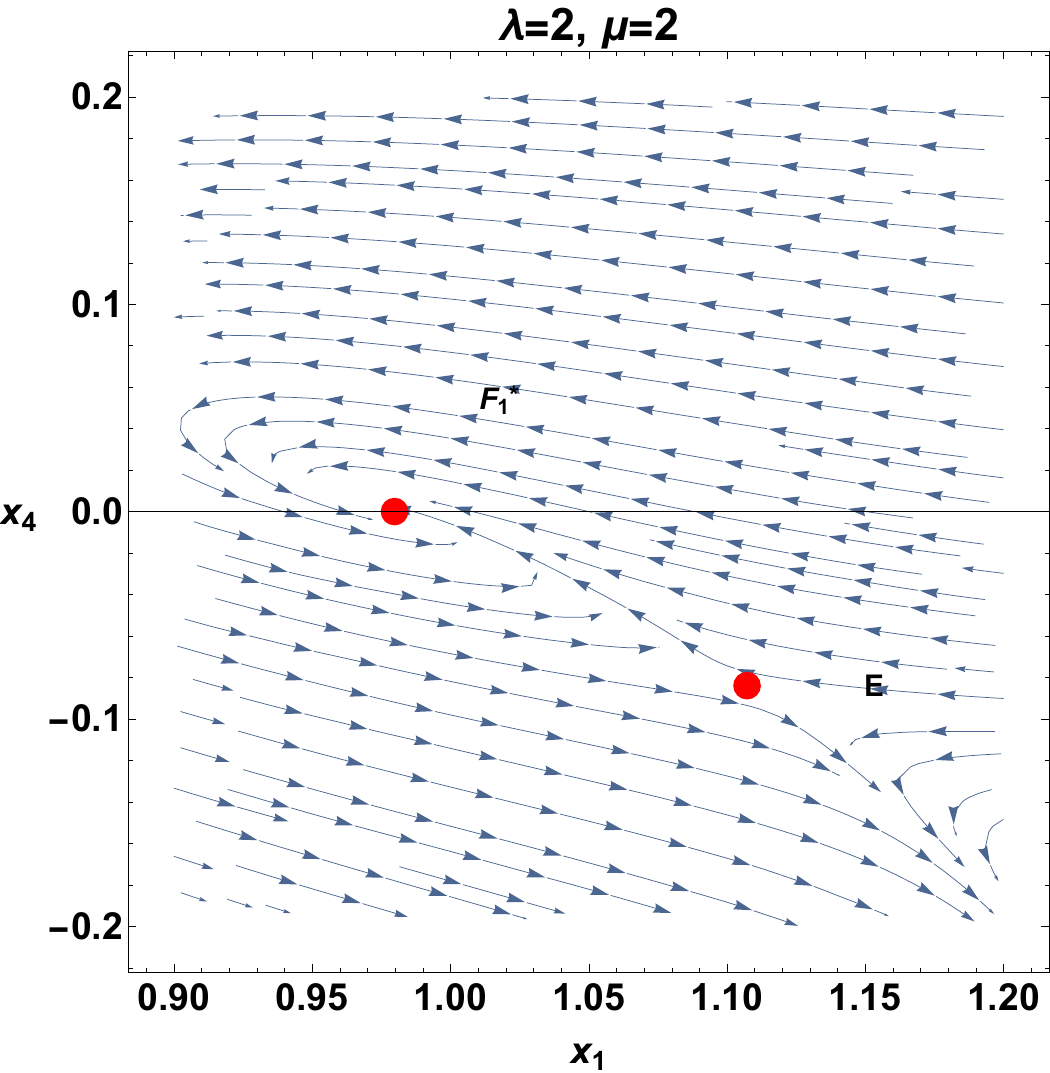}
    \caption{Phase-plot for system \eqref{projection-1}-\eqref{projection-2} for the fixed values $\mu=\lambda=2$,  where the black-dashed parabola is $x_4=-\frac{-\mu ^4+2 x_1^2-4}{2 \sqrt{2} \mu ^2}$. Depicted on the left are the full dynamics of the system, while on the right is a close-up of the nearby points $F_1^*$ and $E$ to confirm the attracting behaviour of $F_1^*$.}
    \label{fig:app1}
\end{figure}

\begin{figure}[h!]
    \centering
    \includegraphics[scale=0.4]{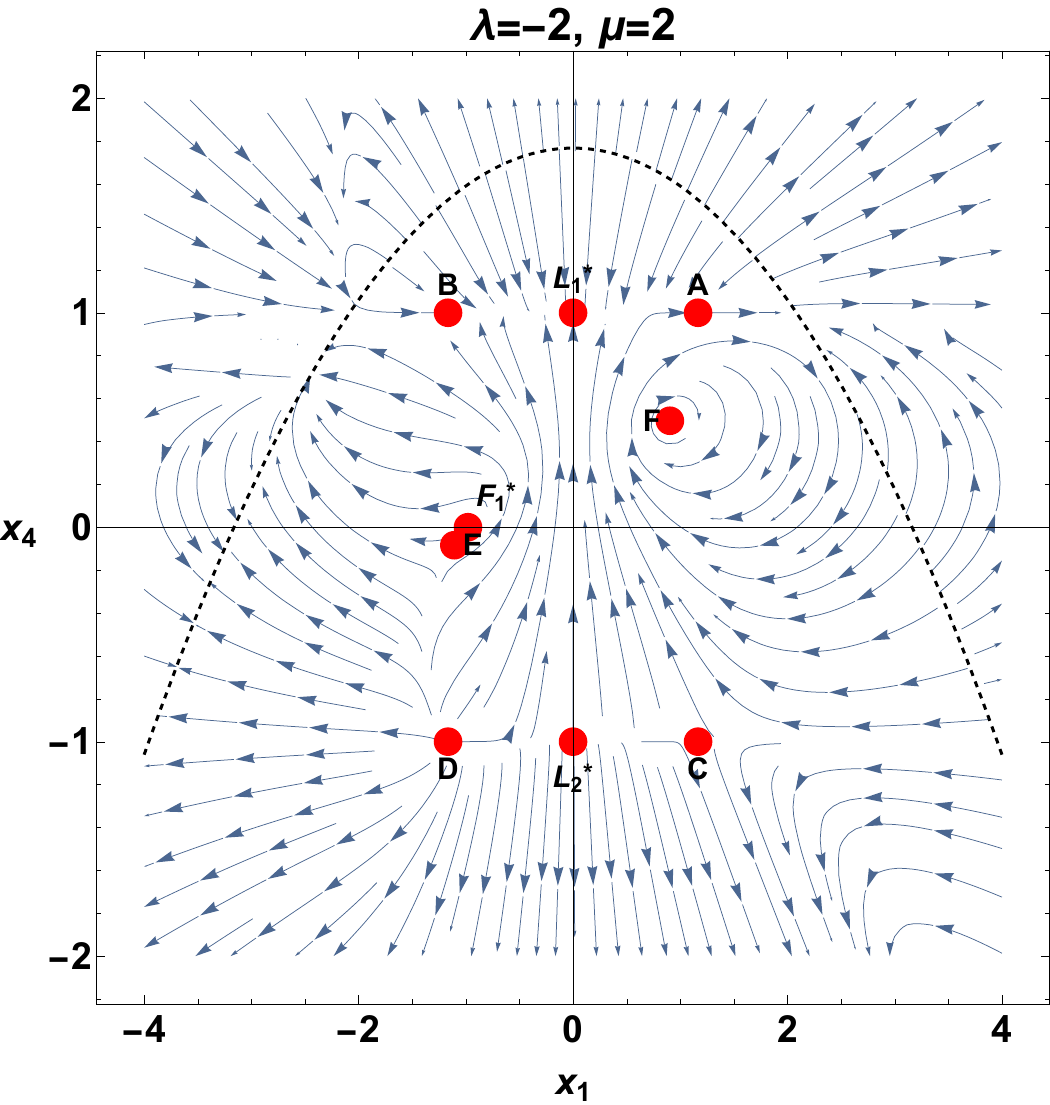}
    \includegraphics[scale=0.4]{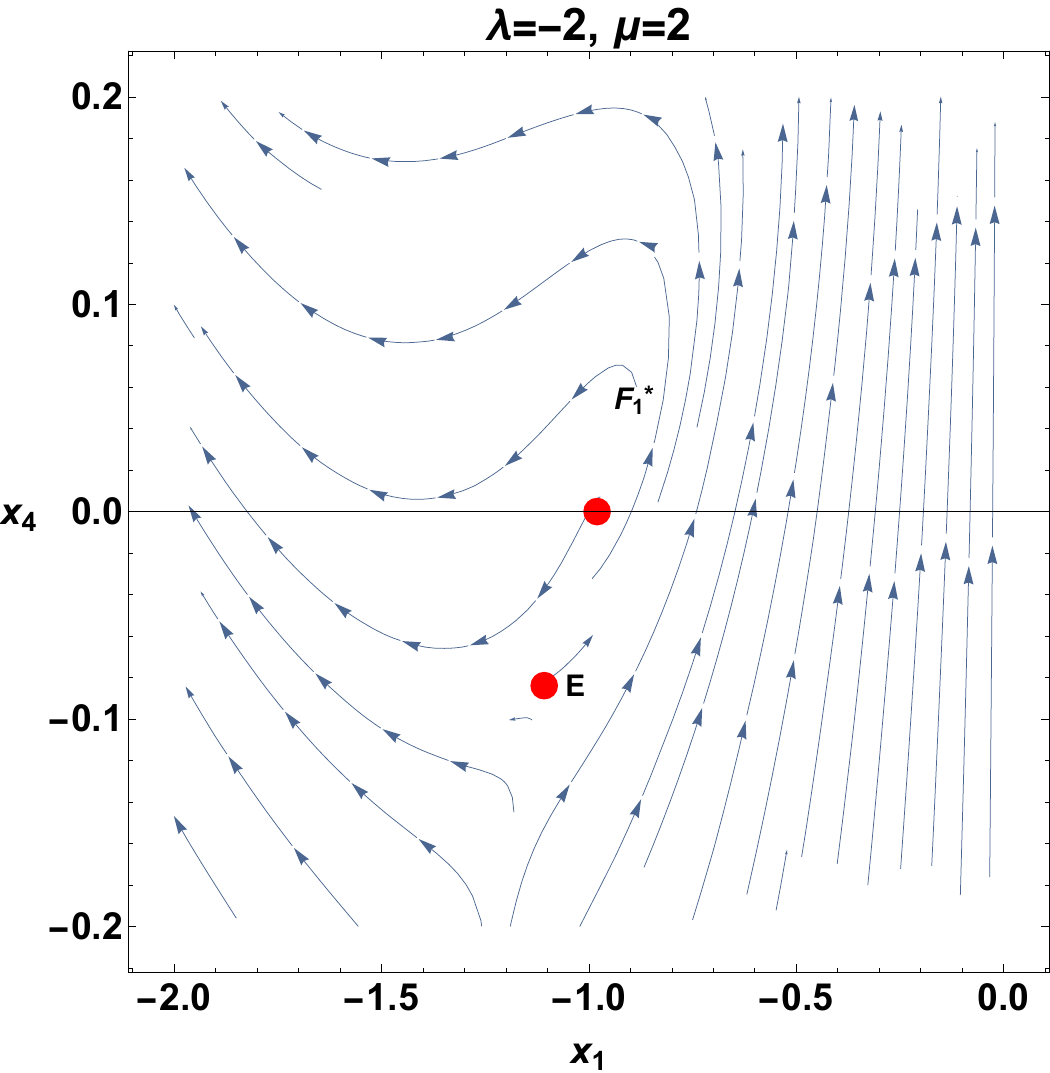}
    \caption{Phase-plot for system \eqref{projection-1}-\eqref{projection-2} for the fixed values $\mu=2, $ and $\lambda=-2$,  where the black-dashed parabola is $x_4=-\frac{-\mu ^4+2 x_1^2-4}{2 \sqrt{2} \mu ^2}$. Depicted on the left are the full dynamics of the system, while on the right is a close-up of the nearby points $F_1^*$ and $E$ to be able to confirm the repelling behaviour of $F_1^*$.}
    \label{fig:app2}
\end{figure}

\begin{figure}[h!]
    \centering
    \includegraphics[scale=0.4]{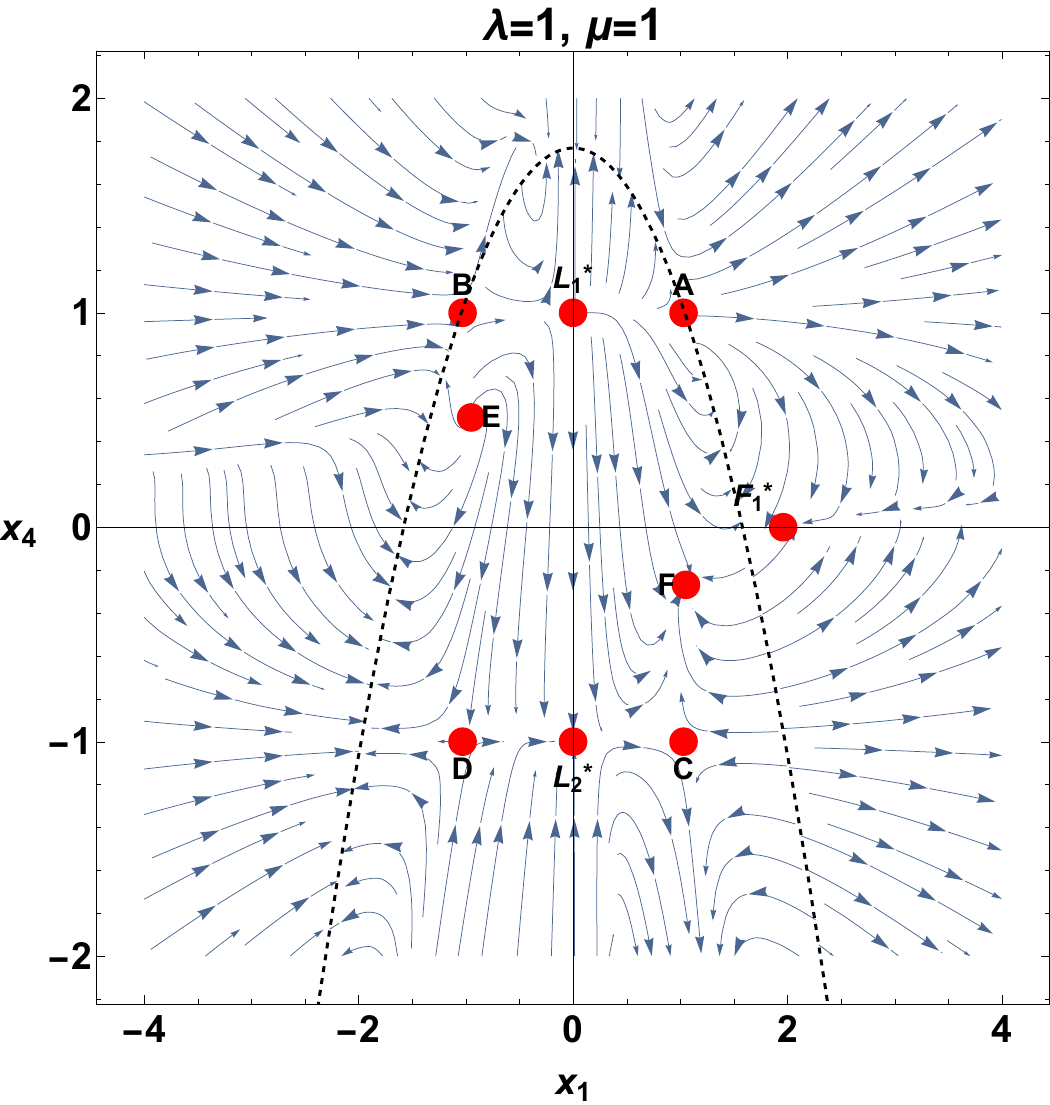}
    \includegraphics[scale=0.4]{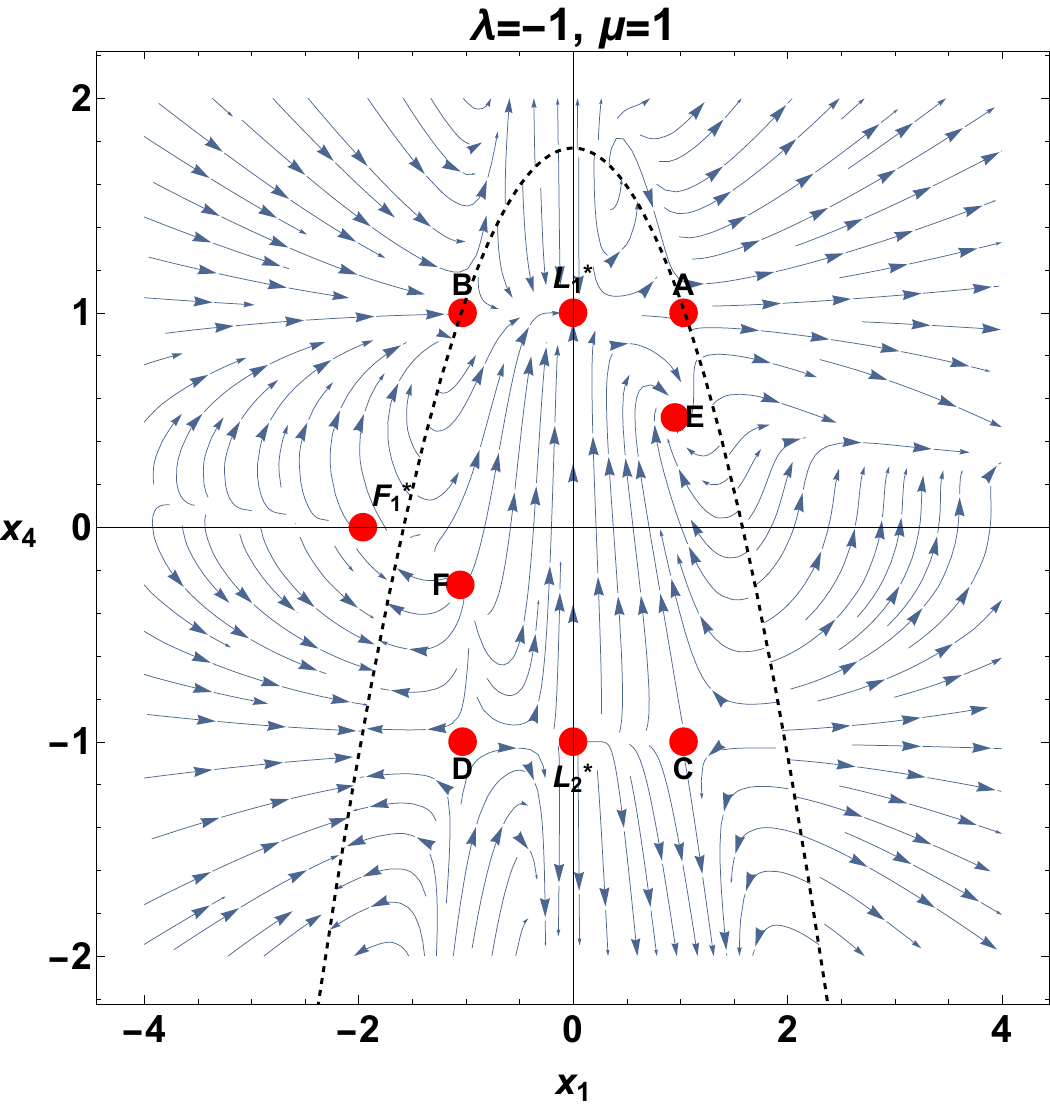}
    \caption{Phase-plot for system \eqref{projection-1}-\eqref{projection-2} for the fixed values $\mu=1$,  and $ \lambda=\pm 1$,  where the black-dashed parabola is $x_4=-\frac{-\mu ^4+2 x_1^2-4}{2 \sqrt{2} \mu ^2}$. We observe that $F_1^*$ lies outside the parabola, but it still behaves as an attractor for $\lambda=1$ and a source for $\lambda=-1$.}
    \label{fig:app3}
\end{figure}

Contrary to the study of the deceleration parameter (summarised in table \ref{tab:4}), where the dependence was mostly on the value of $\mu$ except for the last two points. The stability analysis is more complicated since most eigenvalues depend on $\lambda$ and $\mu$.  We perform numerical stability analysis on the eigenvalues of the Jacobian matrix \eqref{Jacobian-1} of system \eqref{projection-1}-\eqref{projection-2} evaluated at each equilibrium point and present a summary of the results for $\mu=1,2$ and some combinations with $\lambda=\pm 1, \pm 2$ in table \ref{tab:5}. Additionally, in figures \ref{fig:app1}, \ref{fig:app2}, \ref{fig:app3}, we present phase-plane diagrams for the same system and the previously mentioned values of the parameters.

 \begin{table}[ht]
\begin{tabular}{|c|c|c|}
\hline
Label & $(x_1,x_4)$ & Acceleration? \\ \hline
$F_1^*$ & $\left(\frac{4\sqrt{6}\lambda}{5\mu^{2}},0\right)$  & Yes, $q=-1$\\ \hline
$L_1^*$ & $\left(0,1\right)$  & Yes, see text \\ \hline
$L_2^*$ & $(0,-1)$  & No \\ \hline
$A$ & $\left(\sqrt{\frac{\mu ^4+6 \sqrt{2} \mu ^2+32}{5 \sqrt{2} \mu ^2+32}},1\right)$  & Yes, see text\\ \hline
$B$ & $\left(-\sqrt{\frac{\mu ^4+6 \sqrt{2} \mu ^2+32}{5 \sqrt{2} \mu ^2+32}},1\right)$ &  Yes, see text\\ \hline
$C$ & $\left(\sqrt{\frac{\mu ^4+6 \sqrt{2} \mu ^2+32}{5 \sqrt{2} \mu ^2+32}},-1\right)$  & No\\ \hline
$D$ & $\left(-\sqrt{\frac{\mu ^4+6 \sqrt{2} \mu ^2+32}{5 \sqrt{2} \mu ^2+32}},-1\right)$ & No \\ \hline
$E$ & See text  & Yes, see text \\ \hline
$F$ & See text  & Yes, see text \\ \hline
\end{tabular}
\caption{Summary of the equilibrium points of system \eqref{projection-1}-\eqref{projection-2} and the analysis of the deceleration parameter \eqref{q-app}. The asterisk means that the equilibrium point was originally part of one of the families from section \ref{sect-3-2-1}.}
\label{tab:4}
\end{table}

\begin{table}[ht]
\resizebox{\textwidth}{!}{
    \begin{tabular}{|c|c|c|c|c|}
        \hline
         Label & Attractor for $\lambda=\mu=2$ & Attractor for $\lambda=-2, \mu=2$ & Attractor for $\lambda=1, \mu=1$ & Attractor for $\lambda=-1, \mu=1$ \\ \hline
        $F_1^*$ & Yes, $\{-0.1363+0.1858 i,-0.1363-0.1858 i\}$&No, $\{0.1363+0.1858 i,0.1363-0.1858 i\}$ & Yes, $\{-4.078,-0.2416\}$ & No, $\{0.2416,4.078\}$ \\ \hline
        $L_1^*$ &No, $\{2\sqrt{2},0\}$ & Yes, $\{-2\sqrt{2},0\}$ &No, $\{\sqrt{2},0\}$ &Yes, $\{-\sqrt{2},0\}$  \\ \hline
        $L_2^*$ &Yes, $\{-2\sqrt{2},0\}$ & No, $\{2\sqrt{2},0\}$ &Yes, $\{-\sqrt{2},0\}$ &No, $\{\sqrt{2},0\}$  \\ \hline
        $A$ & No, $\{3.142,0.3215\}$& No, $\{-2.515,0.3215\}$& No$, \{31.05,-503.6\}$& No, $\{28.22,-503.6\}$ \\ \hline
        $B$ & No, $\{2.51454,-0.321462\}$ &No$, \{-3.14231,-0.321462\}$ & No, $\{-28.2179,503.621\}$& No, $\{-31.0464,503.621\}$ \\ \hline
        $C$ & Yes, $\{-0.95474,-0.681949\}$ & No, $\{4.70211,-0.681949\}$ & No, $\{0.992799,-1.52955\}$&  No, $\{3.82123,-1.52955\}$\\ \hline
        $D$ & No$, \{-4.70211,0.681949\}$ & No$, 
\{0.95474,0.681949\}$& No, $\{-3.82123,1.52955\}$& No, $

\{-0.992799,1.52955\}$ \\ \hline 
        $E$ &No, $\{-0.4658,0.1335\}$ & No, $\{0.4658,-0.1335\}$ & No, $\{0.7414+0.8029 i,0.7414-0.8029 i\}$& Yes, $\{-0.7414+0.8029 i,-0.7414-0.8029 i\}$ \\ \hline 
          $F$ &Yes, $\{-0.0119+0.9081 i,-0.0119-0.9081 i\}$ &No,  $\{0.0119+0.9081 i,0.0119-0.9081 i\}$ & Yes, $\{-1.038,-0.2459\}$&No, $\{1.038,0.2459\}$  \\ \hline
    \end{tabular}}
    \caption{Numerical stability analysis for the equilibrium points of system \eqref{projection-1}-\eqref{projection-2} for the fixed values of $\mu=1,2$ combined with $\lambda=\pm 1, \pm 2$.  The stability for $L_1^*$ and $L_2^*$ refers to their behaviour along the axis line $x_1=0$. }
    \label{tab:5}
\end{table}
\section{Conclusions}\label{sect-4}
In this section, we summarise the results of this research and present our final comments regarding our findings. This work focused on studying a five-dimensional Gauss-Bonnet with a scalar field coupled to the Gauss-Bonnet term via two different coupling functions. We derived the coupling term $GB_{5D}= 8\dot{a}^3\left(f(\phi)+4a\dot{\phi}\frac{df}{d\phi}\right)$ after integrating by parts. Then we considered three specific models depending on the choice of the coupling function, say 
\begin{enumerate}
    \item $GB_{5D}=0,$ for $f(\phi)=0,$
    \item $GB_{5D}=-8 \alpha \dot{a}^3$,  for $f(\phi)=-\alpha$, 
    \item $GB_{5D}=-8 \alpha  \dot{a}^3 \left(\phi  \dot{a}+4 a \dot{\phi}\right)$,  for $f(\phi)=-\alpha \phi$. 
\end{enumerate}
In the first case, there is no contribution from the Gauss-Bonnet term. The contribution to the field equations is not trivial in the remaining two scenarios, namely constant and linear coupling functions. To begin, in section \ref{sect-2}, we presented a brief review of the tools from the theory of dynamical systems that were used in this research. In section \ref{sect-3}, we derived the field equations for the model using established techniques like integration by parts and the variation method of the point-like Lagrangian obtained from the gravitational action integral.  We explained the reasoning behind choosing the lapse function $N=1$ and briefly discussed the implications of setting the normalisation variable $\chi=0.$  We showed that if that occurs, the models become trivial therefore, in our subsequent analyses, we worked on the assumption that $\chi>0$. We considered a quintessence scalar field, and a phantom scalar field is left for future work.
In section \ref{sect-3-0}, we studied the dynamics of the five-dimensional scalar field model with no contribution of the Gauss-Bonnet term; this means that the coupling function was set to $f(\phi)=0.$  However, the dynamics of this model are described by a straightforward one-dimensional dynamical system with two super-collapsing solutions and one scaling solution as equilibrium points. We formulated Theorem \ref{theo-0} to show that the previously mentioned equilibrium solutions are the late-time attractors for the model and that the $\lambda$ parameter can be chosen so that the only late-time attractor is the scaling solution $C.$
In section \ref{sect-3-1}, we chose the constant coupling function to derive the field equations of the first model studied. We also wrote the Friedmann equation and defined the deceleration parameter. In the following section, section \ref{sect-3-1-1}, we defined $\chi-$normalised variables to derive the first four-dimensional dynamical system. Using Friedmann's equation, we obtained constraints for the new variables that were used to reduce the dimension of the dynamical system. The analysis of the two-dimensional system \eqref{eq-x1-reduced}-\eqref{eq-x4-reduced} showed that this model has equilibrium points that describe super-collapsing scenarios like $P_{1,2,3,4}$ for which the deceleration parameter is $q=3$,  as well as two scaling solutions $P_{5,6}$ for which $q=-1+\frac{3\lambda^2}{4}$.  We formulated Theorem \ref{teo-2} to show that the model has late-time attractors. Furthermore, by restricting the parameter $\lambda$,  only the scaling solution $P_5$ can be the late-time attractor. On the other hand, the other scaling solution $P_6$ is the early time attractor. In section \ref{app-1}, we considered an alternative formulation of system \eqref{eq-x1-reduced}-\eqref{eq-x4-reduced} that takes advantage of the geometrical region described by the constraints on the $x_i$ variables. We defined the new system \eqref{dy-sys-1}-\eqref{dy-sys-2} for the new variables $\theta_1=\arctan \left(\frac{x_2}{x_1}\right)$, $\theta_2=\arctan \left(\frac{x_4}{x_3}\right)$ that represent the poloidal and toroidal directions of a torus defined by the parametrization $x=(R+r\cos(\theta_1-\pi))\cos(\theta_2-\pi),\quad y=(R+r\cos(\theta_1-\pi))\sin(\theta_2-\pi)$,  and $z=r\sin(\theta_1-\pi)$. We showed that under this formulation, the two-dimensional dynamics of system \eqref{dy-sys-1}-\eqref{dy-sys-2} can be depicted on the surface of the previously mentioned torus.
For the other choice of coupling function, in section \ref{sect-3-2}, we derived the field equations of the second model, wrote Friedmann's equation, and redefined the deceleration parameter. In section \ref{sect-3-2-1}, we used the same normalisation variable as before and derived the five-dimensional dynamical system for the model. Using the constraint, we reduced the dimension of the system to finally obtain the three-dimensional dynamical system \eqref{reduced_linear_eq_x1}-\eqref{reduced_linear_eq_x5}. The analysis of the system showed once again that the model has equilibrium points that describe super-collapse that is $Q_{1,2,3,4}$ because $q=3$ and scaling solutions $Q_{5,6}$ since we verify that in this case $q=-1+\frac{3\lambda^2}{4}$. However, for this choice of coupling function, we also obtained three families of equilibrium points. In particular, the family $F_1$ describes a de Sitter solution because $q=1$.  We formulated Theorem \ref{teo-3} to how all the possible attractors for the model but in particular, we can exclude the super-collapse point as late time attractors by setting $\lambda$ in one of the two intervals $-\frac{4}{\sqrt{3}}<\lambda<0$ or $0<\lambda<\frac{4}{\sqrt{3}}$ as stated in corollary \ref{corol-2}. In section \ref{app-2}, we considered a projection to a fixed value of $x_5=\mu$ since we obtained three equilibrium point families that depend on $x_{5}$ in the previous section. To better observe the dynamics, we derived a new two-dimensional system \eqref{projection-1}-\eqref{projection-2} and studied the stability of its equilibrium points numerically for the values $\mu=1,2$ and different values of $\lambda.$ We verified the behaviour described in section \ref{sect-3-2-1} in these two projections.
Comparing our results to those obtained in \cite{Millano:2023czt,Millano:2023gkt} where the authors studied a four-dimensional Gauss-Bonnet cosmology with a linear coupling function and a quintessence scalar field. The authors obtained equilibrium points with $\omega_{\phi}=-\frac{1}{3}$ corresponding to $q=0$ and points with $\omega_{\phi}=q=-1$,  where $\omega_{\phi}$ is the effective equation of state parameter. They also considered a general exponential coupling function for which they obtained scaling solutions with $\omega_{\phi}=\frac{1}{3}(\lambda^2-3)$ corresponding to $q=\frac{1}{3}(\lambda^2-2)$.  As stated before, scaling solutions are present in our model for both choices of the coupling function; we also have the Family $F_1$, which is a late-time de Sitter attractor. It is clear that the dimension of the background space plays an important role in the evolution of the cosmological parameters. 

The main advantage of considering a five-dimensional Gauss-Bonnet model over a four-dimensional one is that in four dimensions, the Gauss-Bonnet term alone does not contribute to the field equations since it can be written as a total derivative. In \cite{Millano:2023czt,
Millano:2023gkt} the authors had to consider a linear coupling function $f(\phi)=f_0\phi$ and an exponential one $f(\phi)=f_0e^{\zeta \phi}$ so that the Gauss-Bonnet term contributed to the field equations. In the five-dimensional case, because it is not a topological invariant, the Gauss-Bonnet term contributes to the field equations irrespective of the nature of the coupling term, which can be settled to one. However, we considered the constant case $f(\phi)=-\alpha$ and (to compare with \cite{Millano:2023czt, Millano:2023gkt}) we also included the linear coupling $f(\phi)=-\alpha \phi$. This means that we were able to obtain relevant information on the evolution of the Gauss-Bonnet cosmological model with simpler versions of the coupling functions compared to the ones used in the four-dimensional case. The main difference between the three models studied in this research is that, in the first model, where there is no contribution from the Gauss-Bonnet term, only one scaling solution and two super-collapsing solutions appear. For the second model, the constant coupling function case, we obtained additional super-collapse solutions, two scaling solutions, and two decelerated solutions. Finally, in the third model, namely the linear coupling function case, we obtained an additional accelerated de Sitter solution together with the scaling, super-collapsing, and decelerated equilibrium points.

We can draw the following conclusions by comparing our five-dimensional model with some relevant works of four-dimensional Gauss-Bonnet. In \cite{Ivanov:2011vy}, the authors use the standard $H$-normalisation leading to a dynamical system of seven ordinary differential equations. They study some particular cases and focus on showing that they have a de Sitter solution. They also consider only a power-law coupling function for the Gauss-Bonnet term while including standard matter in the model.
In \cite{Dialektopoulos:2022kiv}, the authors also include standard matter in their model but only consider positive values of the parameter $\lambda$ in the exponential potential. They use $H$-normalisation and work with a five-dimensional dynamical system. One change is the choice of coupling function as they select an exponential one. They have decelerated, scaling, and de Sitter equilibrium points.
In \cite{Lohakare:2023ocb}, the authors consider an interaction function between the Ricci scalar and the Gauss-Bonnet term defined by $F(R, G)=\alpha R^2G^{\beta}.$ They define a dynamical system of seven equations and, using some relations, reduce the system's dimension to four. They have accelerated, de Sitter, and scaling equilibrium points.

In our work, another advantage of working in five-dimensional gravity together with the choice of normalisation variables is that the number of dimensionless variables and the number of differential equations is reduced. This means we can always display the dynamics of the models using phase plots since every dynamical system can be reduced to one, two, or three ordinary differential equations. Additionally, in our study, we can obtain decelerated, de Sitter, and scaling equilibrium without the need to include standard matter in the gravitational action. We use constant and linear coupling functions for the scalar field and Gauss-Bonnet term, as the five-dimensional contributions do not require more complicated definitions for $f(\phi)$.
The models studied here showed that the scaling solutions can describe the universe's early or late time acceleration phase of the universe; this means that this model can be used to describe inflation and as a dark energy model.
\section*{Funding}
G.L., A.D.M., and A.P. thank the financial support of ANID through Proyecto Fondecyt Regular 2024,  Folio 1240514, Etapa 2024 and of Vicerrectoría de Investigación y Desarrollo Tecnológico (VRIDT) at Universidad Católica del Norte (UCN). VRIDT-UCN funded G.L. through Resolución VRIDT No. 026/2023, Resolución VRIDT No. 027/2023, Proyecto de Investigación Pro Fondecyt Regular 2023 (Resolución VRIDT N°076/2023) and Resolución VRIDT N°09/2024. A.D.M. was supported by Agencia Nacional de Investigación y Desarrollo---ANID  Subdirección de Capital Humano/Doctorado Nacional/año 2020 folio 21200837, Gastos operacionales proyecto de Tesis/2022 folio 242220121 and VRIDT-UCN. C. M. was supported by ANID Subdirección de Capital Humano/Doctorado Nacional/año 2021- folio 21211604.  A.P. acknowledges the funding of VRIDT-UCN  through Resoluci\'{o}n VRIDT No. 096/2022 and Resoluci\'{o}n VRIDT No. 098/2022. 
\section*{Conflict of interest} The authors have no conflict of interests to declare that are relevant to the content of this article.
\section*{Data Availability} Not applicable.

\end{document}